\documentclass[twocolumn,secnumarabic,amssymb, nobibnotes, superscriptaddress, nofootinbib,aps,prd]{revtex4-2}
\usepackage[colorlinks,citecolor=blue,linkcolor=blue,anchorcolor=blue,filecolor=blue, urlcolor=blue]{hyperref}
\usepackage{natbib}
\usepackage{xcolor}

\usepackage{amsmath}
\usepackage{adjustbox}
\usepackage{graphicx}
\usepackage{booktabs}       
\usepackage{amsfonts}       
\usepackage{amsmath}
\usepackage{nicefrac}       
\usepackage{microtype}
\usepackage{hyperref}
\usepackage[capitalise]{cleveref}
\usepackage{comment}

\usepackage{nicematrix}
\usepackage{bigdelim}

\crefname{figure}{Fig.}{Figs.}
\Crefname{figure}{Fig.}{Figs.}
\usepackage{xcolor}

\usepackage[normalem]{ulem}

\usepackage{multirow}
\usepackage{makecell}
\usepackage{comment}
\usepackage{todonotes}
\usepackage{comment}
\usepackage{array}
\usepackage{subfig}

\usepackage{orcidlink}
\usepackage{float}
\usepackage{enumerate}
\usepackage[T1]{fontenc}

\usepackage[most]{tcolorbox} 

\newtcolorbox{mybox}[1][]{
    colback=blue!5!white,    
    colframe=blue!75!black,  
    fonttitle=\bfseries,     
    colbacktitle=blue!75!black, 
    sharp corners,           
    boxrule=0.5mm,           
    #1                       
}

\begin{document}

\title{Ultra-compact twin stars with hybrid equations of state from bosonic dark matter}

\author{Ishfaq Ahmad Rather~\orcidlink{0000-0001-5930-7179}}
\email{rather@astro.uni-frankfurt.de}
\affiliation{Institut f\"{u}r Theoretische Physik, Goethe Universit\"{a}t, 
Max-von-Laue-Str.~1, D-60438 Frankfurt am Main, Germany}

\author{Sarah Louisa Pitz~\orcidlink{0009-0007-4169-4298}}
\email{pitz@itp.uni-frankfurt.de}
\affiliation{Institut f\"{u}r Theoretische Physik, Goethe Universit\"{a}t, 
Max-von-Laue-Str.~1, D-60438 Frankfurt am Main, Germany}

\author{Jürgen Schaffner-Bielich~\orcidlink{0000-0002-0079-6841}}
\email{schaffner@astro.uni-frankfurt.de}
\affiliation{Institut f\"{u}r Theoretische Physik, Goethe Universit\"{a}t, 
Max-von-Laue-Str.~1, D-60438 Frankfurt am Main, Germany}

\begin{abstract}

    The properties of compact stars with a strong first-order phase transition to quark matter and with an additional fluid of self-interacting bosonic dark matter (DM) are studied.
    We find that the inclusion of DM changes the stability of mass-radius configurations relative to the naive one-fluid criterion. 
    For compact star configurations with similar masses and different radii, so-called twin stars, the presence of DM removes the unstable segment between the hadronic and the hybrid branch, so that the stable mass-radius sequence becomes continuous after the onset of the phase transition to quark matter. 
    We furthermore find stable ultra-compact objects (UCOs), defined by a total compactness $C = M_\text{tot}/R_\text{grav} \ge 1/3$. 
    We observe two distinct classes of UCOs: a DM-halo class with $f_\text{DM} \gtrsim 0.9$, and a DM-core class at $f_\text{DM} \lesssim 0.02$.
    The two classes can be separated by the surface redshift of the normal matter, which reaches $z = 0.734 - 0.808$ for the DM-core and extreme values of $z = 2 - 3.468$ for the DM-halo class, allowing us to use their redshift as a way to disentangle the two classes. Furthermore, the high redshift of the DM-halo UCOs could serve as a signal of DM since their redshift exceeds the Buchdahl limit of $z = 2$.
    Finally, we find hybrid star solutions of  ``ultimate twins" with similar masses and visible radii, but different dark matter content, leading to different tidal deformabilities and surface redshifts.
    Future X-ray and gravitational measurements of ultra-compact neutron stars with radii and masses outside the allowed neutron star range 
    can thereby probe the presence and the properties of DM in addition to a first-order phase transition to quark matter.

\end{abstract}

\maketitle

\section{Introduction}

Neutron stars (NSs) are the last stable configuration before the collapse to a black hole, making them the second densest objects in the (visible) universe. Although we have a rather good idea about their composition in the outer layers, the neutron star core remains a mystery. The densities in the cores can most likely reach a few times the nuclear saturation density, allowing for the possibility of a phase transition to quark matter \cite{Alford:2013aca, Glendenning:1998ag, Schertler:2000xq}. This makes neutron stars ideal natural laboratories for exploring the QCD phase diagram \cite{Annala:2019puf, Kurkela:2014vha}. Quark matter, however, is not the only scenario for matter inside a neutron star. Another possibility is dark matter (DM), whose accumulation in compact stars has been reviewed recently \cite{Grippa:2024ach}. Dark matter can be captured by the neutron star as it travels through the galaxy \cite{Barbat:2024yvi, Dengler:2021qcq, Tolos:2015qra}, a process whose rate has been computed in detail for a range of candidates \cite{Bell:2020jou, Busoni:2021zoe}. Another way to include DM in the neutron star is the decay of the neutron into a dark particle, which provides a solution to the unresolved neutron decay anomaly \cite{Fornal:2018eol}. However, both methods generate only a small fraction of dark matter \cite{Ellis:2018bkr, Sagun:2021oml, Shirke:2023ktu}. A way to reach higher dark matter fractions of around $90 \, \%$ could be achieved by an accreting boson star, i.e. a purely dark star made only of bosonic dark matter. These stars can reach extremely high compactness values \cite{Pitz:2023ejc}, which enhances their accretion rate. Once dark matter has accumulated inside the star, it can alter the macroscopic properties such as mass, radius, and tidal deformability. Particularly it can lower the radius of the neutron star, pushing it into a region of the mass-radius space that seems to violate causality \cite{Pitz:2024xvh}, meaning that the speed of sound in the hadronic interior would exceed the speed of light. If one, however, takes the dark matter radius into consideration and considers the total matter radius, the total matter distribution obeys causality. Nevertheless, this pseudo-violation of causality has important implications for observational campaigns of telescopes such as NICER ("Neutron star Interior Composition ExploreR") \cite{Riley:2019yda, Miller:2019cac, Riley:2021pdl, Miller:2021qha, Choudhury:2024xbk, Mauviard:2025dmd} as they are measuring hotspots on the surface of the neutron star, thereby determining the visible radius of the compact star.

A joint NICER-XMM-Newton analysis has provided a new analysis of PSR J1614-2230, returning a mass of $1.937^{+0.012}_{-0.013}\,M_\odot$ with an equatorial radius of $R_\text{eq}= 10.06^{+1.25}_{-0.87}$ km~\cite{Mauviard:2026gzc}, making it the most compact massive pulsar with a measured radius.
Future X-ray telescopes like ATHENA ("The Advanced Telescope for High ENergy Astrophysics") \cite{Matt:2019llr} in 2034 or eXTP ("The enhanced X-ray Timing and Polarimetry") \cite{eXTP:2018anb} in 2027 have a higher sensitivity and thus allow for observations of neutron stars with unusually small (visible) radii. The signals of gravitational waves are also sensitive to the presence of dark matter \cite{Giangrandi:2025rko}. With future detectors like the Einstein Telescope \cite{ET:2025xjr}, the number of observations will increase significantly. Compact stars could then serve as dark matter detectors, since the single-fluid equation of state of ordinary matter alone can not reach certain mass-radius configurations, in particular the two-solar-mass limit and radii below $8$ km at the same time. 

In this work, we present dark matter-admixed hybrid stars, i.e.\ neutron stars that contain a first-order phase transition to quark matter inside their cores, with a second dark matter fluid. We particularly focus on "twin stars", which represent a special case of hybrid stars, where the stars have the same mass but different radii \cite{Kampfer:1981yr, Glendenning:1998ag, Schertler:2000xq, Christian:2017jni, Christian:2023hez, Christian:2025dhe, PhysRevD.107.023012}. The detectability of twin stars, the constraints already placed on them, and the channels by which they might form have been examined in Refs.~\cite{Christian:2021uhd, Christian:2020xwz, Montana:2018bkb, Naseri:2024rby, Espino:2021adh, Haque:2026dre}. We describe quark matter using the constant speed of sound approach proposed by \cite{Alford:2013aca}. For dark matter, we use self-interacting bosonic particles with different stiffness of the EoS \cite{Pitz:2023ejc, Pitz:2024xvh, Colpi:1986ye}. 

A substantial number of papers have studied dark-matter-admixed neutron or hybrid stars, both for fermionic and bosonic dark matter \cite{Das:2025fyf, Bramante:2023djs, Barbat:2024yvi, Arvikar:2025dwl, Grippa:2024ach, Shakeri:2022dwg,  PhysRevD.105.123010, Araujo:2025tlv, sym17101669, Perez-Garcia:2010xlt, PhysRevLett.107.091301, PhysRevD.77.043515, PhysRevD.82.063531, Karkevandi:2024vov, Karkevandi:2021ygv, Koehn:2024gal, Sagun:2022ezx, Giangrandi:2022wht, Sagun:2021oml, Ivanytskyi:2019ojt, Diedrichs:2023trk, Jockel:2023rrm, Hajkarim:2024ecp, Shirke:2023ktu, Cassing:2022tnn, Dengler:2021qcq, Wystub:2021qrn, PhysRevD.92.123002, PhysRevD.93.083009, 2012PhLB..711....6P, PhysRevD.99.063015, 2012APh....37...70L, 2009APh....32..278S, PhysRevD.84.107301, PhysRevC.89.025803, Vikiaris:2026ofd, Dengler:2025ntz, Cano:2026yrl, Li:2026pxw, Thakur:2026pes}. The stability of the resulting configurations requires particular care. For a single fluid, the turning-point criterion $\partial M/\partial \varepsilon_c$ is sufficient, also for a strong enough first-order transition to satisfy the Seidov condition for producing an unstable segment separating the hadronic branch from the twin branch. With two fluids that are only gravitationally coupled, the criterion generalizes to the eigenvalues of the Jacobian of the two conserved particle numbers with respect to the two central densities \cite{Hippert:2022snq}. Earlier studies of dark-matter-admixed compact stars assess stability for each fluid separately, an approximation that ignores the response of one fluid to changes in the other; the full two-fluid
criterion has since been applied to nucleonic stars admixed with bosonic \cite{Pitz:2024xvh} and fermionic \cite{Barbat:2024yvi} dark matter. Ref.~\cite{Biesdorf:2024dor} studied dark matter in hybrid stars using the modified MIT bag model \cite{Lopes:2020dvs}, which did not produce a strong enough first-order phase transition to generate twin stars. The consistent two-fluid criterion for stability also determines how a phase transition affects the stability of the mass-radius sequence with dark matter. In the single-fluid limit, an energy-density jump above the Seidov limit forces $dM/dp^c_\text{NM} < 0$ just above $p_t$, and an unstable segment separates the hadronic from the twin branch. Dark matter contributes a second conserved particle number, so that the unstable segments vanish, leaving a single continuous stable sequence starting with the onset of the phase transition. The first-order transition is present in all these compact star configurations, i.e.\ these stable compact stars with dark matter still possess quark matter cores. 

The paper is structured in the following way: we first present the equations of state we are using to describe the normal matter and the dark matter fluids in \cref{eos}, and the two-fluid TOV equations in \cref{tov}. Then we discuss the stability criterion of two-fluid objects in \cref{stability}. Subsequently, \cref{sec:results} discusses our results for the mass-radius curves, the two-fluid tidal deformability, as well as contour plots showing the stable regions in the compactness space with the corresponding dark matter fractions. We furthermore study the gravitational redshift of photons emitted from the surface of the normal matter star. The discussion about ultimate twins, stars with the same mass and visible radius but different DM fraction, is presented in \cref{ultimate_twins}. \cref{summary} summarizes our findings. The appendix \ref{app:sos07} discusses the mass-radius and contour plot results for a softer quark matter equation of state with a speed-of-sound of $c^2_\text{QM} =0.7$.

\section{Methodology}\label{NS1}
\subsection{Equation of state}
\label{eos}
Both fluids are specified by a barotropic relation $\varepsilon(p)$ and are coupled only through the metric. We describe the two sectors in turn.

\subsection{Normal matter}
For the normal or ordinary matter (NM), we use a hybrid equation of state (EoS) where the hadronic matter (HM) is represented by a piecewise polytrope, as discussed in Ref. \cite{Kurkela:2014vha}, with the following form
\begin{equation}\label{polytropic_eos}
    \varepsilon = \frac{1}{\Gamma - 1} \, p + \left(\mu_0 n_0 - \frac{\Gamma}{\Gamma - 1}p_0 \right) \left( \frac{p}{p_0}\right)^{1/\Gamma} ,
\end{equation}
where $\mu_0$ is the chemical potential at which the two monotropes are matched, $n_0 = n(\mu_0)$ is the corresponding number density, $p_0 = p(\mu_0)$ is the corresponding pressure. $\Gamma$ is a parameter that needs to satisfy the condition $\Gamma > 1$. In this work, we are using the parameter set known as "EoS2" \cite{Kurkela:2014vha} which has a maximum mass of $M_{max} = 2.45 \, M_\odot$ at a radius of $R = 13.3$ km, resulting in a compactness of $C = M / R = 0.27$ (with $G = 1$).

The number density of the hadronic equation of state, EoS2, is given by
\begin{eqnarray}
    n_{polytrope} = \left( \frac{p_{polytrope}}{K}\right)^{1/\Gamma} ,
\end{eqnarray}
with $K = 3828.62 \text{ MeVfm}^{3(\Gamma-1)}$ and $\Gamma = 4.021$ for the first monotrope, and $K = 353.62 \text{ MeVfm}^{3(\Gamma-1)}$ and $\Gamma = 1.195$ for the second one. To get the total particle number $N$, one needs to integrate the number density over the volume times the Schwarzschild factor:
\begin{eqnarray}\label{eq:baryon_number}
    \frac{d N}{dr} = 4 \pi \left(1 - \frac{2 m}{r} \right)^{-1/2} n r^2 dr.
\end{eqnarray}
Here, $r$ is the radial coordinate and $m = m(r)$ is the mass at a given value of $r$.

For the Quark matter (QM) part, we use the constant speed of sound (CSS) approach \cite{Zdunik:2012dj, Alford:2013aca, Alford:2015gna}, which has been extensively used to describe and study the QM in NS cores \cite{Christian:2017jni, Das:2025fyf, Ranea-Sandoval:2015ldr, Pal:2025chs}. So the complete EoS for normal matter can be expressed as:
\begin{equation}
  \varepsilon(p) =
    \begin{cases}
      \varepsilon_\text{HM} (p) & p<P_t\\
      \varepsilon_\text{HM} (P_t) + \Delta \varepsilon + c_\text{QM}^{-2} (p-P_t) & p>P_t ,
    \end{cases}       
\end{equation}
where $p_t$ corresponds to the pressure at the transition point from hadronic to the quark phase, hence predicting the appearance of quark matter. $\Delta \varepsilon$ represents the jump in the energy density.  $c_\text{QM}$ represents the speed of sound of quark matter, which predicts the nature of the EoS. 

Perturbative QCD (pQCD) predicts that the speed of sound approaches the conformal limit $c_s^2 \to 1/3$ from below at asymptotically high baryon densities \cite{Fraga:2013qra, Kurkela:2014vha, Annala:2019puf}. However, at the baryon densities realised in NS cores ($n_B \sim 2$–$6\,n_0$, where $n_0 = 0.16\,\text{fm}^{-3}$), the system is far from the perturbative regime and the speed of sound is not constrained by pQCD to remain below $1/3$. Indeed, analyses of multi-messenger NS data and nuclear theory constraints consistently find that $c_s^2$ must rise above $1/3$ in the density range relevant for $2\,M_\odot$ stars \cite{Annala:2019puf, Tews:2018kmu, Bedaque:2014sqa}. The speed of sound must therefore overshoot the conformal value inside the star and relax towards it at higher density. 
 The maximum allowed value is set by causality: $c_\text{QM}^2 \leq c^2 = 1$. Setting $c_\text{QM}^2 = 1$ therefore corresponds to the stiffest causal quark-matter EoS and produces the maximum possible support against gravitational collapse, yielding the largest hybrid star radii and the highest twin-branch maximum masses. We include $c_\text{QM}^2 = 0.7$ as a softer but still super-conformal alternative, bracketing the phenomenologically interesting range. The causal case ($c_\text{QM}^2=1$) is an extremum rather than a conservative choice, since it is the most favourable value for producing compact, massive twin configurations, and therefore the possible signatures we quote are upper limits.  The $c_\text{QM}^2 = 0.7$ results serve as a consistency check as they confirm that our qualitative conclusions - twin branch, dark doppelgangers, and the ultra-compact DM-halo configuration - hold for a softer quark-matter EoS, with correspondingly reduced radii and maximum masses. The results for $c_\text{QM}^2 = 0.7$ are discussed in the Appendix.
 
 Since $c_\text{QM}^2 > 1/3$ is assumed over the whole quark branch, the EoSs must still be reconcilable with the perturbative result at higher density. We therefore tested them against the constraint of Komoltsev and Kurkela \cite{Komoltsev:2021jzg, Gorda:2022jvk}, which requires a thermodynamically consistent, causal, and stable interpolation between the central state of the maximum-mass star and the pQCD band at $\mu_B = 2.6\, \text{GeV}$ \cite{Kurkela:2009gj}. Sampling the renormalisation-scale parameter over $X \in [1/2, 2]$, all four transition points with $c_\text{QM}^2 = 0.7$ are fully compatible. For $c_\text{QM}^2 = 1.0$, the high-transition cases $(100, 500)$, $(120, 400)$ and $(120, 600)\, \text{MeV/fm}^3$, are likewise compatible, whereas the two low-$P_t$ cases are marginal, with an allowed interpolation existing only for the smaller values of $X$. 
 
 Phase transitions in the EoS destabilize the mass-radius (MR) sequence close to the transition point. The energy-density jump $\Delta\varepsilon$ at the transition pressure $P_t$ governs stellar stability. The Seidov limit defines the critical threshold \cite{Kampfer:1981yr, seidov1971stability}:
\begin{equation}\label{eq:seidov}
\frac{\Delta\varepsilon}{\varepsilon_t} = \frac{1}{2} + \frac{3}{2} \frac{P_t}{\varepsilon_t}.
\end{equation}
Energy density jumps near or above this limit cause the stellar mass to decrease as central pressure rises. This process creates a local minimum in the MR relation. If the sequence regains stability at higher pressures, twin stars form. These stars share the same mass but possess different radii. Twin stars serve as clear indicators for a first-order phase transition or a rapid crossover in the stellar core \cite{Glendenning:1998ag, Schertler:2000xq, Blaschke:2015uva, Zacchi:2016tjw, Alford:2017qgh, Christian:2017jni, Christian:2023hez, Christian:2025dhe}. 
The Seidov limit is a property of the nuclear EoS; $P_t$ and $\Delta\varepsilon$ do not change when dark matter is added, so a given model always sits at the same point of
\cref{eq:seidov}. Twin stars, on the other hand, are a property of the sequence of stars built from that EoS, and the sequence does change, because
the second fluid contributes to the gravitational field and shifts the point at which the configurations become unstable. Three things can therefore happen to a model that lies above the Seidov limit. The mass decrease that the criterion predicts at $P_t$ can weaken and vanish altogether so the sequence rises monotonically through $P_t$  leaving a single uninterrupted sequence, or the decrease can survive while the two stable branches move apart in mass, so that they share no mass between them and no twin pair exists, or, when the jump is large enough that the quark branch carries no stable configuration at all in the single-fluid limit, the second fluid can create one, so that a stable hybrid branch exists only because dark matter is present. The Seidov condition is derived for a single fluid, and with a second one present, lying above the line no longer implies that the transition destabilises the sequence. We follow all these effects as a function of the central dark matter pressure in
\cref{sec:mr}.

\subsection{Dark Matter}
For the description of dark matter, we are using the bosonic model of \cite{Pitz:2023ejc, Pitz:2024xvh}. Here, the dark matter particle is assumed to be a massive, self-interacting scalar field with a particle mass in the sub-GeV range. The equation of state is agnostic to the microphysics, enabling a more generic study. The most important parameters are the particle mass $m_b$ and the exponent of the self-interaction potential $n$:
\begin{equation} \label{eq:self_int}
    V = \frac{\lambda}{2^{n/2}} \left( \phi^* \phi \right)^{n/2} ,
\end{equation}
where $\lambda$ is the self-interaction strength and $\phi$ is the dark matter field. This potential gives a simple, analytic equation of state:
\\
\begin{equation} 
    \varepsilon_\text{DM} = \frac{n + 2}{n - 2} \, p_\text{DM} + \varepsilon_0 \left(\frac{p_\text{DM}}{\varepsilon_0} \right)^{n/2} ,
\end{equation}
\\
with the dark matter pressure $p_\text{DM}$, the dark matter energy density $\varepsilon_\text{DM}$ and the rescaling factor $\varepsilon_0 = \lambda \left( \frac{n}{2} - 1\right) \, m_b^n$. In order to study the effects of stiffness, one can vary $n$ as shown in \cite{Pitz:2023ejc}. In the following, we are going to use $n=4$ to describe a soft interaction and $n=40$ for a stiff interaction. It was shown that the stiff self-interacting case leads to stable, ultra-compact boson stars \cite{Pitz:2023ejc} and, respectively, dark-matter-admixed neutron stars \cite{Pitz:2024xvh}. The term "ultra-compact" refers to compact objects whose compactness is defined as $C = M/R$ (G = 1), with the mass of the compact object $M$ and its radius $R$, and is greater than or equal to $1/3$. 
Since the photon sphere for the Schwarzschild metric lies at $R  = 3 M$ with a corresponding minimal compactness of $C = 1/3$, we are going to assume that the ultra-compact objects we find all exhibit light-rings.
We furthermore restrict our studies to $m_b \in \{100, 300, 1000\}$ MeV since the maximum mass of the compact object scales like $M_\text{max} \propto 1/m_b$ \cite{Colpi:1986ye} and thus remains within the mass range of neutron stars.

\subsection{Two-fluid Tolman-Oppenheimer-Volkoff Equations}
\label{tov}

The Tolman-Oppenheimer-Volkoff (TOV) equations describe spherically symmetric, non-rotating compact objects in hydrostatic equilibrium. Their two-fluid generalisation, in which the components are coupled only through the metric, is the standard framework for dark-matter-admixed stars \cite{Grippa:2024ach, Leung:2022wcf}.
In order to solve them, one needs an equation of state as an input. 
Usually one fixes the pressure in the center of the star $p^c_\text{NM}$ for ordinary/normal matter and $p^c_\text{DM}$ for dark matter.
For the two-fluid case, these equations are modified due to the gravitational coupling of the fluids. The metric of a static, spherically symmetric spacetime is
\begin{equation}
ds^2 = -e^{2\nu(r)}dt^2 + e^{2\lambda(r)}dr^2 + r^2 d\Omega^2 ,
\label{eq:metric}
\end{equation}
with $e^{-2\lambda(r)} = 1 - 2\left(m_\text{NM}(r)+m_\text{DM}(r)\right)/r$.
The two-fluid TOV equations follow as:
\begin{eqnarray}
    \frac{d p_\text{NM}}{dr} &=& - \left( p_\text{NM} + \varepsilon_\text{NM} \right) \frac{d \nu}{dr},\\
    \frac{d p_\text{DM}}{dr} &=& - \left( p_\text{DM} + \varepsilon_\text{DM} \right) \frac{d \nu}{dr},\\
    \frac{d m_\text{NM}}{dr} &=& 4 \pi r^2 \varepsilon_\text{NM}, \\
    \frac{d m_\text{DM}}{dr} &=& 4 \pi r^2 \varepsilon_\text{DM},
\end{eqnarray}
with 
\begin{equation}\label{dnudr}
    \frac{d \nu}{dr} = \frac{\left( m_\text{NM} + m_\text{DM}\right) + 4 \pi r^3 \left( p_\text{NM} + p_\text{DM}\right)}{r (r - 2(m_\text{NM} + m_\text{DM}))},
\end{equation}
where $p_\text{NM}$ and $p_\text{DM}$ represent the normal matter and the dark matter pressure, respectively, with $\varepsilon_\text{NM}$ and $\varepsilon_\text{DM}$ as the corresponding energy densities. $r$ is the radial coordinate, $m_\text{NM}$ and $m_\text{DM}$ correspond to the mass of normal matter and dark matter, respectively. For all our calculations, we use natural units, i.e., $\hbar = c = 1$. For our studies, we are considering purely gravitational interactions of the fluids. 

The second fluid not only modifies the TOV equations but also the tidal deformability. For $l=2$, the tidal Love number $k_2$ is obtained from the static perturbation variable $y(r) = rH'(r)/H(r)$, integrated outwards from the centre alongside the background configuration. Two aspects require care once a second fluid is present. Both components source the perturbation: the single-fluid combination $(\varepsilon + p)\, d\varepsilon/dp$ entering the equation of $y$ is replaced by the sum $\sum_i (\varepsilon_i + p_i)\, d\varepsilon_i/dp_i$ over the two fluids \cite{Diedrichs:2023trk, Barbat:2024yvi}, and the density continutiy at the first-order transitin requires the standard matching of $y$ across the interface \cite{Postnikov:2010yn, Damour:2009vw}. Furthermore, the interior solutions must be matched to the vacuum exterior at the outer boundary of the object, $R_{\rm grav} = max(R_\text{NM}, R_\text{DM})$, with $C=M_\text{tot}/R_\text{grav}$; where the DM forms a halo, the region beyond $R_\text{NM}$ is not vacuum and no matching can be performed there. The dimensionless tidal deformability follows as $\Lambda = 2/3 k_2 C^{-5}$. We stress that $\Lambda$ characterises the response of the object as a whole, and is consequently the only quantity considered here for which the DM radius enters; the mass-radius relations and compactness maps are expressed throughout in terms of $R_\text{NM}$ and $R_\text{grav}$, respectively, where $R_\text{NM}$ is the radius accessible to electromagnetic observation.

\subsection{Two-fluid Stability Analysis}
\label{stability}

Since we are considering two fluids that only interact gravitationally, the usual one-fluid stability analysis is not applicable anymore. Instead, we are studying radial density oscillations for both dark matter and normal matter fluids \cite{Hippert:2022snq}. 
The onset of unstable modes for the two-fluid case is characterized by 

\begin{flalign}
\label{HippertStabCon}
& \begin{pmatrix}
    \delta N_\text{NM} \\
    \delta N_\text{DM}
\end{pmatrix} = \nonumber \\
& \begin{pmatrix}
    \partial N_\text{NM}/\partial\varepsilon_\text{NM}^c & \partial N_\text{NM}/\partial\varepsilon_\text{DM}^c\\
    \partial N_\text{DM}/\partial\varepsilon_\text{NM}^c & \partial N_\text{DM}/\partial\varepsilon_\text{DM}^c
\end{pmatrix}
\begin{pmatrix}
    \delta \varepsilon_\text{NM}^c \\
    \delta \varepsilon_\text{DM}^c
\end{pmatrix} = 0 ,&
\end{flalign}
where $N_\text{NM}$ is the total particle number in the star and $N_\text{DM}$ represents the total DM number, $\varepsilon^c$ denotes the central energy density of the NM or DM. \cref{HippertStabCon} has the trivial solutions $\delta \varepsilon_\text{NM}^c = \delta \varepsilon_\text{DM}^c = 0$, which are of no physical relevance. Therefore, we are going to focus on the variations of the total particle numbers. A matrix is not invertible, i.e., one cannot find an inverse matrix, when the determinant is zero, leading to:
\begin{eqnarray}\label{eqdet}
    \frac{\partial N_\text{NM}}{\partial \varepsilon_\text{NM}^c} 
    \frac{\partial N_\text{NM}}{\partial\varepsilon_\text{DM}^c} - 
    \frac{\partial N_\text{DM}}{\partial\varepsilon_\text{NM}^c} \frac{\partial N_{DM}}{\partial\varepsilon_\text{DM}^c} = 0.
\end{eqnarray}
Expressing this condition through the eigenvalues of the matrix $\kappa_\text{A}$ and $\kappa_\text{B}$, one finds that the onset of instability, i.e., satisfying \cref{HippertStabCon}, is given by having a vanishing or negative real part. We hence refer to compact objects as stable when both eigenvalues are positive. In practice, we locate the boundary as the point where the determinant of \cref{eqdet} changes sign on the  $(\varepsilon_\text{NM}^c, \varepsilon_\text{DM}^c)$ grid. For $p^c_\text{DM} \rightarrow 0$, this reduces to the single-fluid turning point, which we have verified numerically.
In order to determine the stability, we need to calculate the total DM number from \cref{eq:self_int}. This is analogous to \cref{eq:baryon_number}. The dark matter number density can be calculated from the zeroth component of the Noether current:
\begin{equation}
    j^\mu = \frac{\partial \mathcal{L}}{\partial(\partial_\mu \phi^*)} i \phi^* - \frac{\partial \mathcal{L}}{\partial(\partial_\mu \phi)} i \phi ,
\end{equation}
with the Lagrangian $\mathcal{L}$ 
\begin{eqnarray}
    \mathcal{L} = \partial^\mu \phi^* \partial_\mu \phi + m_b^2 \phi^* \phi - V ,
\end{eqnarray}
where $V$ is given by \cref{eq:self_int}. This results in the following dark matter particle number density $j^0 = n_\text{DM}$:
\begin{eqnarray}\label{eq:DM_numDens}
    n_\text{DM} &=& 2 \left( \frac{2^{n/2}}{\lambda} \left(\frac{n}{2}-1 \right)^{-1} p_\text{DM}\right)^{2/n} \\
    &\times& \left[ m_b^2 + \frac{\lambda}{2^{n/2}} \frac{n}{2} \left( \frac{2^{n/2}}{\lambda} \left(\frac{n}{2} -1 \right)^{-1} p_\text{DM}\right)^{1 - \frac{2}{n}}\right]^{1/2}.
\end{eqnarray}


\section{Results and Analysis}
\label{sec:results}

\subsection{Mass-Radius}
\label{sec:mr}

\begin{figure*}[t]
   \centering
    \resizebox{1.0\textwidth}{!}{
    \begin{tabular}{cc}
        \subfloat[]{%
            \includegraphics[width=0.5\textwidth]{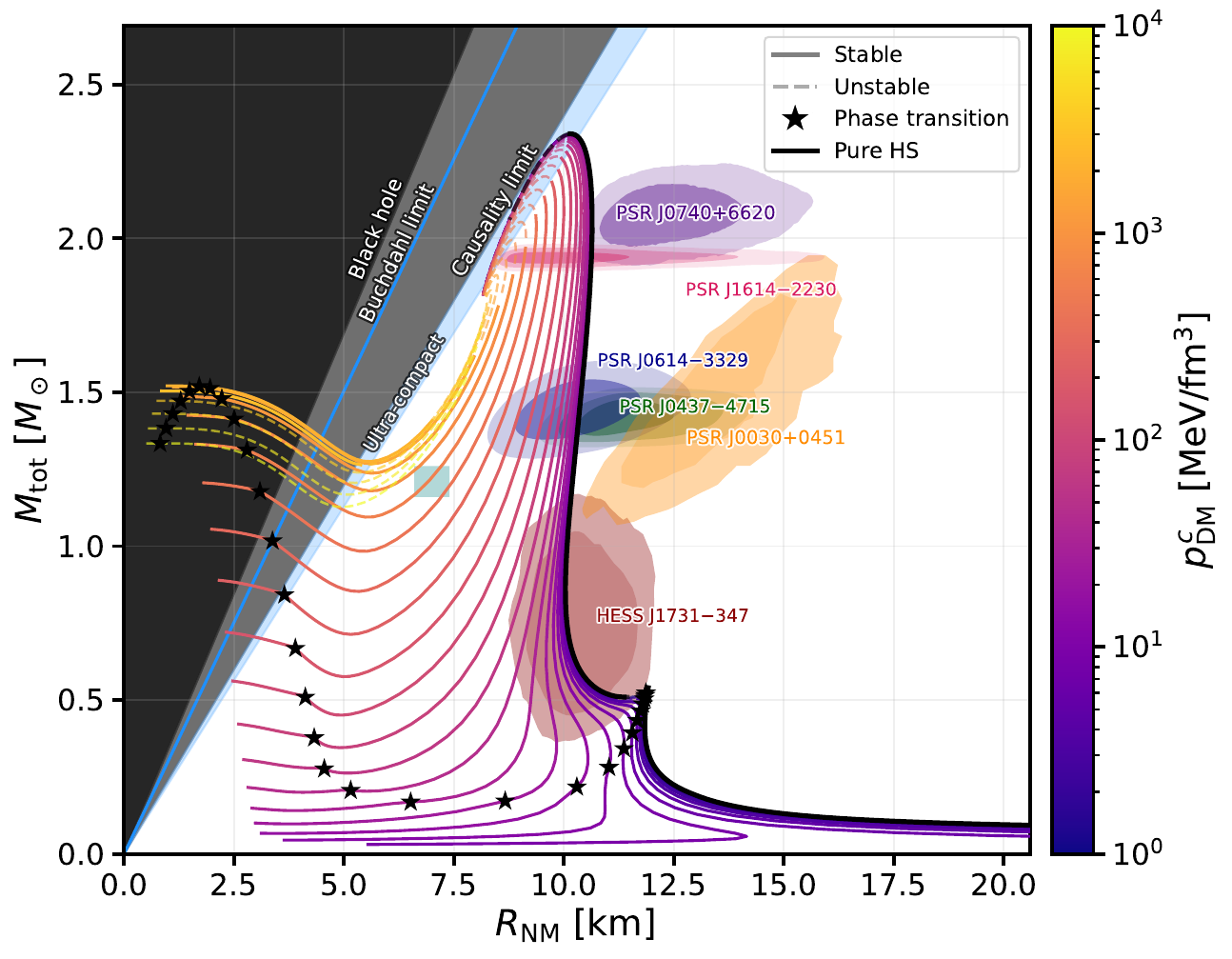}%
            \label{fig:mr_P10_mb300}%
        } &
        \subfloat[]{%
            \includegraphics[width=0.5\textwidth]{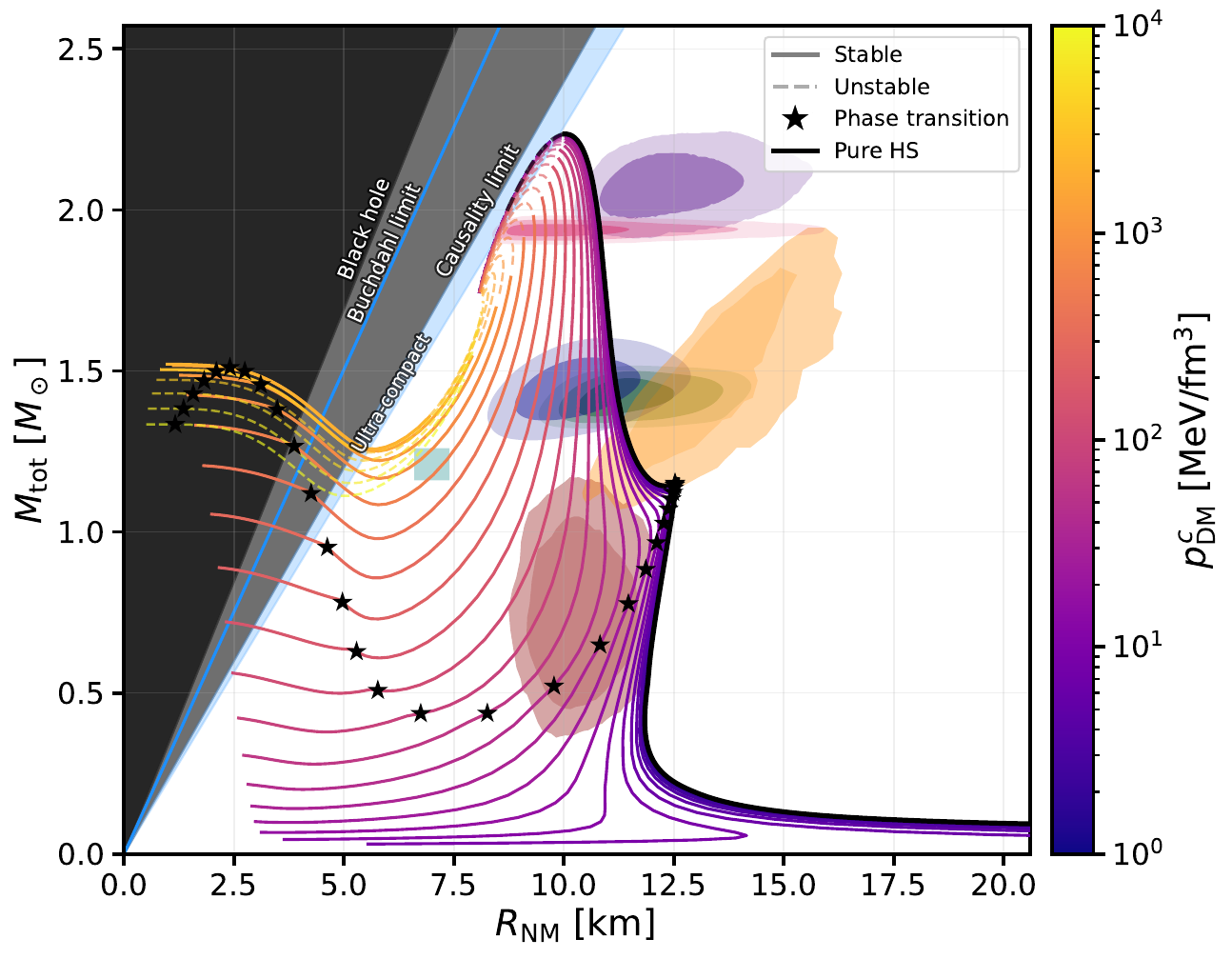}%
            \label{fig:mr_P30_mb300}%
            } \\
        \subfloat[]{%
            \includegraphics[width=0.5\textwidth]{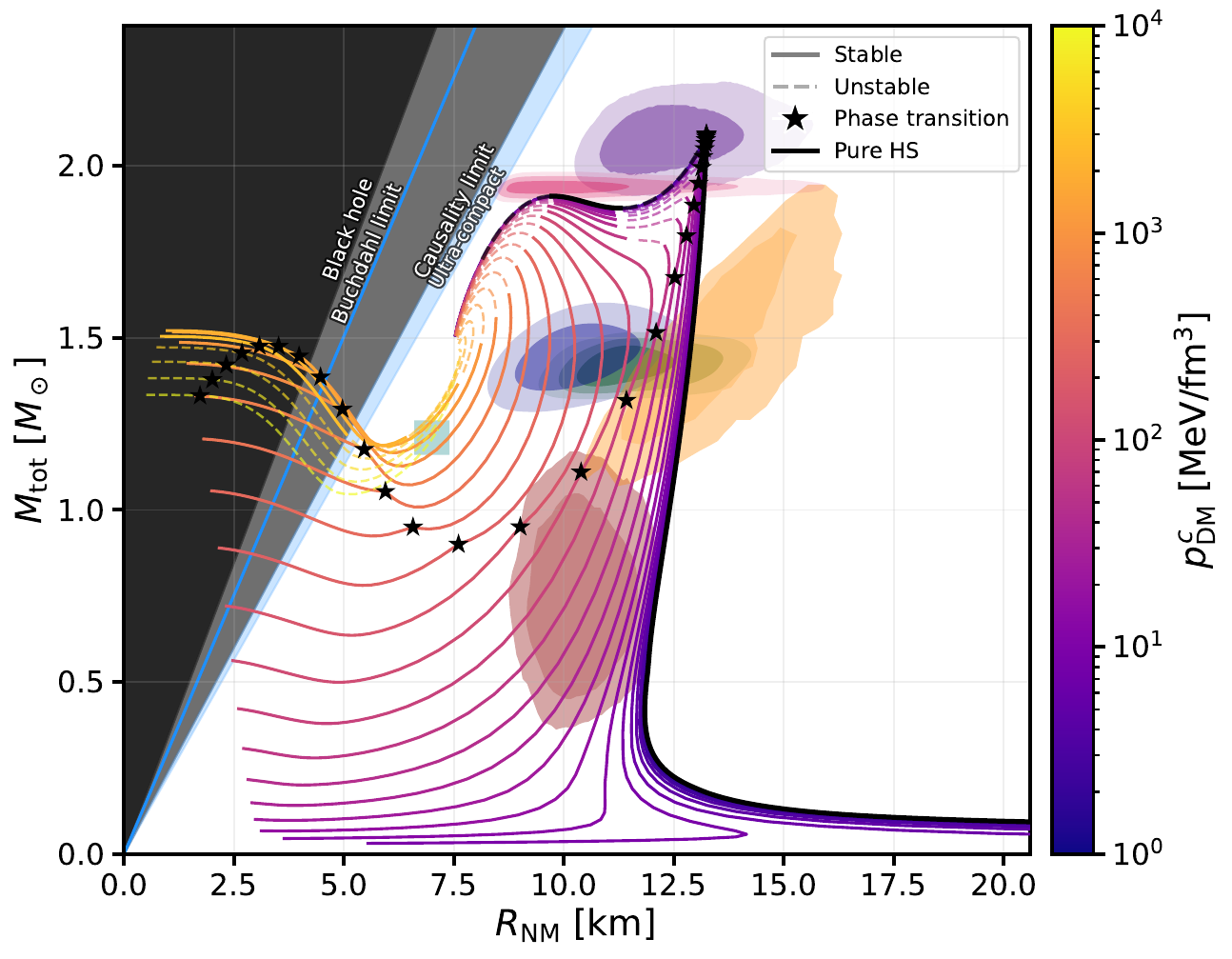}%
            \label{fig:mr_P100_mb1000}%
        } &
        \subfloat[]{%
            \includegraphics[width=0.5\textwidth]{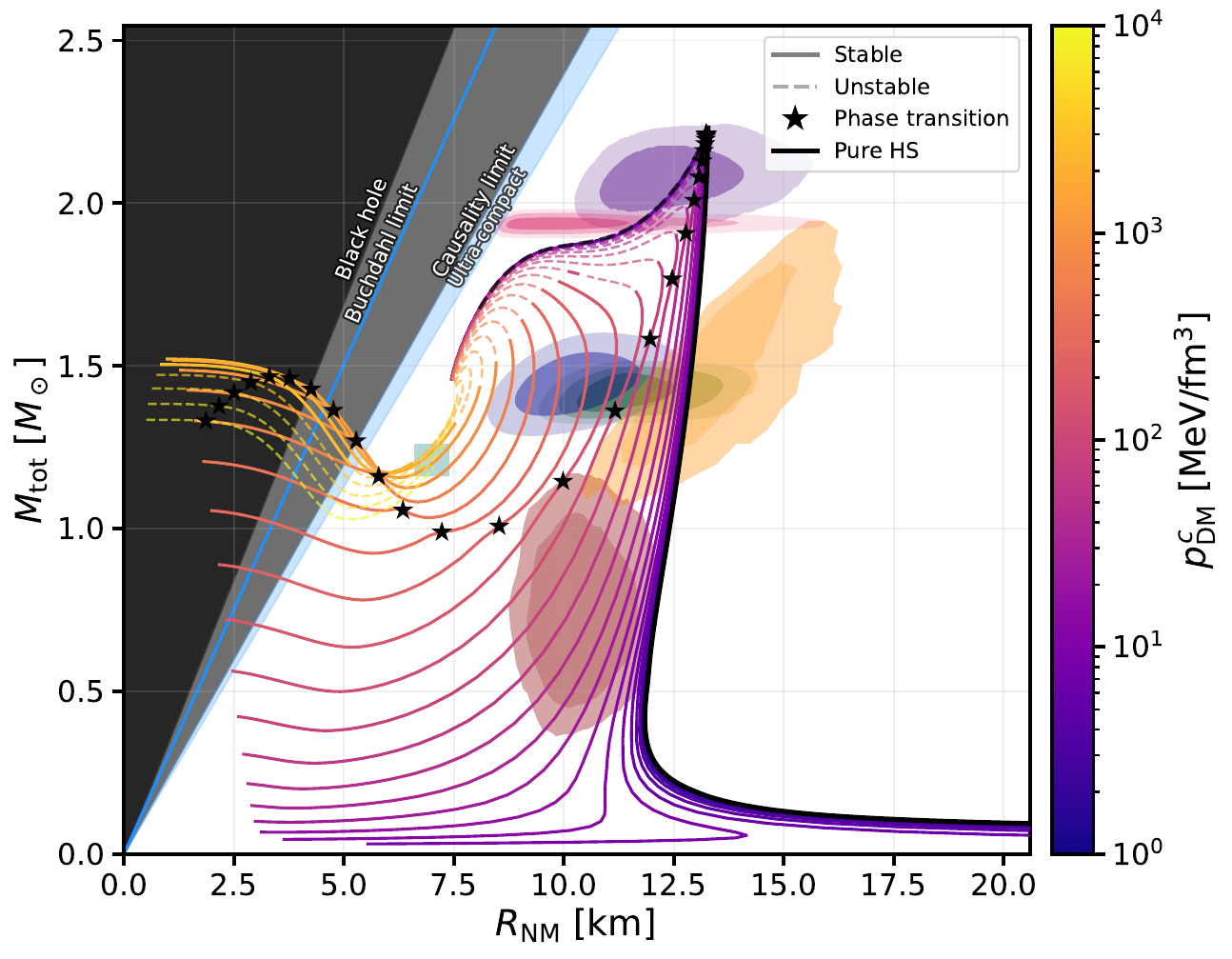}%
            \label{fig:mr_P120_mb1000}%
        }
    \end{tabular}}
    \captionsetup{justification=justified, singlelinecheck=false}
        \caption{Mass-Radius plots for hybrid/twin star configurations with $(P_t,\,\Delta\varepsilon)$ =  (a) (10, 250), (b) (30, 250), (c) (100, 500),   and (d) (120, 600) $\mathrm{MeV}/\mathrm{fm}^3$, at the DM parameters $m_b$ = 300 MeV, $n$ = 40. The colorful regions black correspond to the 68\% and 95\% confidence intervals of observational constraints from several measurements of mass and radius including the very recent PSR J1614-2230~\cite{Riley:2021pdl, Miller:2019cac, Miller:2021qha, Riley:2019yda, Mauviard:2025dmd, Choudhury:2024xbk, 2022NatAs...6.1444D, NANOGrav:2023hde, Mauviard:2026gzc} as well as XTE J1814-338 (blue rectangle) \cite{Kini:2024ggu}. The shaded regions represent the causality (grey shaded), black hole (black shaded), Buchdahl (blue line), and ultra-compact region (lightblue region).}
    \label{fig:MRcurves1}
\end{figure*}

\begin{figure*}
   \centering
    \resizebox{1.0\textwidth}{!}{
    \begin{tabular}{cc}
        \subfloat[]{%
            \includegraphics[width=0.5\textwidth]{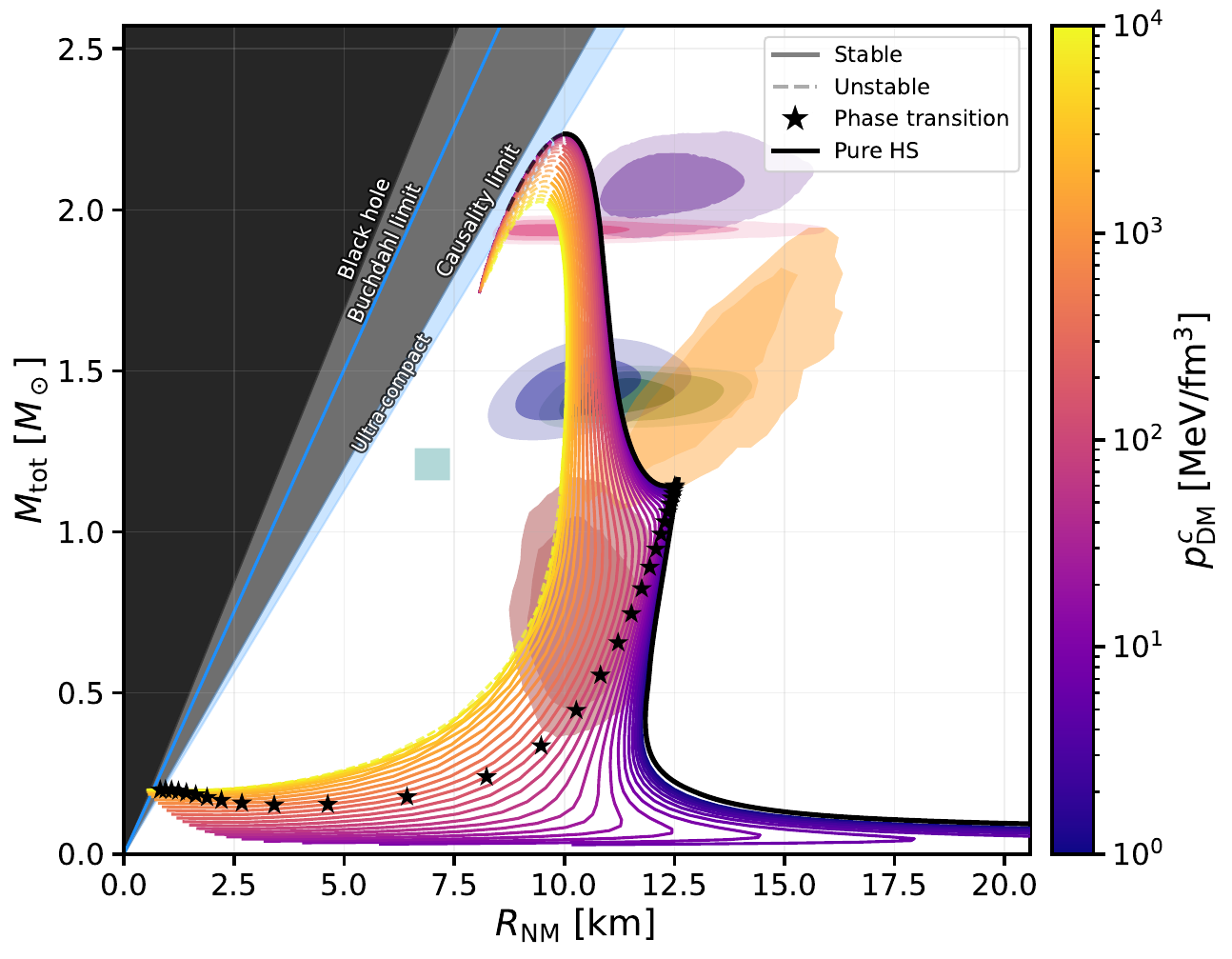}
            \label{fig:mr_P10_mb1000}
        } &
        \subfloat[]{%
            \includegraphics[width=0.5\textwidth]{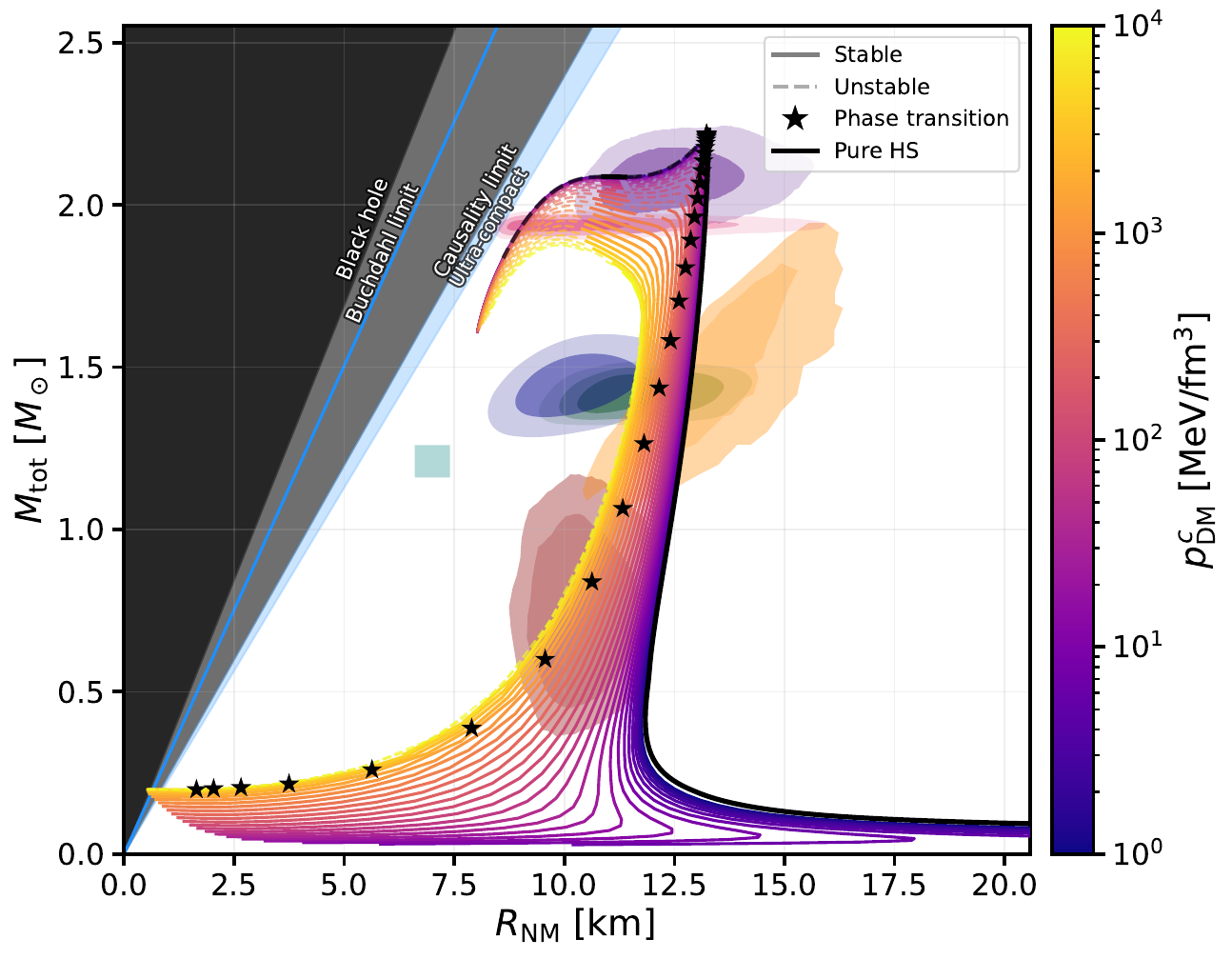}
            \label{fig:mr_P30_mb1000}
        } 
    \end{tabular}}
    \captionsetup{justification=justified, singlelinecheck=false}
    \caption{Same as \cref{fig:MRcurves1}, but for $(P_t,\,\Delta\varepsilon)$ =  (a) (30, 250), (b) (120, 400) at the DM parameters $m_b$ = 1000 MeV, $n$ = 4.}
    \label{fig:MRcurves2}
\end{figure*}

\cref{fig:MRcurves1,fig:MRcurves2} show the mass–radius (MR) sequences for the NM radius $R_{\rm NM}$ computed across the full two-dimensional grid of central pressures $(p^c_{\rm NM},\, p^c_{\rm DM})$, for two representative DM parameter sets: $(m_b,\,n) = (300\,\text{MeV},\,40)$ and $(1000\,\text{MeV},\,4)$, respectively. Each curve within a panel corresponds to a fixed $p^c_{\rm DM}$, with $p^c_{\rm NM}$ varying from sub-nuclear to supra-nuclear densities. The color scale encodes $\log_{10}(p^c_{\rm DM})$, so that the progression from dark to light traces the deformation of the sequence with increasing DM content. The two-fluid stability boundary, obtained from the vanishing determinant of the $2\times 2$ Jacobian of \cref{HippertStabCon}, divides each curve into stable (solid) and unstable (dashed) segments. 

 Panel (a) with $\Delta\varepsilon = 250\,\text{MeV/fm}^3$ at the comparatively low transition pressure $P_t = 10\,\text{MeV/fm}^3$ places this configuration well above the Seidov threshold of \cref{eq:seidov}. The hadronic branch is destabilized as soon as $p^c_{\rm NM}$ exceeds $P_t$: the gravitational mass decreases with increasing central pressure, creating a local minimum in the MR sequence. The hybrid (quark-matter) branch then re-emerges at higher $p^c_{\rm NM}$, forming the classic third-family (twin-star) branch \cite{Glendenning:1998ag, Schertler:2000xq}. For pure NM (lowest-$p^c_{\rm DM}$ curves, dark purple), the twin branch maximum mass reaches $M_{\rm max} \approx 2.2\,M_\odot$, consistent with the NICER constraints from PSR~J0740+6620 \cite{Miller:2021qha, Riley:2021pdl}, and enters the ultra-compact region. Increasing $p^c_{\rm DM}$ adds a gravitational mass without providing pressure support to the NM fluid, systematically reducing $R_{\rm NM}$ and the maximum mass of the stable sequence. At the highest $p^c_{\rm DM}$ values, the DM self-gravity is sufficient to make the NM radius arbitrarily small, producing the family of ultra-compact configurations ($C \geq 1/3$) discussed in Ref.~\cite{Pitz:2024xvh}. The two-fluid stability criterion proves more restrictive than the single-fluid counterpart throughout this regime: the Jacobian stability boundary terminates the twin branch at lower central pressures relative to the naive $dM/dp^c > 0$ criterion, a feature that has not been explored for hybrid-star EoS previously. 
 
\begin{figure*}
   \centering
    \resizebox{1.0\textwidth}{!}{
    \begin{tabular}{cc}
        \subfloat[]{%
            \includegraphics[width=0.5\textwidth]{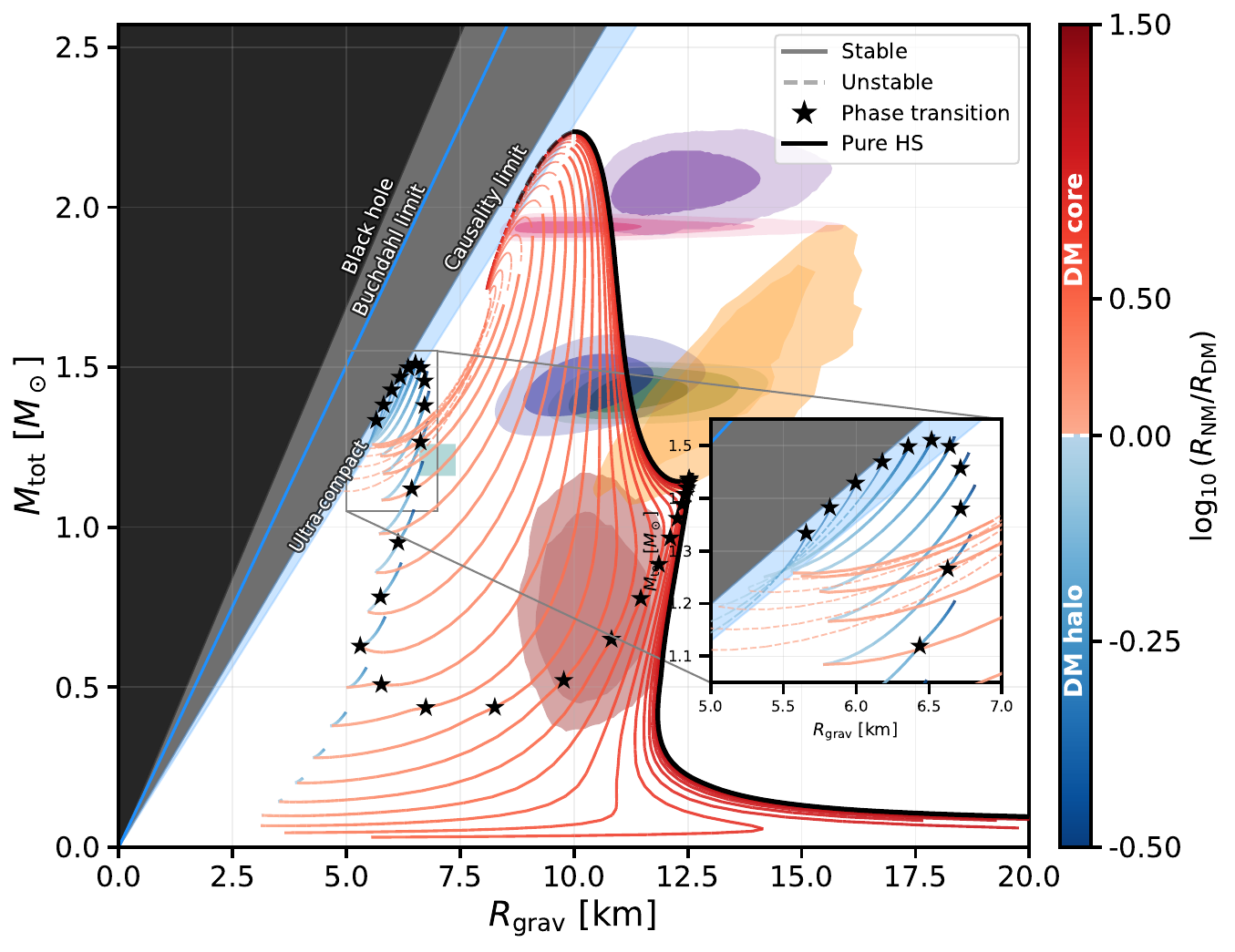}
            \label{fig:mr_total_P30_mb300}
        } &
        \subfloat[]{%
            \includegraphics[width=0.5\textwidth]{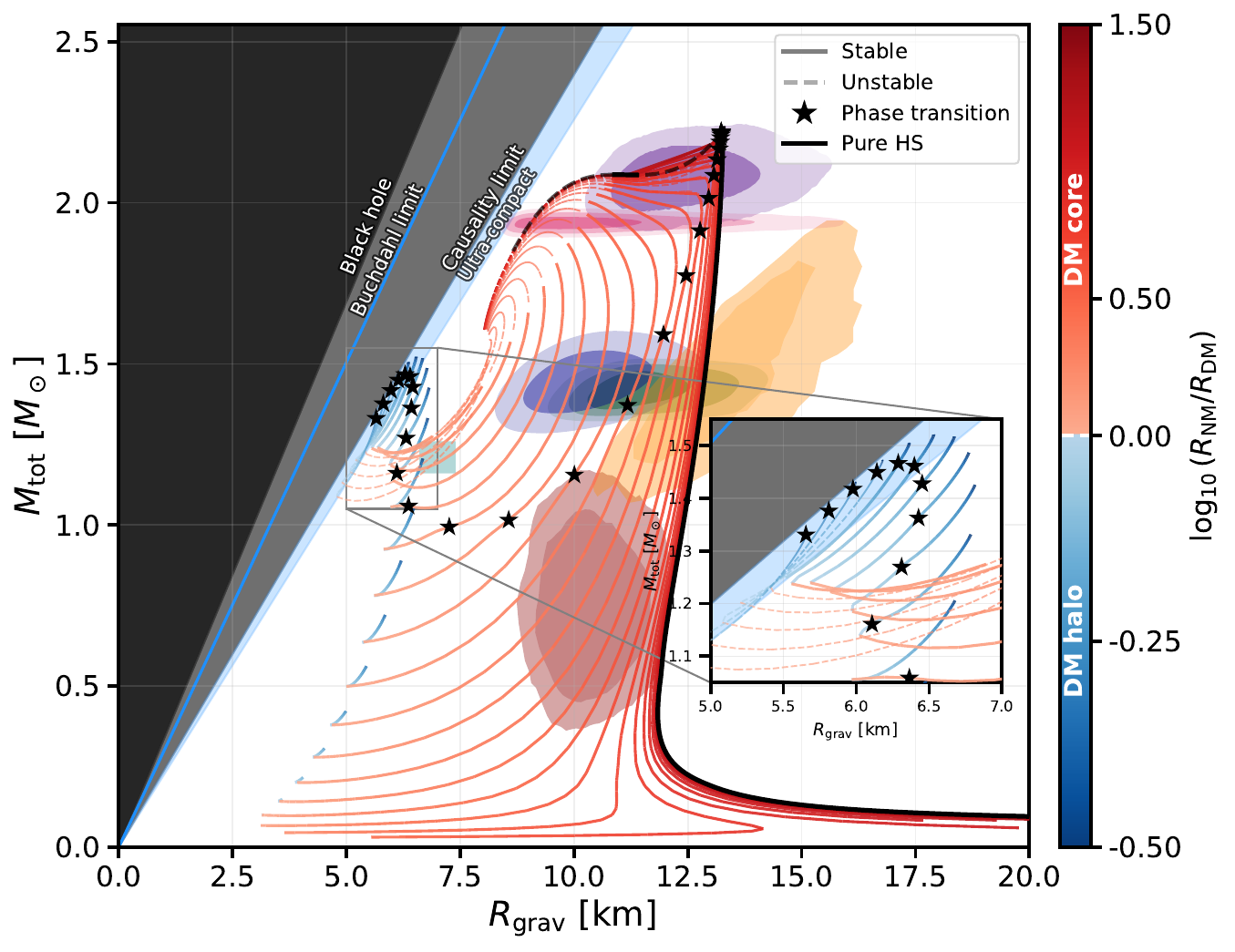}
            \label{fig:mr_total_P120_mb300}
        }
    \end{tabular}}
    \captionsetup{justification=justified, singlelinecheck=false}
        \caption{Mass-gravitational radius plot for hybrid/twin star configurations with (a) $(P_t,\,\Delta\varepsilon)$ = (30, 250) and (b) $(P_t,\,\Delta\varepsilon)$ = (120, 400) $\text{MeV/fm}^3$, both at DM parameters $m_b$ = 300 MeV, $n$ = 40. The colormap represents the ratio of radii showing the core and halo configurations. The inset shows a zoomed version of the stable ultra-compact configurations.}
    \label{fig:MRgrav}
\end{figure*}

 The curves in \cref{fig:MRcurves1} enter the black hole ($R< 2GM/c^2$) and causality-limited regions at high $p^c_{\rm DM}$. This is not a physical violation, since the radius plotted is $R_{\rm NM}$, the radius of the nuclear-matter component alone, and not the total gravitational radius of the object, $R_{\rm grav} = \max(R_{\rm NM}, R_{\rm DM})$, which governs those limits and stays outside the excluded regions for all stable solutions, as shown in \cref{fig:MRgrav}.
 We keep $R_\text{NM}$ because it is the radius accessible to electromagnetic observation and because $R_\text{NM}/R_\text{grav}$ distinguishes an NM core in a DM halo from the converse~\cite{Pitz:2024xvh}. As the transition point is changed to $(P_t,\, \Delta\varepsilon)$ =  $(30,250)$, $(100,500)$, and $(120,600)\,\text{MeV/fm}^3$, quark matter appears at progressively higher density. In panel (b), the quark phase is reached only by the more massive configurations, and the radius separation between the hadronic and the hybrid family shrinks to $\Delta R_\text{NM} \approx 0.35$ km. The two stable branches are separated by an unstable segment but do not overlap in mass at any $p^c_\text{DM}$, so this sequence is gapped rather than twinned in the sense defined below. For panel~(c), the larger $P_t$ delays the deconfinement beyond $2\,M_\odot$ and the twin star branch appears (even for lower values of $p^c_\text{DM}$) at lower mass, inside the 95\% but outside the 68\% credible region of the new mass-radius inference for PSR J1614-2230 \cite{Mauviard:2026gzc}. The unstable region between two stable branches narrows down with increasing $p^c_\text{DM}$ and eventually disappears, so the corresponding MR curve becomes continuous. Panel~(c), with $P_t = 100$, $\Delta\varepsilon = 500$ satisfies the Seidov criterion and produces a narrow twin branch in the vicinity of $M \approx 2\,M_\odot$. The extent to which the branch survives is controlled by $c_\text{QM}$: the causal quark EoS sustains it to higher masses, whereas $c_\text{QM}^2 = 0.7$ softens the branch and can eliminate it entirely for moderate $\Delta\varepsilon$, consistent with the findings of \cite{Alford:2013aca, Alford:2015gna}. 

With $\Delta\varepsilon= 600\,\text{MeV/fm}^3$ and $P_t = 120\,\text{MeV/fm}^3$, the jump relative to the transition energy density is the largest, and the hybrid branch carries no stable configurations at all at low DM content. Stable hybrid configurations appear only above $p^c_{\rm DM} \approx 17\,\text{MeV/fm}^3$, where the DM fraction is $f_{\rm DM}= 0.9\%$ and form a continuous branch rather than a disconnected one. Dark matter therefore acts in the opposite direction to the one seen in the other three panels: instead of removing the unstable segment that the transition would otherwise produce, it stabilizes the hybrid configurations that do not exist without it.

All panels satisfy the $2\,M_\odot$ lower bound \cite{Demorest:2010bx, Antoniadis:2013pzd, Fonseca:2021wxt} on the hadronic branch at low $p^c_{\rm DM}$. The DM contributes a second conserved particle number $N_{\rm DM}$ that dilutes the fractional weight of the NM discontinuity in the Jacobian determinant \cite{Hippert:2022snq}. Beyond a threshold value of $p^c_{\rm DM}$, the determinant remains non-negative across $p_t$, and the sequence passes through $p_t$ continuously, with the quark core still present. For ($P_t, \Delta\varepsilon$) = (10, 250) MeV/fm$^3$ with stiff DM, the unstable segment spans $\Delta M$ = 0.014$\,M_\odot$ at $p^c_\text{DM}$ = 1 MeV/fm$^3$, narrows to 0.013, 0.007, and 0.002$\,M_\odot$ at 1.4, 1.9, and 2.6 MeV/fm$^3$, and has closed by 3.6 MeV/fm$^3$, where the dark matter mass fraction is still only $f_\text{DM} \approx 1\%$. 

Whether the two stable branches are twins is a separate question from whether they are disconnected. We measure the mass overlap $\Delta M_\text{twin}= M^{(1)}_\text{max}-M^{(2)}_\text{min}$ between the last stable configuration of the hadronic branch and the first stable configuration of the hybrid branch. Because the Jacobian terminates the lower branch before its maximum and restarts the upper branch after its minimum, disconnection no longer guarantees mass degeneracy. For ($P_t, \Delta\varepsilon$) = (100, 500)\,\text{MeV/fm}$^3$ with stiff DM, the overlap falls from $\Delta M_\text{twin}=0.14\,M_\odot$ with ($\Delta R_\text{NM}=2.2$ km) at $p^c_\text{DM}= 1\,\text{MeV/fm}^3$ to $0.02\,M_\odot$ at 45 $\text{MeV/fm}^3$, corresponding to $f_\text{DM} \approx 2\%$; it changes sign at 62 $\text{MeV/fm}^3$, and the unstable segment closes beyond 85 $\text{MeV/fm}^3$. Each fixed-$p^c_\text{DM}$ sequence is therefore one of three kinds- \textit{twins} ($\Delta M_\text{twin}>0$), \textit{gapped} ($\Delta M_\text{twin}<0$, disconnected but no equal-mass pair), or \textit{connected} (no unstable segment)- and we use these three terms throughout. Applied to \cref{fig:MRcurves1}: panel (c) is twinned up to $p^c_\text{DM}= 45\,\text{MeV/fm}^3$ ($\Delta M_\text{twin}$ falling from 0.14 to 0.02$M_\odot$), gapped at 62 $\text{MeV/fm}^3$, and connected beyond 85 $\text{MeV/fm}^3$. Panel (d) is twinned at low $p^c_\text{DM}$ and gapped by 45 $\text{MeV/fm}^3$. Panel (b) is gapped wherever it is disconnected. For panel (a), the overlap is smaller than the mass spacing of our $p^c_\text{NM}$ grid, so its classification is not resolved, and we do not quote one. Dark matter therefore acts in two opposite directions, depending on the transition. In panels (a) to (c) it removes the twin character in two steps, closing first the mass overlap and then removing the unstable segment, so that a sequence which begins twinned or gapped ends as connected. In panel (d), it works the other way, supplying a stable hybrid branch where none exists without it. All EoSs lie above the Seidov line and stay there at every $p^c_\text{DM}$, since the nuclear EoS is unchanged, so the criterion predicts neither behaviour; nor does the margin above the line order them.  Hence, in the two-fluid case, the Seidov criterion identifies the EoSs that can produce an unstable segment, not sequences that support twin stars. Which of the two twin branches is dynamically favoured has been studied through hydrodynamic evolutions in~\cite{Haque:2026ero}, where the preference follows the comparison of binding energies.

To our knowledge, this is the first use of the new mass-radius inference for PSR J1614$-$2230~\cite{Mauviard:2026gzc} in the context of hybrid and dark matter-admixed stars. For the early-onset transition, $P_t= 10$ and $30\,\text{MeV/fm}^3$, the DM-free hybrid sequence passes through the $68\%$ credible region on its own, at $R_\text{NM}= 10.6$ and $10.8$ km at $M=1.94\,M_\odot$, so the source is reproduced without any DM and is naturally interpreted as a hybrid star with a deconfined core, or, for ($P_t, \Delta\varepsilon$) = (100, 500)\,\text{MeV/fm}$^3$, as a member of the twin branch.  For the delayed transitions, the star is still purely hadronic at this mass, lying outside the $68\%$ region. 

The heavier and softer DM model of \cref{fig:MRcurves2},  $(m_b = 1000\,\text{MeV}$, $n = 4$),  behaves quite differently. Since the maximum mass of an isolated boson star scales as $M \propto m_b^{-1}$ \cite{Colpi:1986ye}, a self-gravitating DM core is strongly suppressed, and the same range of $p^c_\text{DM}$ deforms $R_{\rm NM}$ far less; we therefore restrict the figure to two representative cases for comparison with \cref{fig:MRcurves1}. At low $p^c_{\rm DM}$, the sequences are indistinguishable from the single-fluid hybrid result, departing from it appreciably only above $p^c_{\rm DM} \sim 10^2\,\text{MeV/fm}^3$. The disparity traces back to the factor $\varepsilon_0 \propto m_b^n$ in the DM EoS, which renders the stiffness a steep function of both the boson mass and the self-interaction exponent. The twin-star structure itself derives entirely from the baryonic sector and is present for every DM parameter set considered; what differs between the two families is the extent of the sequence that is simultaneously a twin configuration and appreciably DM-admixed. With the soft DM model, the delayed transition of ($P_t, \Delta\varepsilon$) = (120, 400)\,\text{MeV/fm}$^3$ leaves the star purely hadronic at this mass, and only a dark matter core with $f_\text{DM}= 5$--$6\%$ compresses it back inside the $68\%$ region.  PSR J1614$-$2230 is therefore consistent with a hybrid star, a twin, or a dark matter-admixed star.

\cref{fig:MRgrav} replaces the NM radius with the gravitational radius $R_{\rm grav} = \max(R_{\rm NM}, R_{\rm DM})$, which defines the physical size of the compact object as seen by an external observer. It is restricted to the stiff DM model ($m_b, n$) = (300 MeV, 40), which gives the widest spread in $R_{\rm grav}$, and to the transition points (30, 250) and (120, 400) MeV/fm$^3$, representative of a well-developed and a marginal twin branch.
The colormap encodes $\log_{10}(R_{\rm NM}/R_{\rm DM})$: red-to-white tones ($R_{\rm NM}/R_{\rm DM} > 1$) indicate a DM-core geometry in which the baryonic matter extends beyond the DM sphere, while blue tones ($R_{\rm NM}/R_{\rm DM} < 1$) indicate a DM-halo in which a dilute DM cloud envelops the NM core. For $(P_t,\Delta\varepsilon) = (30,\,250)$ $\text{MeV/fm}^3$, the hadronic branch at low $p^c_{\rm DM}$ lies firmly in the DM-core regime, the baryonic component extending to ($R_{\rm NM} \sim 10$–$13$~km) while the DM remains confined well within it. As $p^c_{\rm DM}$ increases, $R_{\rm DM}$ grows and $R_{\rm NM}$ contracts simultaneously, and the configuration crosses over into the DM-halo regime; the location of this crossover is set by  $m_b$ and $n$ through the stiffness of the DM EoS. The inset resolves the ultra-compact region, $C \geq 1/3$, or equivalently $M/R_{\rm grav} \gtrsim 0.16\,M_\odot/\text{km}$, within which the stable UCO branch resides. 

Two routes to ultra-compactness operate in this model, and they populate different parts of the plane. In the first, the dark matter forms an extended halo that fixes the $R_{\rm grav}$ and supplies almost all of the gravitational mass; these configurations carry $f_\text{DM} \gtrsim 0.9$ and are the ones visible in the insets of \cref{fig:MRgrav}, where they occupy $M= 1.25$ - $1.52\,M_\odot$ at $R_{\rm grav}$ = $5.4$ - $6.7$ km with $C$ between $0.33$ and $0.35$. In the second, the compactness is generated by the baryonic sector alone; once the quark-matter branch is entered, that star contracts sharply, and a hybrid configuration can cross $C=1/3$ while the dark matter remains a subdominant core with $f_\text{DM}$ of order a per cent. This second class owes its existence to the phase transition rather than to the dark matter, and is therefore confined to early-onset transitions. In the DM-free limit, the $c^2_\text{QM}=1$ sequence with $(P_t,\Delta\varepsilon) = (10,\,250)\,\text{MeV/fm}^3$ already reaches $C=0.338$ at $M= 2.34\,M_\odot$ and $R_{\rm NM}= 10.21$ km, i.e. it is ultra-compact. Adding a small DM core does not create the population but shifts it by $\Delta C = -0.007$. Thus, a strong early-onset first-order transition with a causal quark phase produces ultra-compact configurations on its own, and these remain stable and ultra-compact when a small bosonic DM core is added; the DM-halo class, by contrast, owes its compactness entirely to the dark matter. The two populations are separated cleanly in the contour plots of \cref{contour_plots} and in the surface redshift, and we defer their discussion to those sections. 

Ultra-compact configurations of either kind possess a photon sphere at $R = 3M$, and the associated phenomenology has been studied extensively for horizonless compact objects \cite{Cardoso:2014sna, Cardoso:2019rvt}. Radiation trapped between the photon sphere and the stellar surface escapes as a train of delayed pulses following the main merger signal, and the spacing of these echoes is set by the light-crossing time of the cavity, so that it encodes the compactness directly \cite{Urbano:2018nrs, Pani:2018flj}. Pani and Ferrari \cite{Pani:2018flj} showed that ordinary hadronic stars are not compact enough for the effect, which requires the surface to lie within the photon sphere, and comparable analyses have been carried out for strange and quark stars \cite{Mannarelli:2018pjb, Kartini:2020ffp}. A spacetime with a light ring generically has a second, stable one, around which the long-lived modes accumulate and may drive a nonlinear instability on a timescale that is not well established \cite{Cardoso:2014sna, Cunha:2017qtt}; and the same objects mimic a black hole in the electromagnetic channel as well \cite{Olivares:2018abq}. A quantitative treatment of the echo signal for the configurations obtained here lies beyond the scope of this work. 
 The DM-halo morphology persists at high $p^c_{\rm DM}$ for both transition points, so it is insensitive to the phase structure of the baryonic sector.

\subsection{Radial Profiles of DM-Admixed UCOs}

\begin{figure}
    \centering
    \includegraphics[width=\columnwidth]{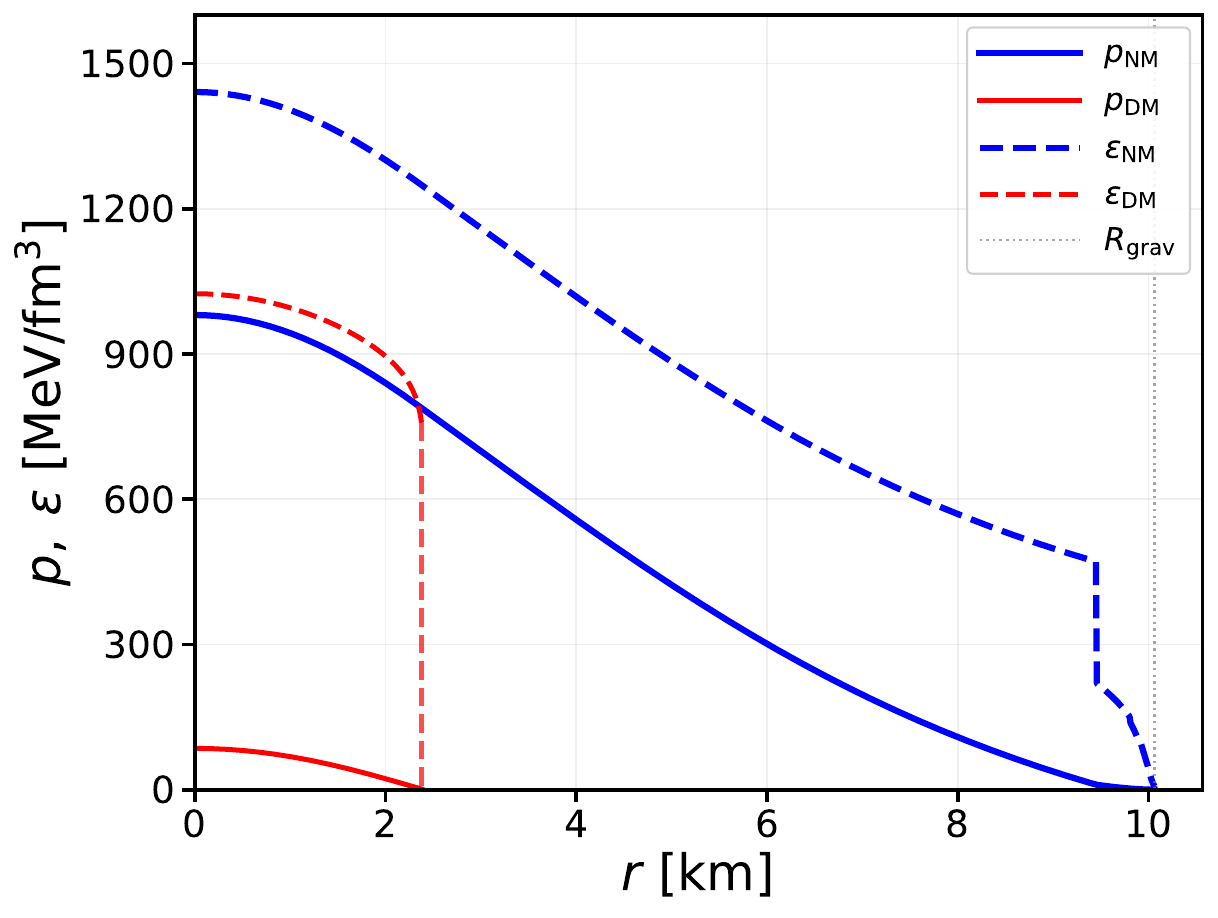}
    \includegraphics[width=\columnwidth]{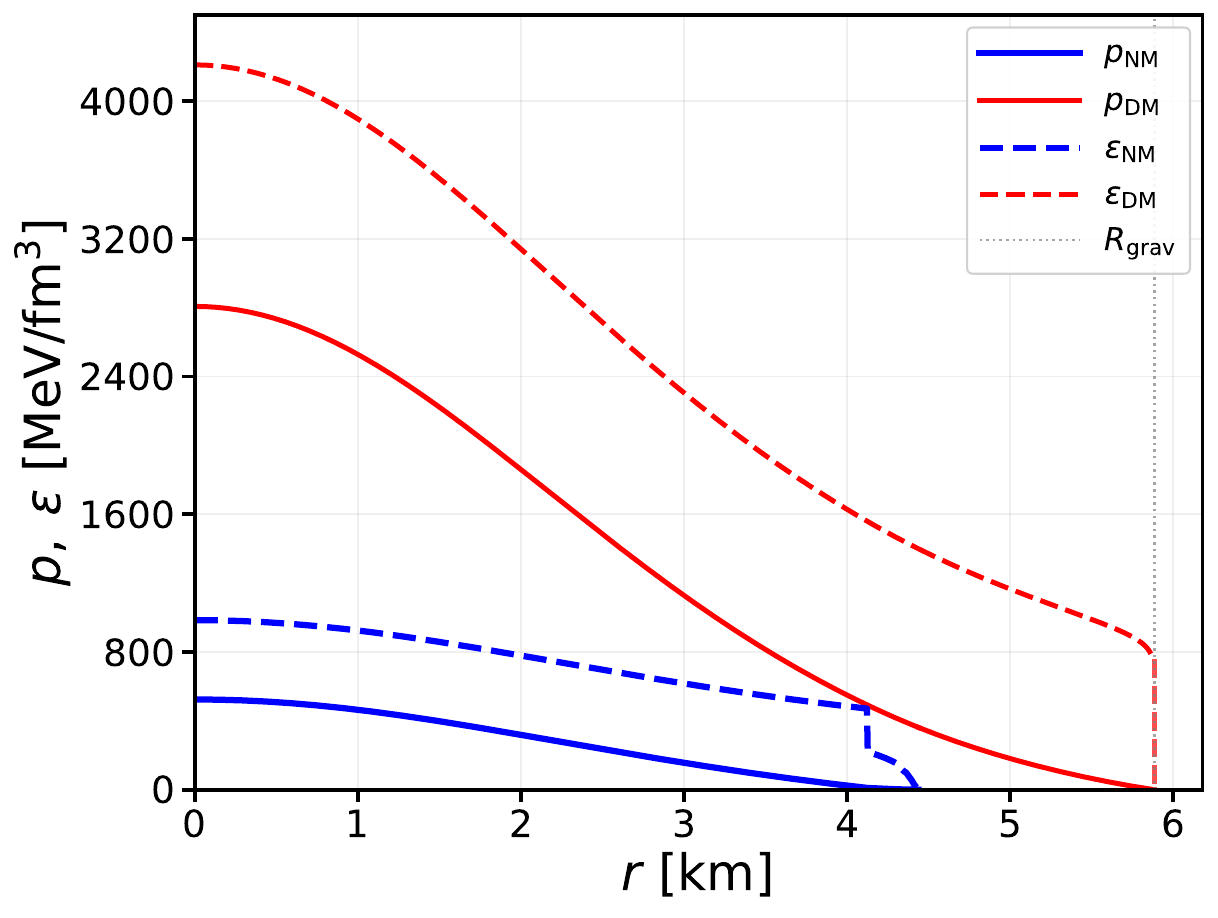}
    \captionsetup{justification=justified, singlelinecheck=false}
    \caption{Radial profiles of pressure $p$ and energy density $\varepsilon$ for both NM (blue) and DM (red) fluids in two representative UCO configurations with $(P_t,\Delta\varepsilon)=(10,250)\,\text{MeV/fm}^3$, $(m_b,n)=(300\,\text{MeV},40)$. Dashed lines represent the energy densities and solid ones the pressures. The upper panel represents the DM-core UCO ($f_{\rm DM}\approx0.02$), while the lower panel represents the DM-halo UCO ($f_{\rm DM}\approx0.87$) configuration. Vertical dotted lines mark the gravitational radius $R_{\rm grav} = \max(R_{\rm NM},R_{\rm DM})$ for each configuration.}
    \label{fig:radial_profiles}
\end{figure}

\cref{fig:radial_profiles} shows the pressure and energy-density profiles of two configurations at $(P_t,\Delta\varepsilon)=(10,250)\,\text{MeV/fm}^3$ with $(m_b,n)=(300\,\text{MeV},40)$, chosen to represent the two morphologies at comparable compactness, $C=0.336$ and $0.334$. The dark matter energy density (dashed lines) and pressure (solid lines) are depicted in red, and the normal matter in blue accordingly. In the halo configuration (lower panel), the central DM pressure exceeds the baryonic one by a factor of five, $2807$ versus $524\text{ MeV/fm}^3$, and the dark matter carries $87\%$ of the gravitational mass. The DM extends $1.45$ km beyond the baryonic surface at $R_{\rm NM} = 4.44$ km, so that $R_{\rm grav} = 5.89$ km: even in this DM-dominated case the halo is compact rather than diffuse, which is why such configurations gain compactness relative to their DM-free counterparts instead of losing it, as discussed in \cref{sec:tidal}. The DM-core configuration (upper panel) inverts the ordering, $p^c_{\rm DM}= 85$ against $980 \text{ MeV/fm}^3$; the dark matter is confined within $2.4$ km, well inside the baryonic surface, so $R_{\rm grav} = R_{\rm NM} =10.06$ km and a DM fraction of $f_{DM} = 0.02$. 
Its compactness is generated entirely by the quark matter branch. The resulting hybrid star is furthermore more massive than the one in the DM halo case: $2.28\,M_\odot$ versus $1.5 \, M_\odot$, despite the fact that it contains nearly no DM.  In both configurations, the low transition pressure places the hadron-quark interface at $93\%$ and $94\%$ of $R_{\rm NM}$, respectively, so the star is quark matter almost throughout and retains only a thin hadronic mantle.

\subsection{Tidal deformability}
\label{sec:tidal}
\begin{figure*}
   \centering
    \resizebox{1.0\textwidth}{!}{
    \begin{tabular}{cc}
        \subfloat[]{%
            \includegraphics[width=0.5\textwidth]{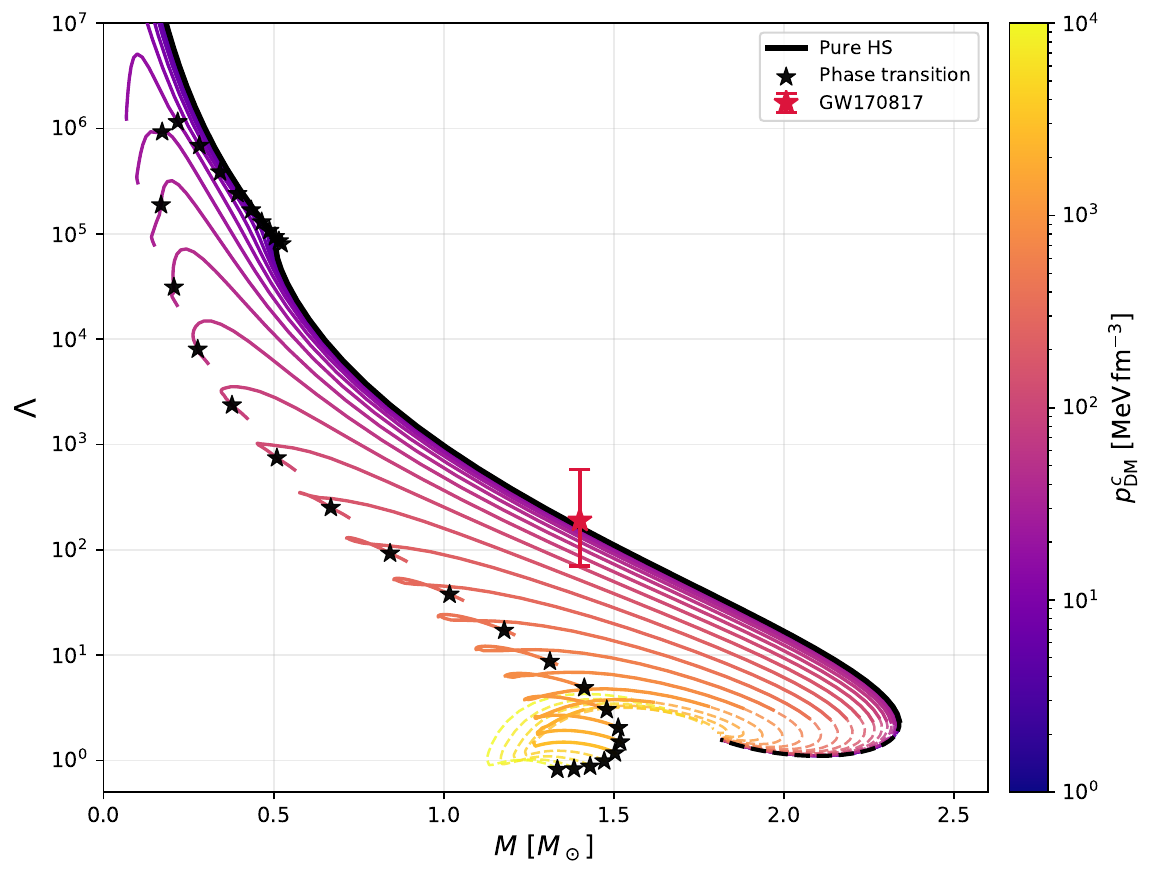}
            \label{fig:lambda_P10_mb300}
        } &
        \subfloat[]{%
            \includegraphics[width=0.5\textwidth]{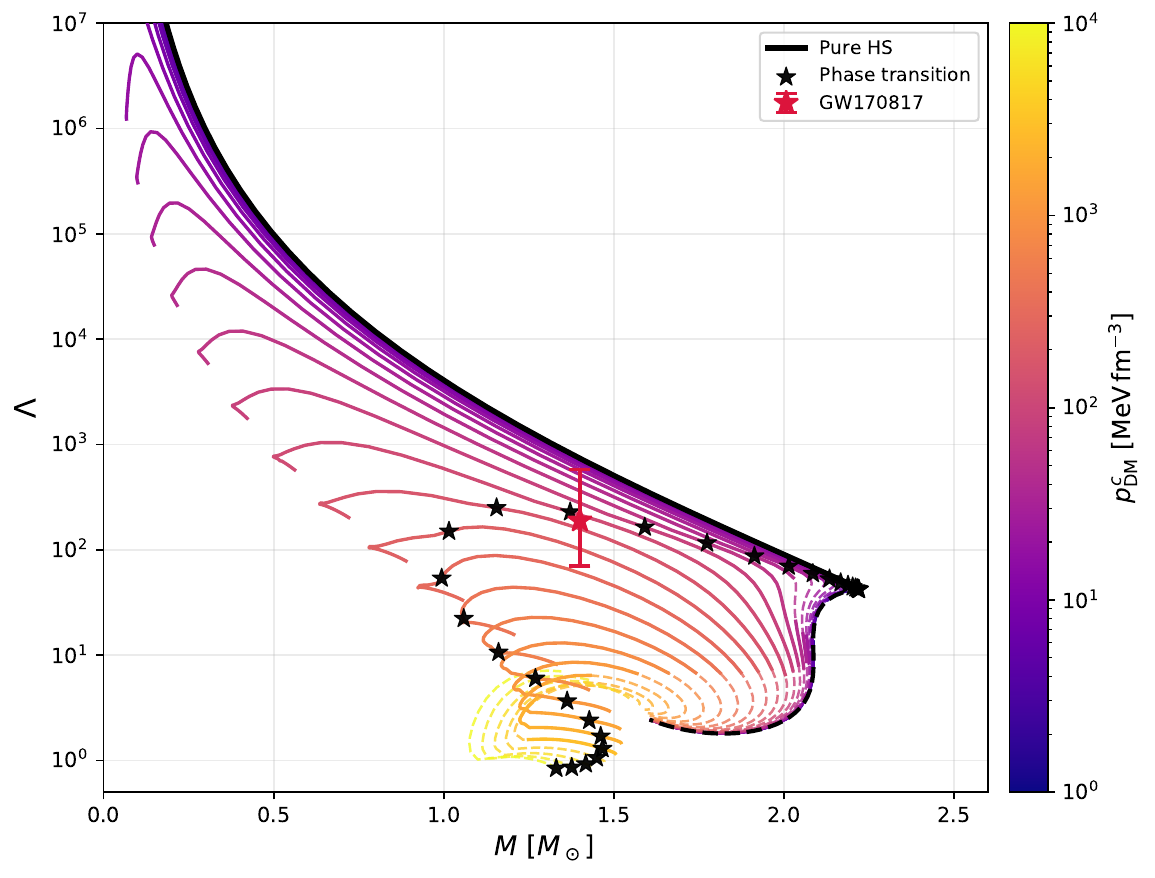}
            \label{fig:lambda_P120_mb300}
        } \\
        \subfloat[]{%
            \includegraphics[width=0.5\textwidth]{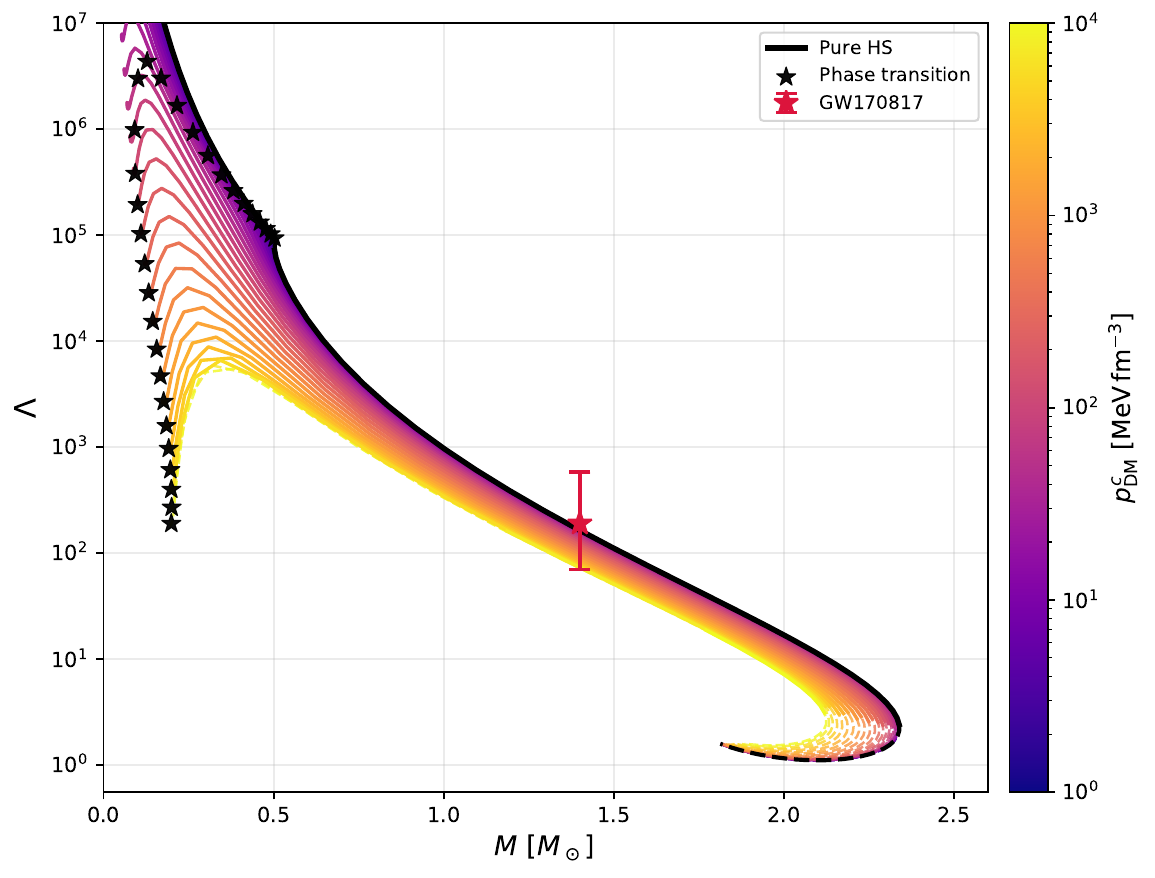}
            \label{fig:lambda_P10_mb1000}
        } &
        \subfloat[]{%
            \includegraphics[width=0.5\textwidth]{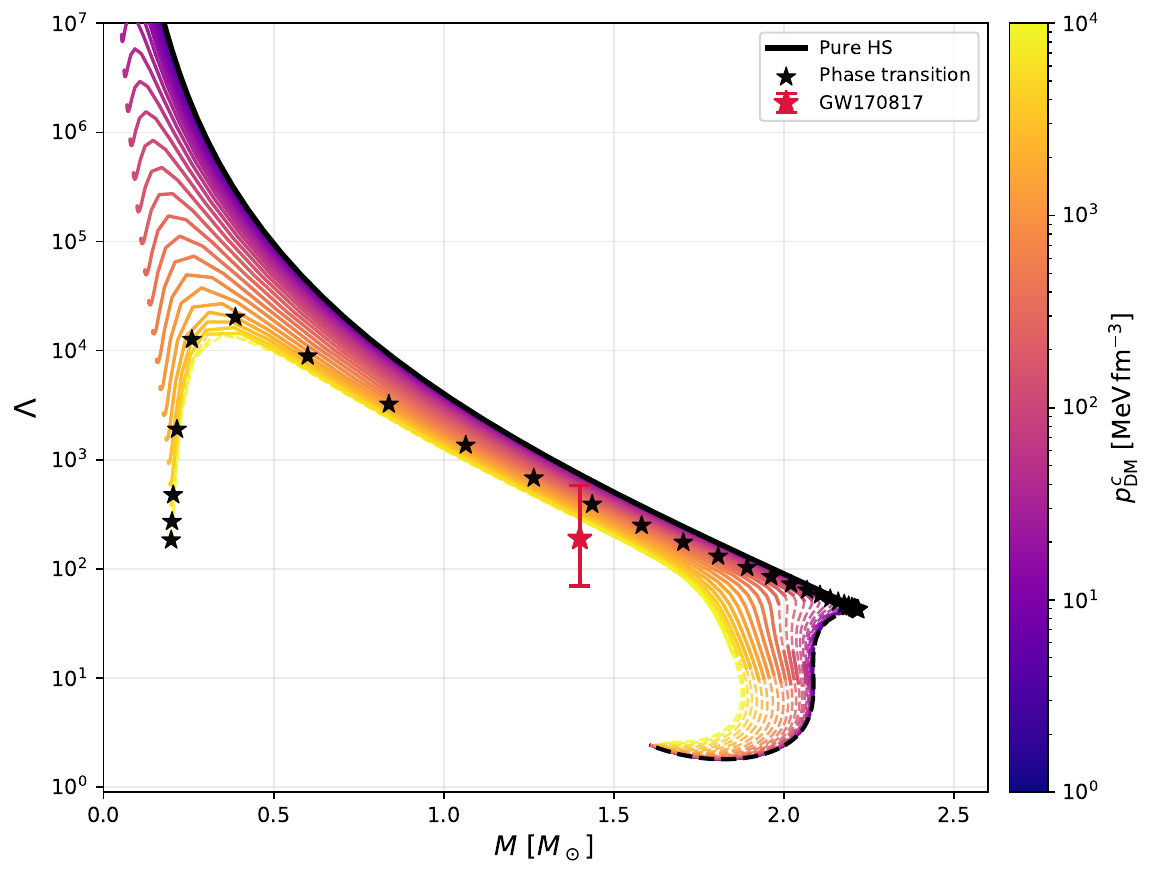}
            \label{fig:lambda_P120_mb1000}
        }
    \end{tabular}}
    \captionsetup{justification=justified, singlelinecheck=false}
        \caption{Dimensionless tidal deformability $\Lambda(M)$ against total gravitational mass $M$ for hybrid star configurations with (a,\,c) $(P_t,\,\Delta\varepsilon)$ = (10,\,250) and (b,\,d) $(P_t,\,\Delta\varepsilon)$ = (120,\,400) $\text{MeV/fm}^3$, at DM parameters $m_b$ = 300 MeV, $n$ = 40 (upper panels) and $m_b$ = 1000 MeV, $n$ = 4 (lower panels). The two phase-transition cases bracket the strongest and most marginal twin-branch $\Lambda$-drop in our grid; the two DM models bracket the stiff–soft DM range. The red star with error bars gives the tidal constraint from the GW170817 measurement $\Lambda_{1.4} = 190^{+390}_{-120}$ \cite{LIGOScientific:2017vwq, LIGOScientific:2018cki}.}
    \label{fig:lambdacurves1}
\end{figure*}

\cref{fig:lambdacurves1} shows the dimensionless tidal deformability as a function of total gravitational mass for $(P_t,\,\Delta\varepsilon)$ = (10,\,250) $\text{MeV/fm}^3$ (a,\,c panels) and $(P_t,\,\Delta\varepsilon)$ = (120,\,400) $\text{MeV/fm}^3$ (b,\,d panels). Upper panels correspond to the DM parameter ($m_b = 300$~MeV, $n=40$), while the lower panel corresponds to ($m_b = 1000$~MeV, $n=4$).  The constraint inferred from GW170817 \cite{LIGOScientific:2017vwq, LIGOScientific:2018cki} ($\Lambda_{1.4} = 190^{+390}_{-120}$) is indicated by the red marker at the canonical mass. For the stiff DM model ($m_b = 300$~MeV, $n=40$) (upper panels), the dominant feature is a pronounced drop in $\Lambda$ at the mass where quark matter first appears in the stellar core. Along the hadronic branch, $\Lambda_{1.4}$ takes values of a few hundred, in accord with GW170817, whereas at the onset of the twin branch it falls by more than an order of magnitude at fixed mass, the characteristic tidal signature of a first-order transition \cite{Han:2018mtj, Most:2018hfd}. 

Configurations on the hybrid branch are considerably more compact and correspondingly harder to deform; $\Lambda$ approaches unity near the twin maximum, and the smallest value encountered anywhere in our stable set is $\Lambda = 1.2$. 
Deconfinement sets in at a fixed central pressure along every sequence, but the mass and deformability at which it occurs vary strongly with the DM content. For panel (a), the onset mass first falls from $0.52\,M_\odot$ to $0.17\,M_\odot$ as $p^c_{DM}$ increases to $\sim 20 \text{ MeV/fm}^3$, the DM compressing the baryonic core faster than it adds to the total mass, and thereafter rises to $1.52\,M_\odot$ once the DM contribution to $M_\text{tot}$ dominates. The deformability at onset behaves similarly, rising to $\Lambda \approx 1.2 \times 10^6$ in the same interval before falling steeply to order unity at the highest DM pressures, a range of some six orders of magnitude. The tidal signature of deconfinement is therefore progressively weakened as the DM content grows, while the transition itself is untouched. Two competing effects operate as $p^c_\text{DM}$ increases. The additional DM mass raises the compactness and thereby suppresses $\Lambda$, while the altered density profile modifies $k_2$. In the halo regime, the extended DM distribution enlarges the volume coupling to the external tidal field, but the increase in compactness dominates, and $\Lambda$ falls below the DM-free hybrid value at the same total mass. Quantitatively, the ratio between each stable halo configuration and the DM-free sequence has a median of $0.024$, and none of the configurations exceeds unity. This behaviour is opposite to that reported for light fermionic DM, for which the halo is dilute and inflates $R_\text{grav}$ without contributing appreciably to the compactness, and consequently enhances $\Lambda$ \cite{Barbat:2024yvi, Nelson:2018xtr}. The origin of the discrepancy is the stiffness of the DM sector: for $(m_b, n) = (300 \text{ MeV}, 40)$ the DM component is itself close to ultra-compact, so that the halo extends to only $\approx$ 6 km while carrying the bulk of the gravitational mass, and the total compactness increases rather than decreases.

For the same transition point (10, 250) MeV/fm$^3$, lowering the boson mass to 100 MeV, so ($m_b, n$) = (100 MeV, 40), the halo at $1.4\,M_\odot$ reaches $R_\text{DM}= 29.5$ km already at $p^c_\text{DM}= 1.6\,\text{MeV/fm}^3$ and the tidal deformability rises to $\Lambda= 2.5\times 10^4$, against $\Lambda= 165$ for the same star without an extended halo. 
The suppression we find and the enhancement reported for light fermionic DM are thus the same mechanism evaluated at opposite ends of one parameter: a compact halo adds more to $M$ than to $R_\text{grav}$ and lowers $\Lambda$, a halo that is dilute does the reverse \cite{Dietrich:2020efo, Sun:2023cqr, Arvikar:2025dwl}. For the soft DM model ($m_b = 1000$~MeV, $n=4$) (lower panel), the spread in $\Lambda$ across DM sequences is much reduced, as anticipated from the weak DM deformation noted in \cref{fig:MRcurves2}. The GW170817 constraint is satisfied for all DM parameter choices at canonical mass. The twin-branch $\Lambda$-drop is present but slightly narrower in mass extent, reflecting the quantitatively unchanged (but qualitatively identical) NM phase-transition structure. This near-degeneracy between the DM-free and DM-admixed $\Lambda(M)$ curves for the heavy, soft DM model highlights an important observational limitation: gravitational-wave measurements alone cannot distinguish DM-admixed from pure hybrid stars when the DM effect on $\Lambda$ is sub-percent. Multi-messenger campaigns combining precise NICER radius measurements with GW tidal data are required to break this degeneracy \cite{Essick:2019ldf, Capano:2019eae, Dietrich:2020efo}. Whether an imprint of this size survives into the merger waveform has begun to be addressed with numerical-relativity simulations of dark-matter-admixed binaries \cite{Giangrandi:2025rko}.

\subsection{DM Fraction and Core-Halo Contour Maps}
\label{contour_plots}

\begin{figure*}
   \centering
    \resizebox{1.0\textwidth}{!}{
    \begin{tabular}{cc}
        \subfloat[]{%
            \includegraphics[width=0.5\textwidth]{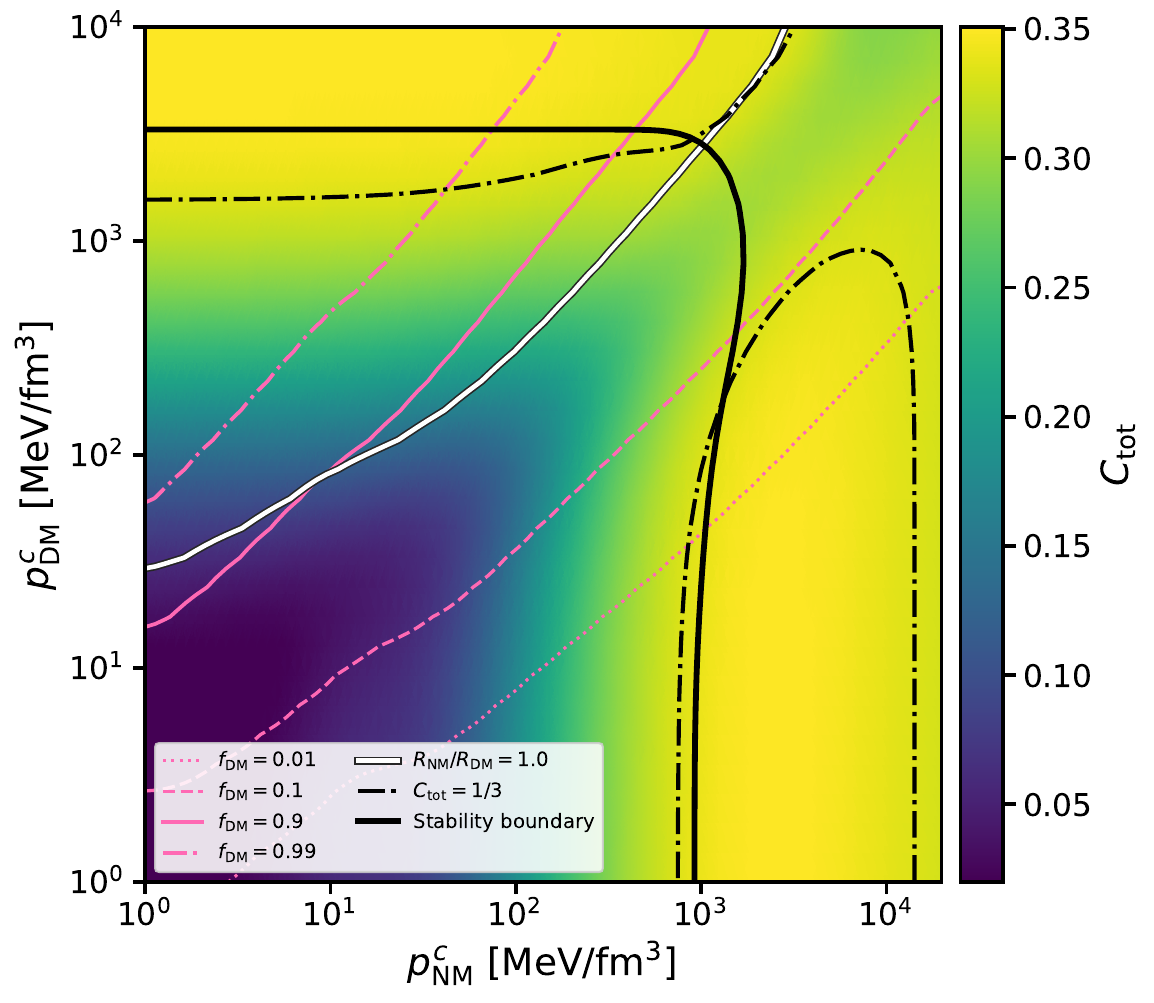}
            \label{fig:fdm_P10_mb300}
        } &
        \subfloat[]{%
            \includegraphics[width=0.5\textwidth]{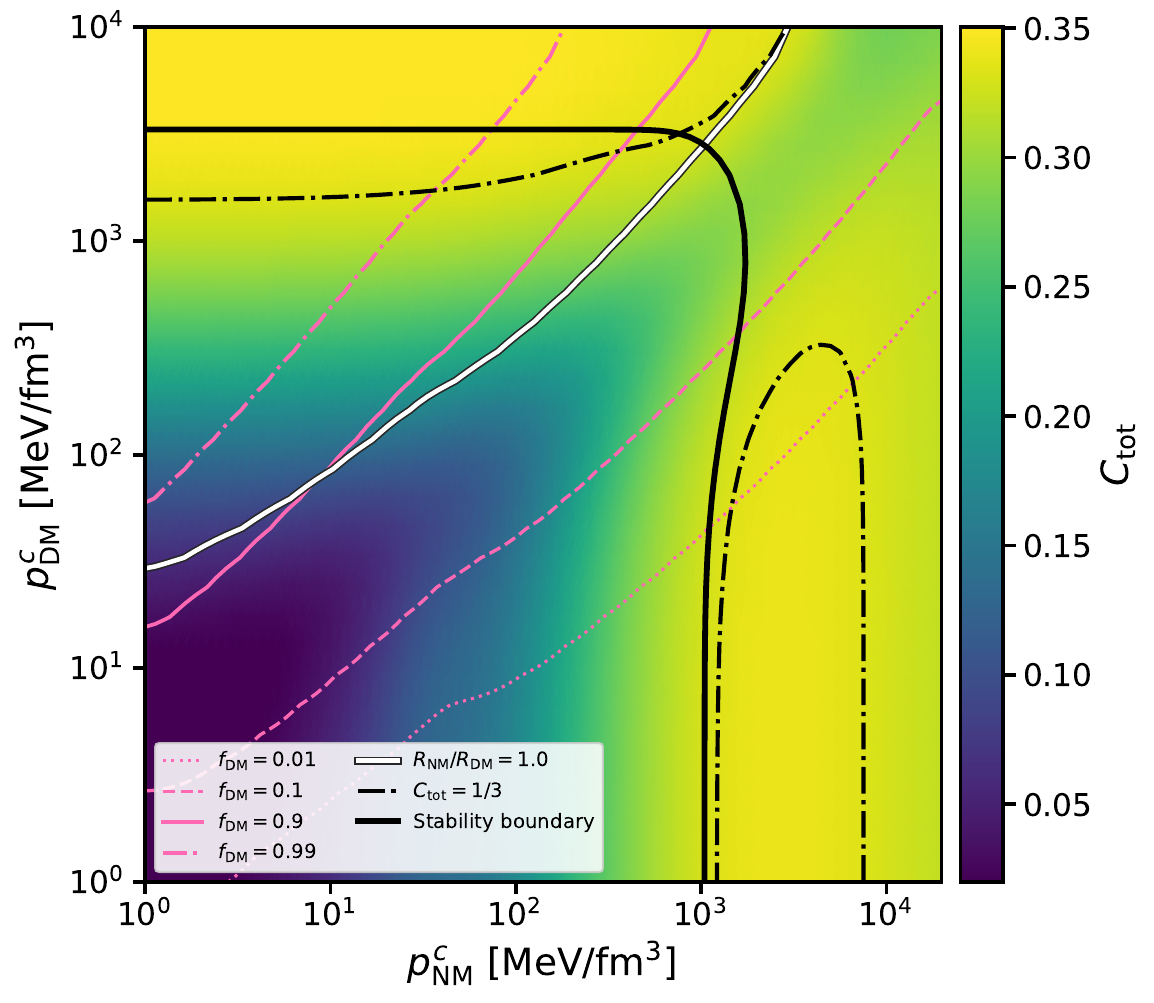}
            \label{fig:Rratio_P10_mb300}
            }
    \end{tabular}}
    \captionsetup{justification=justified, singlelinecheck=false}
        \caption{Total compactness $C_{\rm tot} = M_{\rm tot}/R_{\rm grav}$ (colormap) in the plane of central pressures for hybrid star configurations with  (a) $(P_t,\,\Delta\varepsilon)$ = (10, 250) and (b) $(P_t,\,\Delta\varepsilon)$ = (30, 250) $\mathrm{MeV}/\mathrm{fm}^3$, at DM parameters $m_b$ = 300 MeV, $n$ = 40. Pink lines are iso-lines of the DM fraction $f_\text{DM}$; the white line is $R_{\rm NM}/R_{\rm DM}=1$, which separates the DM-core (below) from DM-halo (above) configurations. The black dash-dotted line is $C=1/3$, and the thick black line is the two-fluid stability boundary.}
    \label{fig:contour1}
\end{figure*}

\cref{fig:contour1} presents the $(P_t,\Delta\varepsilon) = (10,250)$ and $(30,250)\,\text{MeV/fm}^3$ cases as two contour maps in the $(\log_{10} p^c_{\rm NM},\,\log_{10} p^c_{\rm DM})$ plane with $(m_b,n) = (300 \text{ MeV},40)$. The background colormap in both panels encodes the total compactness $C_{\rm tot} = M_{\rm tot}/R_{\rm grav}$, with $R_{\rm grav} = \max(R_{\rm NM}, R_{\rm DM})$ the radius of the outermost fluid, as introduced in Ref.~\cite{Pitz:2024xvh}. For a DM core, the dark matter lies inside the baryonic surface, so $R_{\rm grav} = R_{\rm NM}$ and $C_{\rm tot}$ coincides with the compactness inferred from the visible radius. For a DM halo, it is the dark matter surface that sets $R_{\rm grav}$, so that $C_{\rm tot}$ is smaller than the value $M_{\rm tot}/R_{\rm NM}$ that an electromagnetic radius measurement alone would suggest. Since the $C = 1/3$ contour marks the appearance of a light ring, it is $R_{\rm grav}$ that has to enter, and the ultra-compact configurations discussed below are compact in this full sense. Superimposed on each panel are iso-lines of the DM mass fraction $f_{\rm DM} = M_{\rm DM}/M_{\rm tot}$ at 0.01, 0.1, 0.9, and 0.99 (pink), the line $R_{\rm NM}/R_{\rm DM}=1$ separating a DM core from a DM halo (white), the $C=1/3$ boundary (black dash-dotted), and the two-fluid stability boundary (thick black solid), the stable region laying to the left of the latter. This representation provides a global map of stellar morphology across the entire central-pressure plane. 

In panel (a), corresponding to the low transition pressure $(P_t,\Delta\varepsilon)=(10,250)\,\text{MeV/fm}^3$, the $f_{\rm DM}$ iso-lines run diagonally across the plane. DM-dominated configurations with ($f_{\rm DM} \gtrsim 0.9$) occupy the upper-left region where ($p^c_{\rm DM} \gg p^c_{\rm NM}$), while the lower-right region is baryon-dominated with ($f_{\rm DM} \lesssim 0.02$). The stability boundary traces the edge of the stable domain and exhibits a small region along the $p^c_{\rm NM}$ direction, corresponding to the unstable segment separating the hadronic and hybrid branches; this region narrows with increasing $p^c_{\rm DM}$ for the reason set out in \cref{sec:mr}. The $C=1/3$ contour intersects the stable region in two distinct locations.

(i) DM-halo UCOs: At high $p^c_{\rm DM}$, the DM self-gravity compresses $R_{\rm NM}$ while the DM distribution extends well beyond the baryonic core, so that $R_{\rm grav} = R_{\rm DM}$ and the compactness is DM-driven. This class was identified in Ref.~\cite{Pitz:2024xvh} for purely nucleonic NM and is recovered here.

(ii) DM-core UCOs: At high $p^c_{\rm NM}$, the quark-matter twin branch attains $C \geq 1/3$ through the compactness of the baryonic sector alone, with the DM forming a subdominant core with $f_{\rm DM} \lesssim 0.02$. These are compact hybrid stars carrying a small embedded DM component that plays no significant dynamical role. They have no analogue in Ref.~\cite{Pitz:2024xvh}, where the purely nucleonic EoS never reaches $C \geq 1/3$ without DM assistance. In panel (a), this population is the narrow strip at $p^c_{\rm DM} \lesssim 10^2\,\text{MeV/fm}^3$ in which the $C=1/3$ contour has already been crossed while the stability boundary has not. Both classes of UCOs are themselves not new; however, we find that both of them can be described within a single model. They are furthermore separated by a gap in the DM fraction, where, to our knowledge, no UCOs can exist, neither in this work nor in others. 

In panel (b) with a higher transition pressure $(P_t,\Delta\varepsilon)=(30,250)\,\text{MeV/fm}^3$, the morphological structure is qualitatively unchanged: the $f_{\rm DM}$ contour remains diagonal, the DM-halo region continues to occupy high $p^c_{\rm DM}$, and the stability region persists, although the twin branch is displaced to higher $p^c_{\rm NM}$. The DM-core population (ii) disappears entirely, since the quark-matter branch no longer attains $C=1/3$  within the stable domain once deconfinement is delayed to densities at which the maximum mass is lower, leaving only the DM-halo class. The two contours that bound the strip in panel (a) coincide in panel (b): the $C=1/3$ line and the stability boundary both fall at $p^c_{\rm NM} \approx 10^3\,\text{MeV/fm}^3$, leaving no stable configurations between them. DM-cores therefore require an early-onset transition, strongly super-Seidov at low $P_t$, and are absent for moderate or marginal transitions. The coexistence of two ultra-compact populations with opposite DM morphology, high-$f_{\rm DM}$ halos and low-$f_{\rm DM}$ cores, carries direct observational consequences. Both possess a photon sphere at $R = 3M$ and would produce gravitational-wave echoes in the post-merger signal \cite{Cardoso:2019rvt, Pani:2018flj, Urbano:2018nrs}, rendering them black-hole mimickers in the time domain, and for the halo class, in the electromagnetic one as well \cite{Olivares:2018abq}. They remain distinguishable, however. DM-halo configurations have $\Lambda \ll 1$ on account of the extended DM distribution, whereas DM-core configurations have $\Lambda$ characteristic of a compact quark star modified only slightly by the embedded DM. An object with $f_\text{DM} \lesssim 0.02\%$, a deformability typical of quark matter, and a residual echo signal would therefore constitute evidence for a hybrid star with an embedded bosonic DM core, a configuration inaccessible to single-fluid description. Its electromagnetic identification, combined with a gravitational-wave determination of the tidal deformability, would constrain $P_t$ and $m_b$ simultaneously.

\subsection{Surface Redshift}
\label{redshift}

\begin{figure}
    \centering
    \includegraphics[width=\columnwidth]{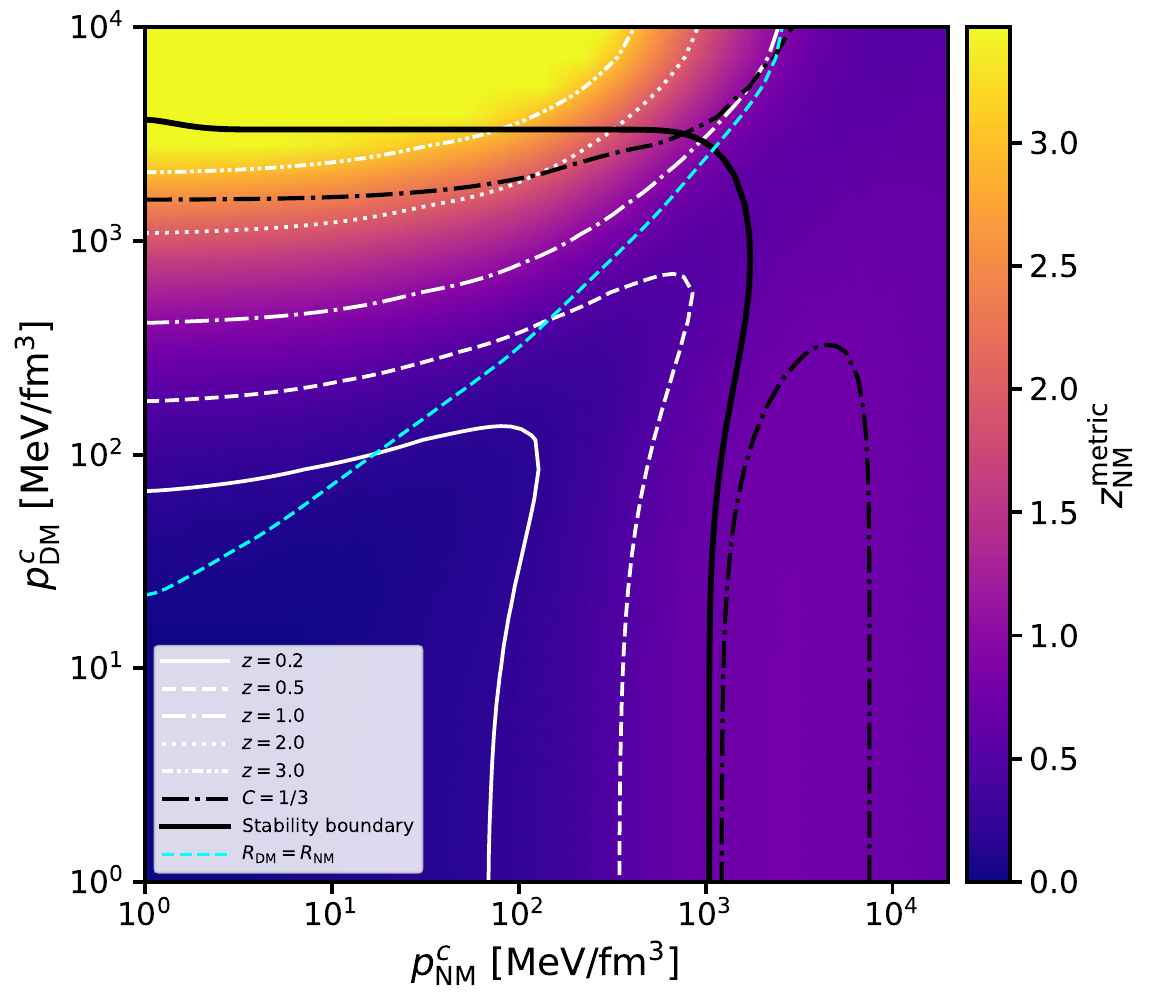}
    \captionsetup{justification=justified, singlelinecheck=false}
    \caption{Surface gravitational redshift in the $(p^c_{\rm NM},\,p^c_{\rm DM})$ plane for $(P_t,\Delta\varepsilon)=(10,\,250)\,\text{MeV/fm}^3$, $(m_b,n)=(300\,\text{MeV},40)$. White iso-redshift contours mark $z=0.2$ (solid), $0.5$ (dashed),$1.0$ (dashed-dotted), $2.0$ (dotted), and $3.0$ (dash-dot-dotted). The thick solid black line is the two-fluid stability boundary; the black dash-dot line is the $C=1/3$ boundary, and the blue dashed lines mark the transition from a DM core (below the line) to a DM halo (above the line).}
    \label{fig:contour_redshift}
\end{figure}

The gravitational redshift of photons emitted at the neutron star surface is set by the time-time component of the metric at the emitting radius,

\begin{equation}
        1+z = \frac{1}{\sqrt{-g_{tt}(R_\text{NM})}} = e^{-\nu(R_\text{NM})} ,
\end{equation} 

with $ds^2= -e^{-\nu(r)}dt^2+e^{2\Lambda(r)}dr^2+r^2d\Omega^2$ and $\nu$ following the two-fluid TOV equation given by \cref{dnudr}. 
 The TOV equations fix only ${d\nu}/{dr}$, so $\nu$ is determined up to an additive constant. That constant follows from requiring the metric to be asymptotically flat, which in practice means matching to the Schwarzschild exterior at the outer edge of the object,

\begin{equation}
    \nu(R_\text{grav}) = \frac{1}{2}\Bigg( 1-\frac{2M_\text{tot}}{R_\text{grav}}\Bigg).
\end{equation}
Integrating \cref{dnudr} inward from $R_\text{grav}$ to $R_\text{NM}$ gives
\begin{equation}\label{redshift}
    1+z = \Bigg( 1-\frac{2M_\text{tot}}{R_\text{grav}}\Bigg)^{-1/2} e^{I_\nu},
\end{equation}
with 
\begin{equation}\label{integral}
    I_\nu = \int_{R_\text{NM}}^{R_\text{grav}} \frac{m_\text{tot} +4\pi r^3 P_\text{tot}}{r(r-2m_\text{tot})} dr ,  
\end{equation}
in agreement with the two-fluid treatment of Ref.~\cite{Shawqi:2024jmk}.

The two morphologies differ in whether $I_\nu$ survives. For a DM core, the dark matter is contained within the neutron star, so $R_\text{grav}= R_\text{NM}$, the emitting surface is itself the vacuum boundary, and the integral vanishes identically. \cref{integral} then reduces to the enclosed-mass expression
\begin{equation}\label{zenc}
    \begin{split}
        z_{\rm enc} &= (1-2\,(m_\text{DM}(r\leq R_\text{NM})\\
    &+ m_\text{NM}(r \leq R_\text{NM}))/ R_\text{NM})^{-1/2}-1 .
    \end{split}
\end{equation} 
which is exact in that case. For a DM Halo, the surface lies beneath a further layer of dark matter, $I_\nu >0$, and \cref{zenc} underestimates the redshift because it counts only the mass interior to $R_\text{NM}$ while assuming vacuum above it. Equivalently, $1-2m/r$ is the inverse radial metric function $1/g_{rr}$, which coincides with $-g_{tt}$ only in vacuum.
Here $R_\text{NM}$ is the radius of the neutron star and $m(r\leq R_\text{NM})$ is the mass of both fluids contained inside it. 
The gravitational redshift thus provides an electromagnetic observable complementary to tidal deformability: it depends only on the total compactness and is in principle measurable from spectral line shifts of surface-emitted radiation~\cite{Lattimer:2000nx, Ozel:2016oaf}, as claimed for the X-ray burst spectra of EXO 0748-676 \cite{Cottam:2002cu}. 

\cref{fig:contour_redshift} shows the redshift of \cref{redshift} across the $(p^c_{\rm NM},\,p^c_{\rm DM})$ plane for the $(P_t,\Delta\varepsilon)=(10,250)\,\text{MeV/fm}^3$, EoS with $(m_b,n)=(300\,\text{MeV},40)$. 
In addition, the limit on a surface redshift (using the vacuum Schwarzschild metric as the outside solution) from causality of $z\approx 0.851$ for a compactness of $C=0.354$, of $z=2$ for the Buchdahl limit of $C=4/9$ \cite{Buchdahl:1959zz}, and of $z\approx 4.83$ for the Bondi limit of $C=6\sqrt{2}-8\approx 0.485$ 
for arbitrary energy density profiles \cite{Bondi:1964zz} are indicated by horizontal lines.
The map rises from $z\lesssim0.2$ in the low-density corner (purple) in two directions: toward the NM-dominated lower-right region, where stable stars reach $z\eqsim0.8$ through the quark-matter branch alone, and far more steeply toward the DM dominated upper-left, where the emitting surface is buried under the halo, and $z$ exceeds 3. The enclosed-mass expression reproduces the first rise and misses the second entirely.
The redshift iso-lines develop a characteristic kink at the phase-transition pressure: as $p^c_{\rm NM}$ crosses $P_t$, the NM star enters the quark-matter branch and the compactness increases abruptly, shifting the $z$-contour toward lower $p^c_{\rm NM}$ at fixed $p^c_{\rm DM}$. The two UCO populations identified in \cref{fig:contour1} occupy distinct redshift ranges. DM-core UCOs (lower-right corner with $f_{\rm DM}\lesssim0.02$) reach $z=0.734-0.808$, bounded by the causality limit $z\leq0.851$ for a compactness of C=0.354, because their surface coincides with the vacuum boundary. DM-halo UCOs (upper-left, enclosed by the $C=1/3$ contour at high $p^c_{\rm DM}$) are not bounded in this way, since the emitting surface is an interior point of a more compact configuration, and they span redshifts $z=2-3.468$. These two populations are therefore separated in the opposite sense to the enclosed-mass estimate, which would place the halo class at $z\leq 0.771$. The amplification is carried entirely by $I_\nu$: in the extreme halo configurations, almost none of the gravitational mass lies inside $R_\text{NM}$, so the redshift is produced by dark matter above the emitting surface rather than by the mass beneath it.

\begin{figure}
    \centering
    \includegraphics[width=\columnwidth]{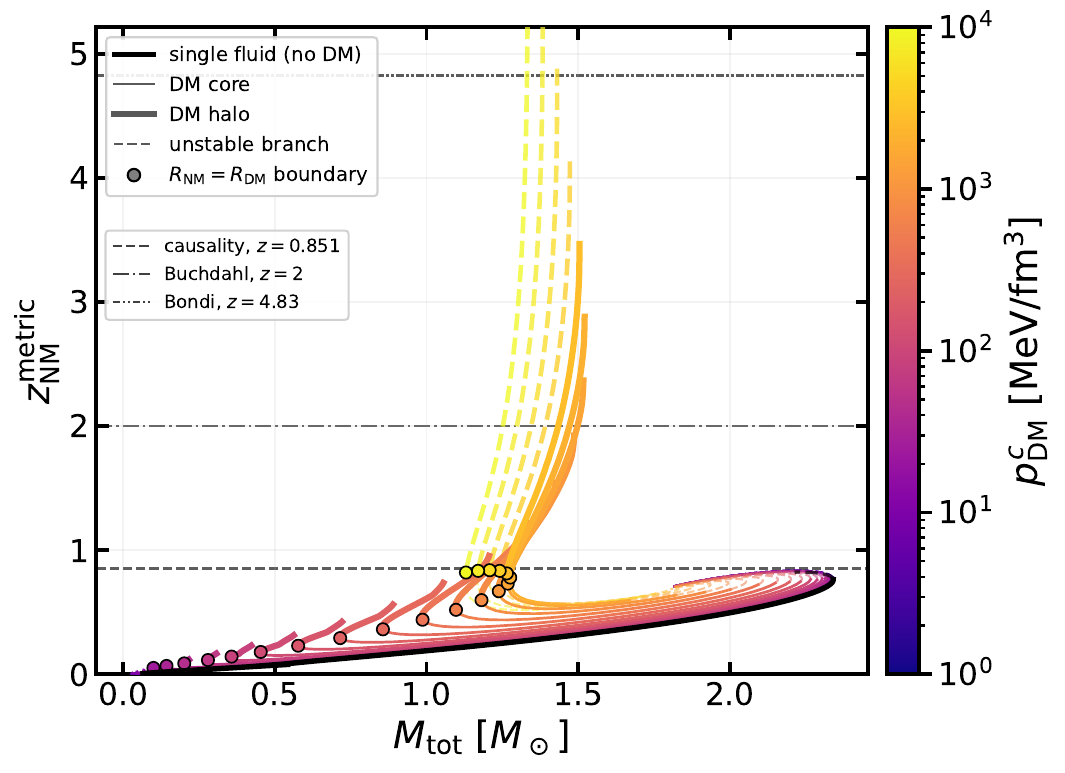}
    \includegraphics[width=\columnwidth]{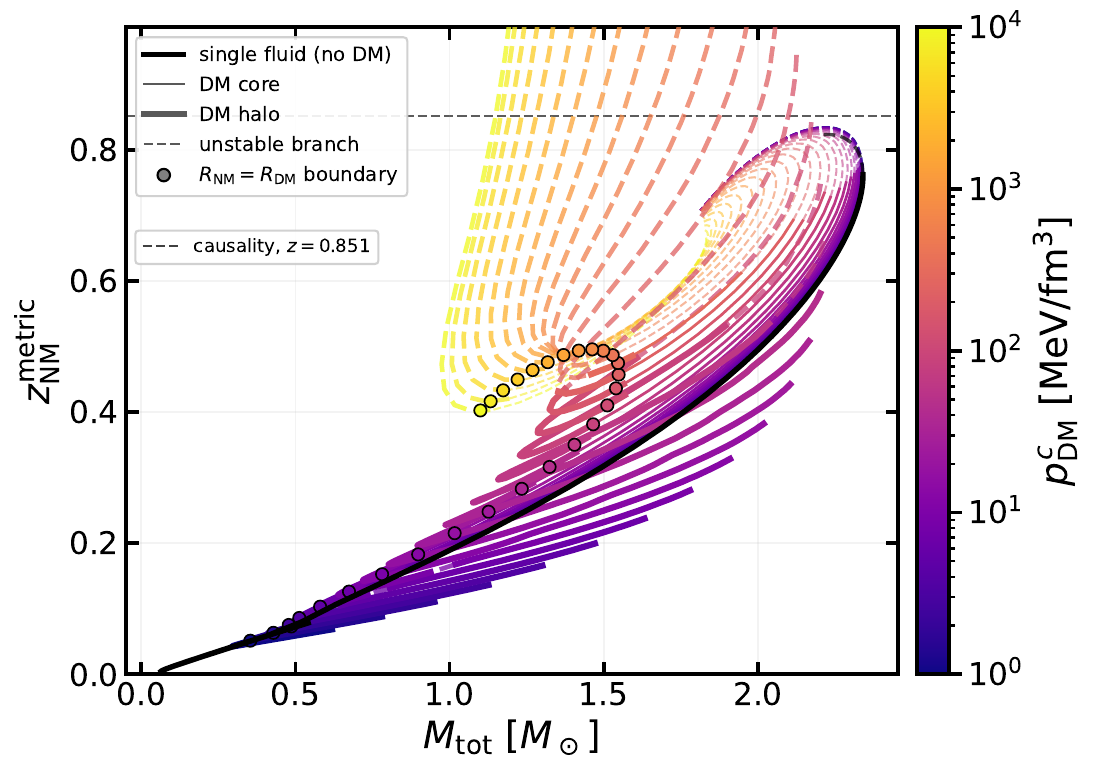}
    \captionsetup{justification=justified, singlelinecheck=false}
    \caption{Surface redshift as a function of total mass for $(P_t,\Delta\varepsilon)=(10,250)\,\text{MeV/fm}^3$ with $m_b=300\,\text{MeV}$ and (a) $n=40$, (b) $n=4$. The black curve is the DM-free hybrid sequence, while the coloured curves are sequences of fixed $p^c_\text{DM}$; thin lines mark the DM core and thick lines mark the DM halo configurations. Solid (dashed) lines mark the stable (unstable) region; circles indicate the $R_{\rm NM}=R_{\rm DM}$ transition for DM halo and core configurations. The three horizontal lines mark the redshift limit from the causal bound $z=0.851$ (dashed line), the Buchdahl limit $z=2$ (dashed-dot line), and the Bondi limit $z\approx4.83$ (dashed-dot-dot line)~\cite{Haensel:1999gj, Bondi:1964zz, Arrechea:2024vxp}, for a surface matched to vacuum.}
    \label{fig:redshift_mass}
\end{figure}

Since both classes of UCOs are small, which makes determining properties such as the radius challenging, measuring their redshift is the more promising route: a radius below 8 km from NICER or a comparable X-ray mission, together with $z \lesssim 0.9$, points to the halo scenario, whereas $z \approx 0.73$–$0.81$ at a radius near 10 km favors a DM-core hybrid star. No core-class configuration in our grid exceeds $z=0.808$, so a larger value identifies a halo unambiguously.
Chatterjee and Nath \cite{Chatterjee:2025pkx} find that ordinary neutron stars satisfying current astrophysical constraints cannot exceed $z=0.763$, a value that lies below the causal bound $z_\text{max} = 0.851$ obtained for static configurations \cite{Haensel:1999gj}. Our DM-free hybrid sequence already reaches $z=0.759$, essentially saturating this bound through the stiff quark-matter branch alone. However, both bounds apply to a surface matched to vacuum. A DM halo places the emitting surface inside the configuration, where no such limit holds, and the stable halo sequences reach $z=3.47$ while the outer surface itself stays well below the Buchdahl value, $M_\text{tot}/R_\text{grav}= 0.33-0.34$ against $4/9$ and the Bondi limit, $z\approx 4.83$~\cite{Bondi:1964zz, Arrechea:2024vxp}. The signature of a dark matter halo is therefore a redshift that exceeds what any single-fluid EoS can produce at the measured radius, rather than the high compactness alone.

\cref{fig:redshift_mass} shows the surface redshift as a function of the total mass for the same EoS configuration as \cref{fig:contour_redshift} with  $m_b=300\,\text{MeV}$, $n=40$ (upper panel) as well as for $n=4$ (lower panel) to see the effect of DM and the softening of the DM self-interaction more clearly. Whether a halo raises or lowers $z$ at fixed $M_\text{tot}$ depends on how concentrated it is, and the two self-interaction strengths behave oppositely. The comparison is made at the same total mass, so what matters is the size of the object relative to a DM-free star of equal mass.
For $n=40$, the dark matter is confined to $R_\text{DM}\leq 6.8$ km and the whole configuration is smaller than the DM-free star of the same mass. The potential at the surface is correspondingly deeper, and none of the stable halo configurations falls below the DM-free curve. The highest redshift of the model, $z=3.47$, is reached in this regime, above the causal and Buchdahl limiting values but below the one for the Bondi limit. For $n=4$, the halo extends to $R_\text{DM}\leq 28$ km, making the object about twice as large as the DM-free star of equal mass, giving the shallower potential at $R_\text{NM}$, and about half of the stable halo configurations lie below the DM-free curve, with the maximum redshift retaining the single-fluid value $0.759$ set by the quark-matter branch.
The redshift of photons $z$ can be determined using fewer model assumptions compared to the inference for mass and radius. A gravitationally redshifted absorption line gives $z$ from the line position alone, with no modeling of hotspot geometry, emission pattern, and inclination that pulse profile analysis requires~\cite{Cottam:2002cu}. No such surface spectral line has yet been uniquely identified for a neutron star spectrum. Large magnetic fields, of  $B\sim 4\times 10^{12}$ G and more, can lead to a condensed surface so that plasma effects wash out spectral features, as discussed for the spectrum of the isolated neutron star RX J1856.5-3754 \cite{Ho:2006uk}.
Accreting neutron stars, spun up and carrying lower magnetic fields of $B\sim10^{8}$--$10^{9}$ G, would then be the more natural targets, and the sources observed by NICER belong to this latter class.
\subsection{Ultimate Twins}
\label{ultimate_twins}

 Because the dark matter is electromagnetically invisible, an X-ray measurement returns the pair ($M, R_\text{NM}$) but says nothing about the dark matter content. Two stable stars may therefore agree in ($M, R_\text{NM}$) while carrying different dark matter fractions, and hence differ in tidal deformability and in surface redshift. We call these \textit{ultimate twins}, as they are the inverse of the doppelganger equations of state of Raithel and Most \cite{Raithel:2022aee}, which coincide in $\Lambda(M)$ and are separated by radius. The two members lie on different $p^c_\text{DM}$ sequences with no unstable segment involved and are therefore distinct from the twins of \cref{sec:mr}. Since the sequences of fixed $p^c_\text{DM}$ are nested rather than intersecting in the $(M, R_\text{NM})$ plane, the degeneracy is exact only for vanishing $|\Delta f_\text{DM}|$, and any pairing rests on a finite tolerance, which is in any case what an observation imposes.
 Requiring $|\Delta M| < 0.05\,M_\odot$ and $|\Delta R_\text{NM}| < 0.3$ km, both inside current uncertainties, together with $|\Delta f_\text{DM}| > 0.05$, we find such pairs for most of the EoS combinations considered, over a mass range from 1.2 to $2.33\,M_\odot$. Pairs whose deformabilities differ by less than about a factor of $1.5$ are not
informative, since the matching tolerance alone produces a spread of this size. The contrast in $\Lambda$ is set by the halo extent, and therefore by the dark matter sector alone. A soft self-interaction, $n=4$, produces a dilute halo reaching $R_\text{DM} \approx 26$ km, so that two stars sharing $(M, R_\text{NM})$ differ by a factor of three in total compactness and, since $\Lambda \propto C^{-5}$, by three orders of magnitude in deformability; a stiff self-interaction confines the halo to $\sim 6$ km and the contrast largely disappears. 

The boson mass acts the same way, $m_b = 1000\,$MeV producing no pairs at all, while the transition pressure and $c^2_\text{QM}$ have only a weak influence, early-onset transitions being mildly favored. The largest contrasts are, however, of no observational use, as the halo extends enough to produce them also makes the star easy to deform, giving $\Lambda\sim10^{4}$ at $1.5\,M_\odot$, far above the upper limit inferred from GW170817.  The strongest admissible pair in our grid occurs for $(P_t,\Delta\varepsilon)=(10,250)\,\text{MeV/fm}^3$, EoS with $(m_b,n)=(100 \text{ MeV},40)$: two stable configurations at $M=2.32\,M_\odot$ and $R_\text{NM} = 10.11$ km carry $f_\text{DM}= 0.40$ and 0.02, and have $\Lambda= 489$ and 2.4, a factor of 206. The dark matter halo of the first member extends to $R_\text{DM}= 27.5$ km, so that the two stars have total compactness of $C= 0.12$ and 0.34, although their visible radii agree to within 0.02 km.
For the stiff model of \cref{fig:lambda_P10_mb300}, $(m_b, n)= (300 \text{ MeV}, 40)$, the same search returns contrasts of at most a factor of three, making this the conservative end of the range. A factor of 206 is well within the reach of the Einstein Telescope and Cosmic Explorer, whereas a factor of two at $\Lambda$ of order unity is not a realistic target, the tidal contribution to the inspiral phase being negligible there.  The surface redshift provides an independent handle, the two members of the pair quoted above differing by $z= 0.32$ against 0.76, so that a combined electromagnetic and gravitational wave measurement can resolve configurations that no mass-radius determination alone could separate.

\section{Summary}
\label{summary}

In this paper, we discussed the changing properties of hybrid stars with a strong phase transition when including dark matter as a second fluid. The normal matter was described by a piecewise polytropic EoS combined with a constant speed of sound EoS for the quark matter. The transition between these two phases is a first-order phase transition that is characterized by the parameter set of $( \Delta \varepsilon_t, P_t)$. The second fluid, dark matter, was modeled as a bosonic, self-interacting scalar field generating ultra-compact boson and dark matter-admixed neutron stars, as known from previous work \cite{Pitz:2023ejc, Pitz:2024xvh}. We have performed a stability analysis in the matter of \cite{Hippert:2022snq} by looking at radial density perturbations that arise from varying the central energy densities of both fluids. The resulting mass-radius curves emphasized the importance of this stability analysis since the stable regions differ substantially from the one-fluid stability criterion. The most obvious example is the unstable region that connects two branches of hybrid stars with the same mass at different radii (so-called twin stars). DM closes this unstable segment at DM fractions of a few percent, so that the phase transition remains in the EoS and stars retain their quark cores while the sequence no longer shows an unstable branch. Hence, the Seidov criterion, derived for a single fluid, no longer identifies the stable sequences that support twin stars, and a consistent analysis reveals additional stable compact star sequences. For the strongest transition we consider, $(P_t, \Delta \varepsilon_t) = (120, 600)\,\text{MeV/fm}^3$, the second fluid acts in the opposite direction: below $p_\text{DM}^c\approx 17\,\text{MeV/fm}^3$, the hybrid branch is unstable over its whole extent, and the dark matter restores a stable hybrid branch that does not exist without it.
Furthermore, the maximum mass is significantly reduced with higher DM content. 

We have studied the impact of different values of the parameters $(P_t, \Delta \varepsilon_t)$ as well as the DM parameters $(m_b, n)$. The soft DM case $n = 4$ shows more moderate deviations from the pure NM case. For stiff DM, i.e., $n = 40$, we have seen that the mass-radius curves reach into a region of the mass-radius space that seems to be forbidden by causality. These stable objects we found are so compact that they seem to exceed even the non-rotating black hole limit $C = 1 / 2$. However, this high compactness can be explained by the fact that we are looking at the NM radius and the total mass. The reason for looking at this pseudo-compactness is the fact that these two observables, total mass and radius of normal matter, are determined by gravitational and electromagnetic measurements separately. X-ray telescopes are blind to the DM radius, whereas pulsar mass measurements and gravitational wave detectors measure the gravitational mass, i.e., the total mass. These findings are not new; however, the additional phase transition to quark matter pushes the compact objects to even higher values of $C$. Here, a second island of stable UCOs emerges that has not been seen in the previous study \cite{Pitz:2024xvh}. Both classes of ultra-compact stars are not new; however, they are generated using separate models. Here, both classes are supported by a single model separated only by a gap at intermediate DM fraction where no stable UCO exists. The UCOs with very low DM fractions ($\lesssim \, 10 \, \%$) all possess a DM core and the ones with high fractions ($\gtrsim \, 90 \, \%$) all a DM halo. The high compactness of the UCOs with DM cores is dictated by the phase transition to quark matter rather than by the DM, and they form a narrow population, occupying $M= 2.30 - 2.34\,M_\odot$ at $R_\text{NM}= 10.1 - 10.3$ km. They require an early-onset transition together with a stiff quark phase, and are absent for $c^2_\text{QM} = 0.7$ altogether. The high compactness of the objects with a DM halo is instead set by the stiff DM EoS. 

Additionally, we have studied the gravitational redshift of photons emitted at the surface of the neutron star. Here, another difference between these two classes of stable UCOs appears: UCOs with a DM halo generate significantly higher redshifts than those with a DM core. In the halo case, the redshift range is $z = 2 - 3.468$ and in the core case: $z = 0.734 - 0.808$. So, in the case of the DM halo, the redshift starts at the Buchdahl limit, i.e. the maximum redshift of an incompressible fluid where the energy density does not increase outwards. Since in our case the gravitational radius is given by the DM radius ($R_\text{DM} > R_\text{NM}$) we obey this limit, and it only appears to be violated since X-ray telescopes like NICER or ATHENA would not be able to see the DM halo. In the case of a DM core UCO, the Buchdahl limit is never exceeded, and the redshift stays well below $z < 2$.

The mass-radius measurement of PSR J1614$-2230$ \cite{Mauviard:2026gzc} corroborates our findings from the observational side: the source is reproduced by an early-onset transition with no dark matter, as a hybrid or twin star, or by a delayed transition with a dark matter core of a few percent, and only the surface redshift and the tidal deformability separate the two. The tidal deformability behaves in the opposite sense to what is found for lighter dark matter candidates. Because the stiff self-interaction confines the halo to $\sim$ 6 km while it carries most of the gravitational mass, the total compactness rises and $\Lambda$ falls, reaching values of order unity on the hybrid branch. The sign of this effect is set by the boson mass through the halo size, and reverses at $m_b$ = 100 MeV, where the halo reaches $\approx$ 30 km and $\Lambda_{1.4}$ is enhanced to 2.5 $\times$ 10$^4$, which is the behavior reported for light fermionic dark matter and is excluded by GW170817 \cite{Dietrich:2020efo, Sun:2023cqr, Arvikar:2025dwl}.
For stiff dark matter, $m_b=300\,\text{MeV}$ and $n=40$, the configuration has $\Lambda_{1.4}$ even below the lower limit of $70$ from GW170817
with a prior from the EOS, once $p^c_\text{DM}$ exceeds $ \sim 10^2\,\text{MeV/fm}^3$. 
For $m_b=1000\,\text{MeV}$, the whole sequence is compatible with the limit from GW170817. 

Taken together, our results show that a first-order phase transition and an additional dark matter component are not independent entities to be treated one at a time. 
A high redshift combined with a high total mass ($ \sim 2$ - $2.4 \, M_\odot$) is an indication of a first-order phase transition, whereas an even higher redshift together with a low or moderately high total mass ($\sim 0.5$ - $1.5 \, M_\odot$) is a signature for dark matter. These objects could be targets for future X-ray telescopes like ATHENA or eXTP, as their sensitivity will exceed that of NICER \cite{eXTP:2018anb, Matt:2019llr}. Consequently, observations of ultra-compact neutron stars, combined with measurements of their spectral lines, would indicate whether dark matter and/or a phase transition to quark matter are present inside the star. 

Several extensions suggest themselves for future work: the parametrized EoSs used here could be replaced by microscopic ones, a relativistic mean-field description for the hadronic phase matched to a renormalization-group consistent NJL model for the quark phase \cite{Gholami:2024ety, Christian:2025dhe}, which would tie $P_t$ and $\Delta \varepsilon$ to the underlying couplings instead of treating them as free parameters. The stability analysis could be made dynamical: the criterion of Ref. \cite{Hippert:2022snq} is a static one, obtained from the Jacobian of the conserved particle numbers, whereas solving the two-fluid radial oscillation equations would give the squared eigenfrequencies directly and identify the mode that becomes unstable, rather than only the point at which it does. Beyond these, a non-gravitational coupling between the two sectors, rotation, and finite temperature all remain to be explored.

\section*{Data Availability}

The numerical data underlying this article are publicly available on GitHub \cite{ultra_compact_twin_stars}.

\section*{Acknowledgement}
We thank Anna L. Watts for useful discussions about the NICER measurements and dark matter. We also thank Laura Tolos, Oscar Dario, and Jose Jimenez for the discussion regarding the redshift calculation.
The authors acknowledge support by the
Deutsche Forschungsgemeinschaft (DFG, German Research Foundation) through the CRC-TR 211 ’Strong-interaction matter under extreme conditions’– project number 315477589 – TRR 211. I. A. R. gratefully acknowledges support from the Deutsche Forschungsgemeinschaft (DFG, German Research Foundation) – Project Number 579861443 and also in part by the Alexander von Humboldt Foundation through a Humboldt Research Fellowship. S.L.P. is supported by the Research and Development (R$\&$D) program of GSI Helmholtzzentrum für Schwerionenforschung.

\bibliographystyle{apsrev4-112}
\bibliography{references} 

@article{Grippa:2024ach,
    author = "Grippa, Francesco and Lambiase, Gaetano and Poddar, Tanmay Kumar",
    title = "{Searching for New Physics in an Ultradense Environment: A Review on Dark Matter Admixed Neutron Stars}",
    eprint = "2412.09381",
    archivePrefix = "arXiv",
    primaryClass = "astro-ph.HE",
    doi = "10.3390/universe11030074",
    journal = "Universe",
    volume = "11",
    number = "3",
    pages = "74",
    year = "2025"
}

@article{Bell:2020jou,
    author = "Bell, Nicole F. and Busoni, Giorgio and Robles, Sandra and Virgato, Michael",
    title = "{Improved Treatment of Dark Matter Capture in Neutron Stars}",
    eprint = "2004.14888",
    archivePrefix = "arXiv",
    primaryClass = "hep-ph",
    doi = "10.1088/1475-7516/2020/09/028",
    journal = "JCAP",
    volume = "09",
    pages = "028",
    year = "2020"
}

@article{Busoni:2021zoe,
    author = "Busoni, Giorgio",
    title = "{Capture of Dark Matter in Neutron Stars}",
    eprint = "2201.00048",
    archivePrefix = "arXiv",
    primaryClass = "hep-ph",
    doi = "10.3103/S0027134922020205",
    journal = "Moscow Univ. Phys. Bull.",
    volume = "77",
    number = "2",
    pages = "301--305",
    year = "2022"
}

@article{Dengler:2021qcq,
    author = {Dengler, Yannick and Schaffner-Bielich, J{\"u}rgen and Tolos, Laura},
    title = "{Second Love number of dark compact planets and neutron stars with dark matter}",
    eprint = "2111.06197",
    archivePrefix = "arXiv",
    primaryClass = "astro-ph.HE",
    doi = "10.1103/PhysRevD.105.043013",
    journal = "Phys. Rev. D",
    volume = "105",
    number = "4",
    pages = "043013",
    year = "2022"
}

@article{Tolos:2015qra,
    author = {Tolos, Laura and Schaffner-Bielich, J{\"u}rgen and Dengler, Yannick},
    title = "{Dark Compact Planets}",
    eprint = "1507.08197",
    archivePrefix = "arXiv",
    primaryClass = "astro-ph.HE",
    doi = "10.1103/PhysRevD.92.123002",
    journal = "Phys. Rev. D",
    volume = "92",
    pages = "123002",
    year = "2015",
    note = "[Erratum: Phys.Rev.D 103, 109901 (2021)]"
}

@article{Ellis:2018bkr,
    author = {Ellis, John and H{\"u}tsi, Gert and Kannike, Kristjan and Marzola, Luca and Raidal, Martti and Vaskonen, Ville},
    title = "{Dark Matter Effects On Neutron Star Properties}",
    eprint = "1804.01418",
    archivePrefix = "arXiv",
    primaryClass = "astro-ph.CO",
    reportNumber = "CERN-TH-2018-072, KCL-PH-TH/2018-13, KCL-PH-TH-2018-13",
    doi = "10.1103/PhysRevD.97.123007",
    journal = "Phys. Rev. D",
    volume = "97",
    number = "12",
    pages = "123007",
    year = "2018"
}

@article{Sagun:2021oml,
    author = "Sagun, V. and Giangrandi, E. and Ivanytskyi, O. and Lopes, I. and Bugaev, K. A.",
    title = "{Constraints on the fermionic dark matter from observations of neutron stars}",
    eprint = "2111.13289",
    archivePrefix = "arXiv",
    primaryClass = "astro-ph.HE",
    reportNumber = "Proceedings of PANIC2021 Conference",
    doi = "10.22323/1.380.0313",
    journal = "PoS",
    volume = "PANIC2021",
    pages = "313",
    year = "2022"
}

@article{Christian:2021uhd,
    author = {Christian, Jan-Erik and Schaffner-Bielich, J{\"u}rgen},
    title = "{Confirming the Existence of Twin Stars in a NICER Way}",
    eprint = "2109.04191",
    archivePrefix = "arXiv",
    primaryClass = "astro-ph.HE",
    doi = "10.3847/1538-4357/ac75cf",
    journal = "Astrophys. J.",
    volume = "935",
    number = "2",
    pages = "122",
    year = "2022"
}

@article{Christian:2020xwz,
    author = {Christian, Jan-Erik and Schaffner-Bielich, J{\"u}rgen},
    title = "{Supermassive Neutron Stars Rule Out Twin Stars}",
    eprint = "2011.01001",
    archivePrefix = "arXiv",
    primaryClass = "astro-ph.HE",
    doi = "10.1103/PhysRevD.103.063042",
    journal = "Phys. Rev. D",
    volume = "103",
    number = "6",
    pages = "063042",
    year = "2021"
}

@article{Montana:2018bkb,
    author = "Montana, Gloria and Tolos, Laura and Hanauske, Matthias and Rezzolla, Luciano",
    title = "{Constraining twin stars with GW170817}",
    eprint = "1811.10929",
    archivePrefix = "arXiv",
    primaryClass = "astro-ph.HE",
    doi = "10.1103/PhysRevD.99.103009",
    journal = "Phys. Rev. D",
    volume = "99",
    number = "10",
    pages = "103009",
    year = "2019"
}

@article{Naseri:2024rby,
    author = "Naseri, Mahdi and Bozzola, Gabriele and Paschalidis, Vasileios",
    title = "{Exploring pathways to forming twin stars}",
    eprint = "2406.15544",
    archivePrefix = "arXiv",
    primaryClass = "astro-ph.HE",
    doi = "10.1103/PhysRevD.110.044037",
    journal = "Phys. Rev. D",
    volume = "110",
    number = "4",
    pages = "044037",
    year = "2024"
}

@article{Espino:2021adh,
    author = "Espino, Pedro L. and Paschalidis, Vasileios",
    title = "{Fate of twin stars on the unstable branch: Implications for the formation of twin stars}",
    eprint = "2105.05269",
    archivePrefix = "arXiv",
    primaryClass = "astro-ph.HE",
    doi = "10.1103/PhysRevD.105.043014",
    journal = "Phys. Rev. D",
    volume = "105",
    number = "4",
    pages = "043014",
    year = "2022"
}

@article{Leung:2022wcf,
    author = "Leung, Kwing-Lam and Chu, Ming-chung and Lin, Lap-Ming",
    title = "{Tidal deformability of dark matter admixed neutron stars}",
    eprint = "2207.02433",
    archivePrefix = "arXiv",
    primaryClass = "astro-ph.HE",
    doi = "10.1103/PhysRevD.105.123010",
    journal = "Phys. Rev. D",
    volume = "105",
    number = "12",
    pages = "123010",
    year = "2022"
}

@article{Haque:2026dre,
    author = "Haque, Shamim and Shinde, Atharva and Saha, Asim Kumar and Malik, Tuhin and Mallick, Ritam",
    title = "{Investigating Twin Star Equation of States in Light of Recent Astrophysical Observations}",
    eprint = "2601.16674",
    archivePrefix = "arXiv",
    primaryClass = "astro-ph.HE",
    month = "1",
    year = "2026"
}

@article{Haque:2026ero,
    author = "Haque, Shamim and Rezzolla, Luciano and Mallick, Ritam",
    title = "{Dynamical response of twin stars to perturbations}",
    eprint = "2602.02654",
    archivePrefix = "arXiv",
    primaryClass = "gr-qc",
    doi = "10.1103/99c1-2p2y",
    journal = "Phys. Rev. D",
    volume = "113",
    number = "10",
    pages = "103044",
    year = "2026"
}

@article{Cano:2026yrl,
    author = "Cano, Yaiza and Alarc{\'o}n, Jose Manuel",
    title = "{Analyzing Fermionic Dark Matter scenarios with anomalous compact objects}",
    eprint = "2603.22490",
    archivePrefix = "arXiv",
    primaryClass = "hep-ph",
    month = "3",
    year = "2026"
}

@article{Cottam:2002cu,
    author = "Cottam, J. and Paerels, F. and Mendez, M.",
    title = "{Gravitationally redshifted absorption lines in the x-ray burst spectra of a neutron star}",
    eprint = "astro-ph/0211126",
    archivePrefix = "arXiv",
    doi = "10.1038/nature01159",
    journal = "Nature",
    volume = "420",
    pages = "51--54",
    year = "2002"
}

@article{Biesdorf:2024dor,
    author = {Biesdorf, Carline and Schaffner-Bielich, J{\"u}rgen and Tolos, Laura},
    title = "{Masquerading hybrid stars with dark matter}",
    eprint = "2412.05207",
    archivePrefix = "arXiv",
    primaryClass = "hep-ph",
    doi = "10.1103/PhysRevD.111.083038",
    journal = "Phys. Rev. D",
    volume = "111",
    number = "8",
    pages = "083038",
    year = "2025"
}

@article{Lopes:2020dvs,
    author = "Lopes, Luiz L. and Biesdorf, Carline and Marquez, K. D. and Menezes, D{\'e}bora P.",
    title = "{Modified MIT Bag Models -- part II: QCD phase diagram and hot quark stars}",
    eprint = "2009.13552",
    archivePrefix = "arXiv",
    primaryClass = "hep-ph",
    doi = "10.1088/1402-4896/abef35",
    journal = "Phys. Scripta",
    volume = "96",
    number = "6",
    pages = "065302",
    year = "2021"
}

@article{Zdunik:2012dj,
    author = "Zdunik, J. L. and Haensel, P.",
    title = "{Maximum mass of neutron stars and strange neutron-star cores}",
    eprint = "1211.1231",
    archivePrefix = "arXiv",
    primaryClass = "astro-ph.SR",
    doi = "10.1051/0004-6361/201220697",
    journal = "Astron. Astrophys.",
    volume = "551",
    pages = "A61",
    year = "2013"
}

@article{Alford:2013aca,
    author = "Alford, Mark G. and Han, Sophia and Prakash, Madappa",
    title = "{Generic conditions for stable hybrid stars}",
    eprint = "1302.4732",
    archivePrefix = "arXiv",
    primaryClass = "astro-ph.SR",
    doi = "10.1103/PhysRevD.88.083013",
    journal = "Phys. Rev. D",
    volume = "88",
    number = "8",
    pages = "083013",
    year = "2013"
}

@article{Alford:2015gna,
    author = "Alford, Mark G. and Han, Sophia",
    title = "{Characteristics of hybrid compact stars with a sharp hadron-quark interface}",
    eprint = "1508.01261",
    archivePrefix = "arXiv",
    primaryClass = "nucl-th",
    doi = "10.1140/epja/i2016-16062-9",
    journal = "Eur. Phys. J. A",
    volume = "52",
    number = "3",
    pages = "62",
    year = "2016"
}

@article{Kampfer:1981yr,
    author = "Kampfer, Burkhard",
    title = "{On the Possibility of Stable Quark and Pion Condensed Stars}",
    doi = "10.1088/0305-4470/14/11/009",
    journal = "J. Phys. A",
    volume = "14",
    pages = "L471--L475",
    year = "1981"
}

@article{seidov1971stability,
  title={The stability of a star with a phase change in general relativity theory},
  author={Seidov, ZF},
  journal={Soviet Astronomy, Vol. 15, p. 347},
  volume={15},
  pages={347},
  year={1971}
}

@article{Glendenning:1998ag,
    author = "Glendenning, Norman K. and Kettner, Christiane",
    title = "{Nonidentical neutron star twins}",
    eprint = "astro-ph/9807155",
    archivePrefix = "arXiv",
    reportNumber = "LBL-42080, LBNL-42080",
    journal = "Astron. Astrophys.",
    volume = "353",
    pages = "L9",
    year = "2000"
}

@article{Schertler:2000xq,
    author = "Schertler, K. and Greiner, C. and Schaffner-Bielich, J. and Thoma, M. H.",
    title = "{Quark phases in neutron stars and a 'third family' of compact stars as a signature for phase transitions}",
    eprint = "astro-ph/0001467",
    archivePrefix = "arXiv",
    doi = "10.1016/S0375-9474(00)00305-5",
    journal = "Nucl. Phys. A",
    volume = "677",
    pages = "463--490",
    year = "2000"
}

@article{Blaschke:2015uva,
    author = "Blaschke, David and Alvarez-Castillo, David E.",
    editor = "Andrianov, Alexander and Brambilla, Nora and Kim, Victor and Kolevatov, Sergei",
    title = "{High-mass twins {\&} resolution of the reconfinement, masquerade and hyperon puzzles of compact star interiors}",
    eprint = "1503.03834",
    archivePrefix = "arXiv",
    primaryClass = "astro-ph.HE",
    doi = "10.1063/1.4938602",
    journal = "AIP Conf. Proc.",
    volume = "1701",
    number = "1",
    pages = "020013",
    year = "2016"
}

@article{Zacchi:2016tjw,
    author = {Zacchi, Andreas and Tolos, Laura and Schaffner-Bielich, J{\"u}rgen},
    title = "{Twin Stars within the SU(3) Chiral Quark-Meson Model}",
    eprint = "1612.06167",
    archivePrefix = "arXiv",
    primaryClass = "astro-ph.HE",
    doi = "10.1103/PhysRevD.95.103008",
    journal = "Phys. Rev. D",
    volume = "95",
    number = "10",
    pages = "103008",
    year = "2017"
}

@article{Alford:2017qgh,
    author = "Alford, Mark G. and Sedrakian, Armen",
    title = "{Compact stars with sequential QCD phase transitions}",
    eprint = "1706.01592",
    archivePrefix = "arXiv",
    primaryClass = "astro-ph.HE",
    doi = "10.1103/PhysRevLett.119.161104",
    journal = "Phys. Rev. Lett.",
    volume = "119",
    number = "16",
    pages = "161104",
    year = "2017"
}

@article{Christian:2017jni,
    author = {Christian, Jan-Erik and Zacchi, Andreas and Schaffner-Bielich, J{\"u}rgen},
    title = "{Classifications of Twin Star Solutions for a Constant Speed of Sound Parameterized Equation of State}",
    eprint = "1707.07524",
    archivePrefix = "arXiv",
    primaryClass = "astro-ph.HE",
    doi = "10.1140/epja/i2018-12472-y",
    journal = "Eur. Phys. J. A",
    volume = "54",
    number = "2",
    pages = "28",
    year = "2018"
}

@article{Christian:2023hez,
    author = {Christian, Jan-Erik and Schaffner-Bielich, J{\"u}rgen and Rosswog, Stephan},
    title = "{Which first order phase transitions to quark matter are possible in neutron stars?}",
    eprint = "2312.10148",
    archivePrefix = "arXiv",
    primaryClass = "nucl-th",
    doi = "10.1103/PhysRevD.109.063035",
    journal = "Phys. Rev. D",
    volume = "109",
    number = "6",
    pages = "063035",
    year = "2024"
}

@article{Pitz:2023ejc,
    author = {Pitz, Sarah Louisa and Schaffner-Bielich, J{\"u}rgen},
    title = "{Generating ultracompact boson stars with modified scalar potentials}",
    eprint = "2308.01254",
    archivePrefix = "arXiv",
    primaryClass = "astro-ph.HE",
    doi = "10.1103/PhysRevD.108.103043",
    journal = "Phys. Rev. D",
    volume = "108",
    number = "10",
    pages = "103043",
    year = "2023"
}

@article{Pitz:2024xvh,
    author = {Pitz, Sarah Louisa and Schaffner-Bielich, J{\"u}rgen},
    title = "{Generating ultracompact neutron stars with bosonic dark matter}",
    eprint = "2408.13157",
    archivePrefix = "arXiv",
    primaryClass = "astro-ph.HE",
    doi = "10.1103/PhysRevD.111.043050",
    journal = "Phys. Rev. D",
    volume = "111",
    number = "4",
    pages = "043050",
    year = "2025"
}

@article{Colpi:1986ye,
    author = "Colpi, M. and Shapiro, S. L. and Wasserman, I.",
    title = "{Boson Stars: Gravitational Equilibria of Selfinteracting Scalar Fields}",
    doi = "10.1103/PhysRevLett.57.2485",
    journal = "Phys. Rev. Lett.",
    volume = "57",
    pages = "2485--2488",
    year = "1986"
}

@article{Das:2025fyf,
    author = "Das, H. C.",
    title = "{Influence of dark matter on hybrid and twin stars}",
    eprint = "2509.04831",
    archivePrefix = "arXiv",
    primaryClass = "astro-ph.HE",
    doi = "10.1103/fj15-z1rj",
    journal = "Phys. Rev. D",
    volume = "114",
    number = "2",
    pages = "023046",
    year = "2026"
}

@article{Ranea-Sandoval:2015ldr,
    author = "Ranea-Sandoval, Ignacio F. and Han, Sophia and Orsaria, Milva G. and Contrera, Gustavo A. and Weber, Fridolin and Alford, Mark G.",
    title = "{Constant-sound-speed parametrization for Nambu{\textendash}Jona-Lasinio models of quark matter in hybrid stars}",
    eprint = "1512.09183",
    archivePrefix = "arXiv",
    primaryClass = "nucl-th",
    doi = "10.1103/PhysRevC.93.045812",
    journal = "Phys. Rev. C",
    volume = "93",
    number = "4",
    pages = "045812",
    year = "2016"
}

@article{Urbano:2018nrs,
    author = {Urbano, Alfredo and Veerm{\"a}e, Hardi},
    title = "{On gravitational echoes from ultracompact exotic stars}",
    eprint = "1810.07137",
    archivePrefix = "arXiv",
    primaryClass = "gr-qc",
    reportNumber = "CERN-TH-2018-224",
    doi = "10.1088/1475-7516/2019/04/011",
    journal = "JCAP",
    volume = "04",
    pages = "011",
    year = "2019"
}

@article{Pani:2018flj,
    author = "Pani, Paolo and Ferrari, Valeria",
    title = "{On gravitational-wave echoes from neutron-star binary coalescences}",
    eprint = "1804.01444",
    archivePrefix = "arXiv",
    primaryClass = "gr-qc",
    doi = "10.1088/1361-6382/aacb8f",
    journal = "Class. Quant. Grav.",
    volume = "35",
    number = "15",
    pages = "15LT01",
    year = "2018"
}

@article{Mannarelli:2018pjb,
    author = "Mannarelli, M. and Tonelli, F.",
    title = "{Gravitational wave echoes from strange stars}",
    eprint = "1805.02278",
    archivePrefix = "arXiv",
    primaryClass = "gr-qc",
    doi = "10.1103/PhysRevD.97.123010",
    journal = "Phys. Rev. D",
    volume = "97",
    number = "12",
    pages = "123010",
    year = "2018"
}

@article{Kartini:2020ffp,
    author = "Kartini, D. and Sulaksono, A.",
    title = "{Gravitational wave echoes from quark stars}",
    doi = "10.1088/1742-6596/1572/1/012034",
    journal = "J. Phys. Conf. Ser.",
    volume = "1572",
    pages = "012034",
    year = "2020"
}

@article{Cunha:2017qtt,
    author = "Cunha, Pedro V. P. and Berti, Emanuele and Herdeiro, Carlos A. R.",
    title = "{Light-Ring Stability for Ultracompact Objects}",
    eprint = "1708.04211",
    archivePrefix = "arXiv",
    primaryClass = "gr-qc",
    doi = "10.1103/PhysRevLett.119.251102",
    journal = "Phys. Rev. Lett.",
    volume = "119",
    number = "25",
    pages = "251102",
    year = "2017"
}

@article{Olivares:2018abq,
    author = "Olivares, Hector and Younsi, Ziri and Fromm, Christian M. and De Laurentis, Mariafelicia and Porth, Oliver and Mizuno, Yosuke and Falcke, Heino and Kramer, Michael and Rezzolla, Luciano",
    title = "{How to tell an accreting boson star from a black hole}",
    eprint = "1809.08682",
    archivePrefix = "arXiv",
    primaryClass = "gr-qc",
    doi = "10.1093/mnras/staa1878",
    journal = "Mon. Not. Roy. Astron. Soc.",
    volume = "497",
    number = "1",
    pages = "521--535",
    year = "2020"
}

@article{Diedrichs:2023trk,
    author = {Diedrichs, Robin Fynn and Becker, Niklas and Jockel, C{\'e}dric and Christian, Jan-Erik and Sagunski, Laura and Schaffner-Bielich, J{\"u}rgen},
    title = "{Tidal deformability of fermion-boson stars: Neutron stars admixed with ultralight dark matter}",
    eprint = "2303.04089",
    archivePrefix = "arXiv",
    primaryClass = "gr-qc",
    doi = "10.1103/PhysRevD.108.064009",
    journal = "Phys. Rev. D",
    volume = "108",
    number = "6",
    pages = "064009",
    year = "2023"
}

@article{Barbat:2024yvi,
    author = {Barbat, Mikel F. and Schaffner-Bielich, J{\"u}rgen and Tolos, Laura},
    title = "{Comprehensive study of compact stars with dark matter}",
    eprint = "2404.12875",
    archivePrefix = "arXiv",
    primaryClass = "astro-ph.HE",
    doi = "10.1103/PhysRevD.110.023013",
    journal = "Phys. Rev. D",
    volume = "110",
    number = "2",
    pages = "023013",
    year = "2024"
}

@article{Pal:2025chs,
    author = "Pal, Suman and Chaudhuri, Gargi",
    title = "{Can a Hybrid Star with Constant Sound Speed Parameterization Explain the New NICER Mass{\textendash}Radius Measurements?}",
    eprint = "2605.05042",
    archivePrefix = "arXiv",
    primaryClass = "nucl-th",
    doi = "10.3847/1538-4357/adf6a7",
    journal = "Astrophys. J.",
    volume = "991",
    number = "2",
    pages = "158",
    year = "2025"
}

@article{Nelson:2018xtr,
    author = "Nelson, Ann and Reddy, Sanjay and Zhou, Dake",
    title = "{Dark halos around neutron stars and gravitational waves}",
    eprint = "1803.03266",
    archivePrefix = "arXiv",
    primaryClass = "hep-ph",
    reportNumber = "INT-PUB-18-010",
    doi = "10.1088/1475-7516/2019/07/012",
    journal = "JCAP",
    volume = "07",
    pages = "012",
    year = "2019"
}

@article{Raithel:2022aee,
    author = "Raithel, Carolyn A. and Most, Elias R.",
    title = {{Tidal deformability doppelg{\"a}nger: Implications of a low-density phase transition in the neutron star equation of state}},
    eprint = "2208.04295",
    archivePrefix = "arXiv",
    primaryClass = "astro-ph.HE",
    doi = "10.1103/PhysRevD.108.023010",
    journal = "Phys. Rev. D",
    volume = "108",
    number = "2",
    pages = "023010",
    year = "2023"
}

@article{Komoltsev:2021jzg,
    author = "Komoltsev, Oleg and Kurkela, Aleksi",
    title = "{How Perturbative QCD Constrains the Equation of State at Neutron-Star Densities}",
    eprint = "2111.05350",
    archivePrefix = "arXiv",
    primaryClass = "nucl-th",
    doi = "10.1103/PhysRevLett.128.202701",
    journal = "Phys. Rev. Lett.",
    volume = "128",
    number = "20",
    pages = "202701",
    year = "2022"
}

@article{Gorda:2022jvk,
    author = "Gorda, Tyler and Komoltsev, Oleg and Kurkela, Aleksi",
    title = "{Ab-initio QCD Calculations Impact the Inference of the Neutron-star-matter Equation of State}",
    eprint = "2204.11877",
    archivePrefix = "arXiv",
    primaryClass = "nucl-th",
    doi = "10.3847/1538-4357/acce3a",
    journal = "Astrophys. J.",
    volume = "950",
    number = "2",
    pages = "107",
    year = "2023"
}

@article{Kurkela:2009gj,
    author = "Kurkela, Aleksi and Romatschke, Paul and Vuorinen, Aleksi",
    title = "{Cold Quark Matter}",
    eprint = "0912.1856",
    archivePrefix = "arXiv",
    primaryClass = "hep-ph",
    reportNumber = "BI-TP-2009-30, CERN-PH-TH-2009-229, INT-PUB-09-060, TUW-09-19",
    doi = "10.1103/PhysRevD.81.105021",
    journal = "Phys. Rev. D",
    volume = "81",
    pages = "105021",
    year = "2010"
}

@article{Haensel:1999gj,
    author = "Haensel, P. and Lasota, J. P. and Zdunik, J. L.",
    title = "{Maximum redshift and minimum rotation period of neutron stars}",
    eprint = "astro-ph/9905036",
    archivePrefix = "arXiv",
    journal = "Nucl. Phys. B Proc. Suppl.",
    volume = "80",
    pages = "1110",
    year = "2000"
}

@article{Miller:2021qha,
    author = "Miller, M. C. and others",
    title = "{The Radius of PSR J0740+6620 from NICER and XMM-Newton Data}",
    eprint = "2105.06979",
    archivePrefix = "arXiv",
    primaryClass = "astro-ph.HE",
    doi = "10.3847/2041-8213/ac089b",
    journal = "Astrophys. J. Lett.",
    volume = "918",
    number = "2",
    pages = "L28",
    year = "2021"
}

@article{Riley:2021pdl,
    author = "Riley, Thomas E. and others",
    title = "{A NICER View of the Massive Pulsar PSR J0740+6620 Informed by Radio Timing and XMM-Newton Spectroscopy}",
    eprint = "2105.06980",
    archivePrefix = "arXiv",
    primaryClass = "astro-ph.HE",
    doi = "10.3847/2041-8213/ac0a81",
    journal = "Astrophys. J. Lett.",
    volume = "918",
    number = "2",
    pages = "L27",
    year = "2021"
}

@article{Riley:2019yda,
    author = "Riley, Thomas E. and others",
    title = "{A $NICER$ View of PSR J0030+0451: Millisecond Pulsar Parameter Estimation}",
    eprint = "1912.05702",
    archivePrefix = "arXiv",
    primaryClass = "astro-ph.HE",
    doi = "10.3847/2041-8213/ab481c",
    journal = "Astrophys. J. Lett.",
    volume = "887",
    number = "1",
    pages = "L21",
    year = "2019"
}

@article{Miller:2019cac,
    author = "Miller, M.C. and others",
    title = "{PSR J0030+0451 Mass and Radius from $NICER$ Data and Implications for the Properties of Neutron Star Matter}",
    eprint = "1912.05705",
    archivePrefix = "arXiv",
    primaryClass = "astro-ph.HE",
    doi = "10.3847/2041-8213/ab50c5",
    journal = "Astrophys. J. Lett.",
    volume = "887",
    number = "1",
    pages = "L24",
    year = "2019"
}

@article{Mauviard:2025dmd,
    author = "Mauviard, Lucien and others",
    title = "{A NICER View of the 1.4 M$_{⊙}$ Edge-on Pulsar PSR J0614-3329}",
    eprint = "2506.14883",
    archivePrefix = "arXiv",
    primaryClass = "astro-ph.HE",
    doi = "10.3847/1538-4357/ae145d",
    journal = "Astrophys. J.",
    volume = "995",
    number = "1",
    pages = "60",
    year = "2025"
}

@article{Choudhury:2024xbk,
doi = {10.3847/2041-8213/ad5a6f},
url = {https://dx.doi.org/10.3847/2041-8213/ad5a6f},
year = {2024},
month = {aug},
publisher = {The American Astronomical Society},
volume = {971},
number = {1},
pages = {L20},
author = {Devarshi Choudhury and others},
title = {A NICER View of the Nearest and Brightest Millisecond Pulsar: PSR J0437–4715},
journal = {The Astrophysical Journal Letters}
}

@article{Kini:2024ggu,
    author = "Kini, Yves and others",
    title = "{Constraining the properties of the thermonuclear burst oscillation source XTE J1814{\ensuremath{-}}338 through pulse profile modelling}",
    eprint = "2405.10717",
    archivePrefix = "arXiv",
    primaryClass = "astro-ph.HE",
    reportNumber = "MN-24-1065-MJ.R2",
    doi = "10.1093/mnras/stae2398",
    journal = "Mon. Not. Roy. Astron. Soc.",
    volume = "535",
    number = "2",
    pages = "1507--1525",
    year = "2024"
}

@ARTICLE{2022NatAs...6.1444D,
       author = {{Doroshenko}, Victor and {Suleimanov}, Valery and {P{\"u}hlhofer}, Gerd and {Santangelo}, Andrea},
        title = "{A strangely light neutron star within a supernova remnant}",
      journal = {Nature Astronomy},
         year = 2022,
        month = dec,
       volume = {6},
        pages = {1444-1451},
          doi = {10.1038/s41550-022-01800-1},
       adsurl = {https://ui.adsabs.harvard.edu/abs/2022NatAs...6.1444D}
}

@article{LIGOScientific:2017vwq,
    author = "Abbott, B. P. and others",
    collaboration = "LIGO Scientific, Virgo",
    title = "{GW170817: Observation of Gravitational Waves from a Binary Neutron Star Inspiral}",
    eprint = "1710.05832",
    archivePrefix = "arXiv",
    primaryClass = "gr-qc",
    reportNumber = "LIGO-P170817",
    doi = "10.1103/PhysRevLett.119.161101",
    journal = "Phys. Rev. Lett.",
    volume = "119",
    number = "16",
    pages = "161101",
    year = "2017"
}

@article{LIGOScientific:2018cki,
    author = "Abbott, B. P. and others",
    collaboration = "LIGO Scientific, Virgo",
    title = "{GW170817: Measurements of neutron star radii and equation of state}",
    eprint = "1805.11581",
    archivePrefix = "arXiv",
    primaryClass = "gr-qc",
    reportNumber = "LIGO-P1800115",
    doi = "10.1103/PhysRevLett.121.161101",
    journal = "Phys. Rev. Lett.",
    volume = "121",
    number = "16",
    pages = "161101",
    year = "2018"
}

@article{Demorest:2010bx,
    author = "Demorest, Paul and Pennucci, Tim and Ransom, Scott and Roberts, Mallory and Hessels, Jason",
    title = "{Shapiro Delay Measurement of A Two Solar Mass Neutron Star}",
    eprint = "1010.5788",
    archivePrefix = "arXiv",
    primaryClass = "astro-ph.HE",
    doi = "10.1038/nature09466",
    journal = "Nature",
    volume = "467",
    pages = "1081--1083",
    year = "2010"
}

@article{Antoniadis:2013pzd,
    author = "Antoniadis, John and others",
    title = "{A Massive Pulsar in a Compact Relativistic Binary}",
    eprint = "1304.6875",
    archivePrefix = "arXiv",
    primaryClass = "astro-ph.HE",
    doi = "10.1126/science.1233232",
    journal = "Science",
    volume = "340",
    pages = "6131",
    year = "2013"
}

@article{Fonseca:2021wxt,
    author = "Fonseca, E. and others",
    title = "{Refined Mass and Geometric Measurements of the High-mass PSR J0740+6620}",
    eprint = "2104.00880",
    archivePrefix = "arXiv",
    primaryClass = "astro-ph.HE",
    doi = "10.3847/2041-8213/ac03b8",
    journal = "Astrophys. J. Lett.",
    volume = "915",
    number = "1",
    pages = "L12",
    year = "2021"
}

@article{Cardoso:2014sna,
    author = "Cardoso, Vitor and Crispino, Lu{\'\i}s C. B. and Macedo, Caio F. B. and Okawa, Hirotada and Pani, Paolo",
    title = "{Light rings as observational evidence for event horizons: long-lived modes, ergoregions and nonlinear instabilities of ultracompact objects}",
    eprint = "1406.5510",
    archivePrefix = "arXiv",
    primaryClass = "gr-qc",
    doi = "10.1103/PhysRevD.90.044069",
    journal = "Phys. Rev. D",
    volume = "90",
    number = "4",
    pages = "044069",
    year = "2014"
}

@article{Han:2018mtj,
    author = "Han, Sophia and Steiner, Andrew W.",
    title = "{Tidal deformability with sharp phase transitions in (binary) neutron stars}",
    eprint = "1810.10967",
    archivePrefix = "arXiv",
    primaryClass = "nucl-th",
    doi = "10.1103/PhysRevD.99.083014",
    journal = "Phys. Rev. D",
    volume = "99",
    number = "8",
    pages = "083014",
    year = "2019"
}

@article{Most:2018hfd,
    author = {Most, Elias R. and Weih, Lukas R. and Rezzolla, Luciano and Schaffner-Bielich, J{\"u}rgen},
    title = "{New constraints on radii and tidal deformabilities of neutron stars from GW170817}",
    eprint = "1803.00549",
    archivePrefix = "arXiv",
    primaryClass = "gr-qc",
    doi = "10.1103/PhysRevLett.120.261103",
    journal = "Phys. Rev. Lett.",
    volume = "120",
    number = "26",
    pages = "261103",
    year = "2018"
}

@article{Dietrich:2020efo,
    author = "Dietrich, Tim and Coughlin, Michael W. and Pang, Peter T. H. and Bulla, Mattia and Heinzel, Jack and Issa, Lina and Tews, Ingo and Antier, Sarah",
    title = "{Multimessenger constraints on the neutron-star equation of state and the Hubble constant}",
    eprint = "2002.11355",
    archivePrefix = "arXiv",
    primaryClass = "astro-ph.HE",
    reportNumber = "LA-UR-20-21470",
    doi = "10.1126/science.abb4317",
    journal = "Science",
    volume = "370",
    number = "6523",
    pages = "1450--1453",
    year = "2020"
}

@article{Essick:2019ldf,
    author = "Essick, Reed and Landry, Philippe and Holz, Daniel E.",
    title = "{Nonparametric Inference of Neutron Star Composition, Equation of State, and Maximum Mass with GW170817}",
    eprint = "1910.09740",
    archivePrefix = "arXiv",
    primaryClass = "astro-ph.HE",
    doi = "10.1103/PhysRevD.101.063007",
    journal = "Phys. Rev. D",
    volume = "101",
    number = "6",
    pages = "063007",
    year = "2020"
}

@article{Capano:2019eae,
    author = "Capano, Collin D. and Tews, Ingo and Brown, Stephanie M. and Margalit, Ben and De, Soumi and Kumar, Sumit and Brown, Duncan A. and Krishnan, Badri and Reddy, Sanjay",
    title = "{Stringent constraints on neutron-star radii from multimessenger observations and nuclear theory}",
    eprint = "1908.10352",
    archivePrefix = "arXiv",
    primaryClass = "astro-ph.HE",
    reportNumber = "INT-PUB-19-037, LA-UR-19-28442",
    doi = "10.1038/s41550-020-1014-6",
    journal = "Nature Astron.",
    volume = "4",
    number = "6",
    pages = "625--632",
    year = "2020"
}

@article{Fraga:2013qra,
    author = "Fraga, Eduardo S. and Kurkela, Aleksi and Vuorinen, Aleksi",
    title = "{Interacting quark matter equation of state for compact stars}",
    eprint = "1311.5154",
    archivePrefix = "arXiv",
    primaryClass = "nucl-th",
    reportNumber = "CERN-PH-TH-2013-269, HIP-2013-27-TH",
    doi = "10.1088/2041-8205/781/2/L25",
    journal = "Astrophys. J. Lett.",
    volume = "781",
    number = "2",
    pages = "L25",
    year = "2014"
}

@article{Kurkela:2014vha,
    author = {Kurkela, Aleksi and Fraga, Eduardo S. and Schaffner-Bielich, J{\"u}rgen and Vuorinen, Aleksi},
    title = "{Constraining neutron star matter with Quantum Chromodynamics}",
    eprint = "1402.6618",
    archivePrefix = "arXiv",
    primaryClass = "astro-ph.HE",
    reportNumber = "CERN-PH-TH-2014-032, HIP-2014-02-TH",
    doi = "10.1088/0004-637X/789/2/127",
    journal = "Astrophys. J.",
    volume = "789",
    pages = "127",
    year = "2014"
}

@article{Annala:2019puf,
    author = {Annala, Eemeli and Gorda, Tyler and Kurkela, Aleksi and N{\"a}ttil{\"a}, Joonas and Vuorinen, Aleksi},
    title = "{Evidence for quark-matter cores in massive neutron stars}",
    eprint = "1903.09121",
    archivePrefix = "arXiv",
    primaryClass = "astro-ph.HE",
    reportNumber = "CERN-TH-2019-031, HIP-2019-7/TH",
    doi = "10.1038/s41567-020-0914-9",
    journal = "Nature Phys.",
    volume = "16",
    number = "9",
    pages = "907--910",
    year = "2020"
}

@article{Tews:2018kmu,
    author = "Tews, Ingo and Carlson, Joseph and Gandolfi, Stefano and Reddy, Sanjay",
    title = "{Constraining the speed of sound inside neutron stars with chiral effective field theory interactions and observations}",
    eprint = "1801.01923",
    archivePrefix = "arXiv",
    primaryClass = "nucl-th",
    reportNumber = "INT-PUB-18-001, LA-UR-17-31455",
    doi = "10.3847/1538-4357/aac267",
    journal = "Astrophys. J.",
    volume = "860",
    number = "2",
    pages = "149",
    year = "2018"
}

@article{Bedaque:2014sqa,
    author = "Bedaque, Paulo and Steiner, Andrew W.",
    title = "{Sound velocity bound and neutron stars}",
    eprint = "1408.5116",
    archivePrefix = "arXiv",
    primaryClass = "nucl-th",
    reportNumber = "INT-PUB-14-021",
    doi = "10.1103/PhysRevLett.114.031103",
    journal = "Phys. Rev. Lett.",
    volume = "114",
    number = "3",
    pages = "031103",
    year = "2015"
}

@article{Christian:2025dhe,
    author = "Christian, Jan-Erik and Rather, Ishfaq Ahmad and Gholami, Hosein and Hofmann, Marco",
    title = "{Comprehensive analysis of constructing hybrid stars with a renormalization group-consistent Nambu-Jona-Lasino model}",
    eprint = "2503.13626",
    archivePrefix = "arXiv",
    primaryClass = "astro-ph.HE",
    doi = "10.1051/0004-6361/202555009",
    journal = "Astron. Astrophys.",
    volume = "701",
    pages = "A145",
    year = "2025"
}

@article{Cardoso:2019rvt,
    author = "Cardoso, Vitor and Pani, Paolo",
    title = "{Testing the nature of dark compact objects: a status report}",
    eprint = "1904.05363",
    archivePrefix = "arXiv",
    primaryClass = "gr-qc",
    doi = "10.1007/s41114-019-0020-4",
    journal = "Living Rev. Rel.",
    volume = "22",
    number = "1",
    pages = "4",
    year = "2019"
}

@article{Lattimer:2000nx,
    author = "Lattimer, J. M. and Prakash, M.",
    title = "{Neutron star structure and the equation of state}",
    eprint = "astro-ph/0002232",
    archivePrefix = "arXiv",
    doi = "10.1086/319702",
    journal = "Astrophys. J.",
    volume = "550",
    pages = "426",
    year = "2001"
}

@article{Ozel:2016oaf,
    author = {{\"O}zel, Feryal and Freire, Paulo},
    title = "{Masses, Radii, and the Equation of State of Neutron Stars}",
    eprint = "1603.02698",
    archivePrefix = "arXiv",
    primaryClass = "astro-ph.HE",
    doi = "10.1146/annurev-astro-081915-023322",
    journal = "Ann. Rev. Astron. Astrophys.",
    volume = "54",
    pages = "401--440",
    year = "2016"
}

@article{Fornal:2018eol,
    author = "Fornal, Bartosz and Grinstein, Benjamin",
    title = "{Dark Matter Interpretation of the Neutron Decay Anomaly}",
    eprint = "1801.01124",
    archivePrefix = "arXiv",
    primaryClass = "hep-ph",
    doi = "10.1103/PhysRevLett.120.191801",
    journal = "Phys. Rev. Lett.",
    volume = "120",
    number = "19",
    pages = "191801",
    year = "2018",
    note = "[Erratum: Phys.Rev.Lett. 124, 219901 (2020)]"
}

@article{Giangrandi:2025rko,
    author = {Giangrandi, Edoardo and R{\"u}ter, Hannes R. and Kunert, Nina and Emma, Mattia and Abac, Adrian and Adhikari, Ananya and Dietrich, Tim and Sagun, Violetta and Tichy, Wolfgang and Provid{\^e}ncia, Constan{\c{c}}a},
    title = "{Numerical Relativity Simulations of Dark Matter Admixed Binary Neutron Stars}",
    eprint = "2504.20825",
    archivePrefix = "arXiv",
    primaryClass = "astro-ph.HE",
    month = "4",
    year = "2025"
}

@article{Hippert:2022snq,
    author = "Hippert, Maur{\'\i}cio and Dillingham, Emily and Tan, Hung and Curtin, David and Noronha-Hostler, Jacquelyn and Yunes, Nicol{\'a}s",
    title = "{Dark matter or regular matter in neutron stars? How to tell the difference from the coalescence of compact objects}",
    eprint = "2211.08590",
    archivePrefix = "arXiv",
    primaryClass = "astro-ph.HE",
    doi = "10.1103/PhysRevD.107.115028",
    journal = "Phys. Rev. D",
    volume = "107",
    number = "11",
    pages = "115028",
    year = "2023"
}

@article{Chatterjee:2025pkx,
    author = "Chatterjee, Sagnik and Nath, Kamal Krishna",
    title = "{Insights Into Neutron Stars From Gravitational Redshifts and Universal Relations}",
    eprint = "2502.04943",
    archivePrefix = "arXiv",
    primaryClass = "astro-ph.HE",
    doi = "10.1140/epjc/s10052-025-14611-1",
    month = "2",
    year = "2025"
}

@article{Bramante:2023djs,
    author = "Bramante, Joseph and Raj, Nirmal",
    title = "{Dark matter in compact stars}",
    eprint = "2307.14435",
    archivePrefix = "arXiv",
    primaryClass = "hep-ph",
    doi = "10.1016/j.physrep.2023.12.001",
    journal = "Phys. Rept.",
    volume = "1052",
    pages = "1--48",
    year = "2024"
}

@article{Shakeri:2022dwg,
    author = "Shakeri, Soroush and Karkevandi, Davood Rafiei",
    title = "{Bosonic dark matter in light of the NICER precise mass-radius measurements}",
    eprint = "2210.17308",
    archivePrefix = "arXiv",
    primaryClass = "astro-ph.HE",
    doi = "10.1103/PhysRevD.109.043029",
    journal = "Phys. Rev. D",
    volume = "109",
    number = "4",
    pages = "043029",
    year = "2024"
}

@article{PhysRevD.105.123010,
  title = {Tidal deformability of dark matter admixed neutron stars},
  author = {Leung, Kwing-Lam and Chu, Ming-chung and Lin, Lap-Ming},
  journal = {Phys. Rev. D},
  volume = {105},
  issue = {12},
  pages = {123010},
  numpages = {13},
  year = {2022},
  month = {Jun},
  publisher = {American Physical Society},
  doi = {10.1103/PhysRevD.105.123010},
  url = {https://link.aps.org/doi/10.1103/PhysRevD.105.123010}
}

@article{Araujo:2025tlv,
    author = "Araujo, L. F. and Lugones, G. and Lima, J. A. S.",
    title = "{Dark interactions in neutron star interiors: The interplay of baryons, dark matter, and dark energy}",
    eprint = "2509.16484",
    archivePrefix = "arXiv",
    primaryClass = "astro-ph.HE",
    doi = "10.1103/jdlr-p2z6",
    journal = "Phys. Rev. D",
    volume = "112",
    number = "8",
    pages = "083012",
    year = "2025"
}

@article{Arvikar:2025dwl,
    author = "Arvikar, Payaswinee and Gautam, Sakshi and Venneti, Anagh and Banik, Sarmistha",
    title = "{Fermionic versus Bosonic Dark Matter in Neutron Stars: A bayesian study with multi-density constraints}",
    eprint = "2512.13574",
    archivePrefix = "arXiv",
    primaryClass = "astro-ph.CO",
    doi = "10.1088/1475-7516/2026/03/012",
    journal = "JCAP",
    volume = "03",
    pages = "012",
    year = "2026"
}

@Article{sym17101669,
AUTHOR = {Zhang, Naibo and Li, Bao-An and Zhang, Jiayu and Shen, Weina and Zhang, Hui},
TITLE = {Illuminating Dark Matter Admixed in Neutron Stars with Simultaneous Mass–Radius Constraints},
JOURNAL = {Symmetry},
VOLUME = {17},
YEAR = {2025},
NUMBER = {10},
ARTICLE-NUMBER = {1669},
URL = {https://www.mdpi.com/2073-8994/17/10/1669},
ISSN = {2073-8994},
DOI = {10.3390/sym17101669}
}

@article{Perez-Garcia:2010xlt,
    author = "Perez-Garcia, M. Angeles and Silk, Joseph and Stone, Jirina R.",
    title = "{Dark matter, neutron stars and strange quark matter}",
    eprint = "1007.1421",
    archivePrefix = "arXiv",
    primaryClass = "astro-ph.CO",
    doi = "10.1103/PhysRevLett.105.141101",
    journal = "Phys. Rev. Lett.",
    volume = "105",
    pages = "141101",
    year = "2010"
}

@article{PhysRevLett.107.091301,
  title = {Excluding Light Asymmetric Bosonic Dark Matter},
  author = {Kouvaris, Chris and Tinyakov, Peter},
  journal = {Phys. Rev. Lett.},
  volume = {107},
  issue = {9},
  pages = {091301},
  numpages = {4},
  year = {2011},
  month = {Aug},
  publisher = {American Physical Society},
  doi = {10.1103/PhysRevLett.107.091301},
  url = {https://link.aps.org/doi/10.1103/PhysRevLett.107.091301}
}

@article{PhysRevD.77.043515,
  title = {Compact stars as dark matter probes},
  author = {Bertone, Gianfranco and Fairbairn, Malcolm},
  journal = {Phys. Rev. D},
  volume = {77},
  issue = {4},
  pages = {043515},
  numpages = {9},
  year = {2008},
  month = {Feb},
  publisher = {American Physical Society},
  doi = {10.1103/PhysRevD.77.043515},
  url = {https://link.aps.org/doi/10.1103/PhysRevD.77.043515}
}

@article{PhysRevD.82.063531,
  title = {Can neutron stars constrain dark matter?},
  author = {Kouvaris, Chris and Tinyakov, Peter},
  journal = {Phys. Rev. D},
  volume = {82},
  issue = {6},
  pages = {063531},
  numpages = {9},
  year = {2010},
  month = {Sep},
  publisher = {American Physical Society},
  doi = {10.1103/PhysRevD.82.063531},
  url = {https://link.aps.org/doi/10.1103/PhysRevD.82.063531}
}

@article{Karkevandi:2024vov,
    author = "Karkevandi, Davood Rafiei and Shahrbaf, Mahboubeh and Shakeri, Soroush and Typel, Stefan",
    title = "{Exploring the Distribution and Impact of Bosonic Dark Matter in Neutron Stars}",
    eprint = "2402.18696",
    archivePrefix = "arXiv",
    primaryClass = "astro-ph.HE",
    doi = "10.3390/particles7010011",
    journal = "Particles",
    volume = "7",
    number = "1",
    pages = "201--213",
    year = "2024"
}

@article{Karkevandi:2021ygv,
    author = "Karkevandi, Davood Rafiei and Shakeri, Soroush and Sagun, Violetta and Ivanytskyi, Oleksii",
    title = "{Bosonic dark matter in neutron stars and its effect on gravitational wave signal}",
    eprint = "2109.03801",
    archivePrefix = "arXiv",
    primaryClass = "astro-ph.HE",
    doi = "10.1103/PhysRevD.105.023001",
    journal = "Phys. Rev. D",
    volume = "105",
    number = "2",
    pages = "023001",
    year = "2022"
}

@article{Koehn:2024gal,
    author = "Koehn, Hauke and Giangrandi, Edoardo and Kunert, Nina and Somasundaram, Rahul and Sagun, Violetta and Dietrich, Tim",
    title = "{Impact of dark matter on tidal signatures in neutron star mergers with the Einstein Telescope}",
    eprint = "2408.14711",
    archivePrefix = "arXiv",
    primaryClass = "astro-ph.HE",
    reportNumber = "LA-UR-24-28505",
    doi = "10.1103/PhysRevD.110.103033",
    journal = "Phys. Rev. D",
    volume = "110",
    number = "10",
    pages = "103033",
    year = "2024"
}

@article{Sagun:2022ezx,
    author = "Sagun, Violetta and Giangrandi, Edoardo and Ivanytskyi, Oleksii and Provid{\^e}ncia, Constan{\c{c}}a and Dietrich, Tim",
    title = "{How does dark matter affect compact star properties and high density constraints of strongly interacting matter}",
    eprint = "2211.10510",
    archivePrefix = "arXiv",
    primaryClass = "astro-ph.HE",
    doi = "10.1051/epjconf/202227407009",
    journal = "EPJ Web Conf.",
    volume = "274",
    pages = "07009",
    year = "2022"
}

@article{Giangrandi:2022wht,
    author = "Giangrandi, Edoardo and Sagun, Violetta and Ivanytskyi, Oleksii and Provid{\^e}ncia, Constan{\c{c}}a and Dietrich, Tim",
    title = "{The Effects of Self-interacting Bosonic Dark Matter on Neutron Star Properties}",
    eprint = "2209.10905",
    archivePrefix = "arXiv",
    primaryClass = "astro-ph.HE",
    doi = "10.3847/1538-4357/ace104",
    journal = "Astrophys. J.",
    volume = "953",
    number = "1",
    pages = "115",
    year = "2023"
}

@article{Ivanytskyi:2019ojt,
    author = "Ivanytskyi, O. and P{\'e}rez-Garc{\'\i}a, M. {\'A}ngeles and Sagun, V. and Albertus, C.",
    title = "{Second look to the Polyakov loop Nambu{\textendash}Jona-Lasinio model at finite baryonic density}",
    eprint = "1909.07421",
    archivePrefix = "arXiv",
    primaryClass = "hep-ph",
    doi = "10.1103/PhysRevD.100.103020",
    journal = "Phys. Rev. D",
    volume = "100",
    number = "10",
    pages = "103020",
    year = "2019"
}

@article{Jockel:2023rrm,
    author = "Jockel, C{\'e}dric and Sagunski, Laura",
    title = "{Fermion Proca Stars: Vector-Dark-Matter-Admixed Neutron Stars}",
    eprint = "2310.17291",
    archivePrefix = "arXiv",
    primaryClass = "gr-qc",
    doi = "10.3390/particles7010004",
    journal = "Particles",
    volume = "7",
    number = "1",
    pages = "52--79",
    year = "2024"
}

@article{Hajkarim:2024ecp,
    author = {Hajkarim, Fazlollah and Schaffner-Bielich, J{\"u}rgen and Tolos, Laura},
    title = "{Thermodynamic consistent description of compact stars of two interacting fluids: the case of neutron stars with Higgs portal dark matter}",
    eprint = "2412.04585",
    archivePrefix = "arXiv",
    primaryClass = "hep-ph",
    doi = "10.1088/1475-7516/2025/08/070",
    journal = "JCAP",
    volume = "08",
    pages = "070",
    year = "2025"
}

@article{Shirke:2023ktu,
    author = {Shirke, Swarnim and Ghosh, Suprovo and Chatterjee, Debarati and Sagunski, Laura and Schaffner-Bielich, J{\"u}rgen},
    title = "{R-modes as a new probe of dark matter in neutron stars}",
    eprint = "2305.05664",
    archivePrefix = "arXiv",
    primaryClass = "astro-ph.HE",
    reportNumber = "LIGO-P2300140",
    doi = "10.1088/1475-7516/2023/12/008",
    journal = "JCAP",
    volume = "12",
    pages = "008",
    year = "2023"
}

@article{Cassing:2022tnn,
    author = {Cassing, Marie and Brisebois, Alexander and Azeem, Muhammad and Schaffner-Bielich, J{\"u}rgen},
    title = "{Exotic Compact Objects with Two Dark Matter Fluids}",
    eprint = "2210.13697",
    archivePrefix = "arXiv",
    primaryClass = "gr-qc",
    doi = "10.3847/1538-4357/acb3be",
    journal = "Astrophys. J.",
    volume = "944",
    number = "2",
    pages = "130",
    year = "2023"
}

@article{Wystub:2021qrn,
    author = {Wystub, Stephan and Dengler, Yannick and Christian, Jan-Erik and Schaffner-Bielich, J{\"u}rgen},
    title = "{Constraining exotic compact stars composed of bosonic and fermionic dark matter with gravitational wave events}",
    eprint = "2110.12972",
    archivePrefix = "arXiv",
    primaryClass = "astro-ph.HE",
    doi = "10.1093/mnras/stad633",
    journal = "Mon. Not. Roy. Astron. Soc.",
    volume = "521",
    number = "1",
    pages = "1393--1398",
    year = "2023"
}

@article{PhysRevD.92.123002,
  title = {Dark compact planets},
  author = {Tolos, Laura and Schaffner-Bielich, J\"urgen},
  journal = {Phys. Rev. D},
  volume = {92},
  issue = {12},
  pages = {123002},
  numpages = {5},
  year = {2015},
  month = {Dec},
  publisher = {American Physical Society},
  doi = {10.1103/PhysRevD.92.123002},
  url = {https://link.aps.org/doi/10.1103/PhysRevD.92.123002}
}

@article{PhysRevD.93.083009,
  title = {Quark stars admixed with dark matter},
  author = {Mukhopadhyay, Payel and Schaffner-Bielich, J\"urgen},
  journal = {Phys. Rev. D},
  volume = {93},
  issue = {8},
  pages = {083009},
  numpages = {10},
  year = {2016},
  month = {Apr},
  publisher = {American Physical Society},
  doi = {10.1103/PhysRevD.93.083009},
  url = {https://link.aps.org/doi/10.1103/PhysRevD.93.083009}
}

@ARTICLE{2012PhLB..711....6P,
       author = {{P{\'e}rez-Garc{\'\i}a}, M. {\'A}ngeles and {Silk}, Joseph},
        title = "{Dark matter seeding and the kinematics and rotation of neutron stars}",
      journal = {Physics Letters B},
         year = 2012,
        month = may,
       volume = {711},
       number = {1},
        pages = {6-9},
          doi = {10.1016/j.physletb.2012.03.065},
archivePrefix = {arXiv},
       eprint = {1111.2275},
 primaryClass = {astro-ph.CO},
       adsurl = {https://ui.adsabs.harvard.edu/abs/2012PhLB..711....6P}
}

@article{PhysRevD.99.063015,
  title = {Dark compact objects: An extensive overview},
  author = {Deliyergiyev, Maksym and Del Popolo, Antonino and Tolos, Laura and Le Delliou, Morgan and Lee, Xiguo and Burgio, Fiorella},
  journal = {Phys. Rev. D},
  volume = {99},
  issue = {6},
  pages = {063015},
  numpages = {17},
  year = {2019},
  month = {Mar},
  publisher = {American Physical Society},
  doi = {10.1103/PhysRevD.99.063015},
  url = {https://link.aps.org/doi/10.1103/PhysRevD.99.063015}
}

@ARTICLE{2012APh....37...70L,
       author = {{Li}, Ang and {Huang}, Feng and {Xu}, Ren-Xin},
        title = "{Too massive neutron stars: The role of dark matter?}",
      journal = {Astroparticle Physics},
         year = 2012,
        month = sep,
       volume = {37},
        pages = {70-74},
          doi = {10.1016/j.astropartphys.2012.07.006},
archivePrefix = {arXiv},
       eprint = {1208.3722},
 primaryClass = {astro-ph.SR},
       adsurl = {https://ui.adsabs.harvard.edu/abs/2012APh....37...70L}
}

@ARTICLE{2009APh....32..278S,
       author = {{Sandin}, Fredrik and {Ciarcelluti}, Paolo},
        title = "{Effects of mirror dark matter on neutron stars}",
      journal = {Astroparticle Physics},
         year = 2009,
        month = dec,
       volume = {32},
       number = {5},
        pages = {278-284},
          doi = {10.1016/j.astropartphys.2009.09.005},
archivePrefix = {arXiv},
       eprint = {0809.2942},
 primaryClass = {astro-ph},
       adsurl = {https://ui.adsabs.harvard.edu/abs/2009APh....32..278S}
}

@article{PhysRevD.84.107301,
  title = {Dark-matter admixed neutron stars},
  author = {Leung, S.-C. and Chu, M.-C. and Lin, L.-M.},
  journal = {Phys. Rev. D},
  volume = {84},
  issue = {10},
  pages = {107301},
  numpages = {5},
  year = {2011},
  month = {Nov},
  publisher = {American Physical Society},
  doi = {10.1103/PhysRevD.84.107301},
  url = {https://link.aps.org/doi/10.1103/PhysRevD.84.107301}
}

@article{PhysRevC.89.025803,
  title = {Effects of fermionic dark matter on properties of neutron stars},
  author = {Xiang, Qian-Fei and Jiang, Wei-Zhou and Zhang, Dong-Rui and Yang, Rong-Yao},
  journal = {Phys. Rev. C},
  volume = {89},
  issue = {2},
  pages = {025803},
  numpages = {11},
  year = {2014},
  month = {Feb},
  publisher = {American Physical Society},
  doi = {10.1103/PhysRevC.89.025803},
  url = {https://link.aps.org/doi/10.1103/PhysRevC.89.025803}
}

@article{NANOGrav:2023hde,
    author = "Agazie, Gabriella and others",
    collaboration = "NANOGrav",
    title = "{The NANOGrav 15 yr Data Set: Observations and Timing of 68 Millisecond Pulsars}",
    eprint = "2306.16217",
    archivePrefix = "arXiv",
    primaryClass = "astro-ph.HE",
    doi = "10.3847/2041-8213/acda9a",
    journal = "Astrophys. J. Lett.",
    volume = "951",
    number = "1",
    pages = "L9",
    year = "2023"
}

@article{Sun:2023cqr,
    author = "Sun, Hongyi and Wen, Dehua",
    title = "{New criterion for the existence of dark matter in neutron stars}",
    eprint = "2312.17288",
    archivePrefix = "arXiv",
    primaryClass = "astro-ph.HE",
    doi = "10.1103/PhysRevD.109.123037",
    journal = "Phys. Rev. D",
    volume = "109",
    number = "12",
    pages = "123037",
    year = "2024"
}

@article{Vikiaris:2026ofd,
    author = "Vikiaris, M. and Petousis, V. and Veselsky, M. and Moustakidis, Ch. C.",
    title = "{Neutron dark decay and exotic compact objects}",
    eprint = "2602.04477",
    archivePrefix = "arXiv",
    primaryClass = "nucl-th",
    doi = "10.1103/21p6-sqfh",
    journal = "Phys. Rev. D",
    volume = "114",
    number = "4",
    pages = "043032",
    year = "2026"
}

@article{Matt:2019llr,
    author = "Matt, Giorgio",
    editor = {Strassmeier, K. G. and Brandenburg, A. and Cuntz, M. and Hasinger, G. and Montmerle, T. and Neuh{\"a}user, R.},
    collaboration = "Athena Science Study Team",
    title = "{The advanced telescope for high energy astrophysics}",
    doi = "10.1002/asna.201913555",
    journal = "Astron. Nachr.",
    volume = "340",
    number = "1-3",
    pages = "35--39",
    year = "2019"
}

@article{eXTP:2018anb,
    author = "Zhang, Shuang-Nan and others",
    collaboration = "eXTP",
    title = "{The enhanced X-ray Timing and Polarimetry mission{\textemdash}eXTP}",
    eprint = "1812.04020",
    archivePrefix = "arXiv",
    primaryClass = "astro-ph.IM",
    doi = "10.1007/s11433-018-9309-2",
    journal = "Sci. China Phys. Mech. Astron.",
    volume = "62",
    number = "2",
    pages = "29502",
    year = "2019"
}

@article{ET:2025xjr,
    author = "Abac, Adrian and others",
    collaboration = "ET",
    title = "{The Science of the Einstein Telescope}",
    eprint = "2503.12263",
    archivePrefix = "arXiv",
    primaryClass = "gr-qc",
    reportNumber = "ET-0036C-25",
    doi = "10.1088/1475-7516/2026/03/081",
    journal = "JCAP",
    volume = "03",
    pages = "081",
    year = "2026"
}

@article{Mauviard:2026gzc,
    author = "Mauviard, Lucien and others",
    title = "{A NICER view of PSR J1614$-$2230: a massive and compact millisecond pulsar}",
    eprint = "2609.00172",
    archivePrefix = "arXiv",
    primaryClass = "astro-ph.HE",
    month = "8",
    year = "2026"
}

@article{Postnikov:2010yn,
    author = "Postnikov, Sergey and Prakash, Madappa and Lattimer, James M.",
    title = "{Tidal Love Numbers of Neutron and Self-Bound Quark Stars}",
    eprint = "1004.5098",
    archivePrefix = "arXiv",
    primaryClass = "astro-ph.SR",
    doi = "10.1103/PhysRevD.82.024016",
    journal = "Phys. Rev. D",
    volume = "82",
    pages = "024016",
    year = "2010"
}

@article{Damour:2009vw,
    author = "Damour, Thibault and Nagar, Alessandro",
    title = "{Relativistic tidal properties of neutron stars}",
    eprint = "0906.0096",
    archivePrefix = "arXiv",
    primaryClass = "gr-qc",
    doi = "10.1103/PhysRevD.80.084035",
    journal = "Phys. Rev. D",
    volume = "80",
    pages = "084035",
    year = "2009"
}

@article{Dengler:2025ntz,
    author = "Dengler, Yannick and Kulkarni, Suchita and Maas, Axel and Radl, Kevin",
    title = "{Strongly interacting dark matter admixed neutron stars}",
    eprint = "2503.19691",
    archivePrefix = "arXiv",
    primaryClass = "hep-ph",
    doi = "10.21468/SciPostPhysCore.9.2.019",
    journal = "SciPost Phys. Core",
    volume = "9",
    pages = "019",
    year = "2026"
}

@article{Ho:2006uk,
    author = "Ho, Wynn C. G. and Kaplan, David L. and Chang, Philip and van Adelsberg, Matthew and Potekhin, Alexander Y.",
    title = "{Magnetic Hydrogen Atmosphere Models and the Neutron Star RX J1856.5-3754}",
    eprint = "astro-ph/0612145",
    archivePrefix = "arXiv",
    reportNumber = "SLAC-PUB-12255",
    doi = "10.1111/j.1365-2966.2006.11376.x",
    journal = "Mon. Not. Roy. Astron. Soc.",
    volume = "375",
    pages = "821--830",
    year = "2007"
}

@article{PhysRevD.107.023012,
  title = {Twin stars as probes of the nuclear equation of state: Effects of rotation through the PSR J0952-0607 pulsar and constraints via the tidal deformability from the GW170817 event},
  author = {Tsaloukidis, Lazaros and Koliogiannis, P. S. and Kanakis-Pegios, A. and Moustakidis, Ch. C.},
  journal = {Phys. Rev. D},
  volume = {107},
  issue = {2},
  pages = {023012},
  numpages = {18},
  year = {2023},
  month = {Jan},
  publisher = {American Physical Society},
  doi = {10.1103/PhysRevD.107.023012},
  url = {https://link.aps.org/doi/10.1103/PhysRevD.107.023012}
}

@article{Li:2026pxw,
    author = "Li, Zhi-yao and Jiang, Wei-zhou and Sun, Dian",
    title = "{Radii and Scaling Relations of Dark-Matter-Admixed Neutron Stars}",
    eprint = "2610.06934",
    archivePrefix = "arXiv",
    primaryClass = "nucl-th",
    month = "10",
    year = "2026"
}

@article{Shawqi:2024jmk,
    author = "Shawqi, Shafayat and Morsink, Sharon M.",
    title = "{Interpreting Mass and Radius Measurements of Neutron Stars with Dark Matter Halos}",
    eprint = "2406.03332",
    archivePrefix = "arXiv",
    primaryClass = "astro-ph.HE",
    reportNumber = "INT-PUB-24-025",
    doi = "10.3847/1538-4357/ad77c1",
    journal = "Astrophys. J.",
    volume = "975",
    number = "1",
    pages = "123",
    year = "2024"
}

@article{Thakur:2026pes,
    author = "Thakur, Prashant and Rather, Ishfaq Ahmad and Lim, Y.",
    title = "{Dark matter in neutron stars: two-fluid $f$-mode oscillations in full general relativity}",
    eprint = "2610.04282",
    archivePrefix = "arXiv",
    primaryClass = "astro-ph.HE",
    month = "10",
    year = "2026"
}

@article{Gholami:2024ety,
    author = {Gholami, Hosein and Rather, Ishfaq Ahmad and Hofmann, Marco and Buballa, Michael and Schaffner-Bielich, J{\"u}rgen},
    title = "{Astrophysical constraints on color-superconducting phases in compact stars within the RG-consistent NJL model}",
    eprint = "2411.04064",
    archivePrefix = "arXiv",
    primaryClass = "hep-ph",
    doi = "10.1103/PhysRevD.111.103034",
    journal = "Phys. Rev. D",
    volume = "111",
    number = "10",
    pages = "103034",
    year = "2025"
}

@article{Bondi:1964zz,
    author = "Bondi, H.",
    title = "{Massive spheres in general relativity}",
    doi = "10.1098/rspa.1964.0234",
    journal = "Proc. Roy. Soc. Lond. A",
    volume = "282",
    pages = "303--317",
    year = "1964"
}

@article{Buchdahl:1959zz,
    author = "Buchdahl, Hans A.",
    title = "{General Relativistic Fluid Spheres}",
    doi = "10.1103/PhysRev.116.1027",
    journal = "Phys. Rev.",
    volume = "116",
    pages = "1027",
    year = "1959"
}

@article{Arrechea:2024vxp,
    author = "Arrechea, Julio and Barcel{\'o}, Carlos and Garc{\'\i}a-Moreno, Gerardo and Polo-G{\'o}mez, Jos{\'e}",
    title = "{Beyond Buchdahl{\textquoteright}s limit: Bilayered stars and thin-shell configurations}",
    eprint = "2411.14018",
    archivePrefix = "arXiv",
    primaryClass = "gr-qc",
    doi = "10.1103/dqbd-9z5f",
    journal = "Phys. Rev. D",
    volume = "111",
    number = "12",
    pages = "124017",
    year = "2025"
}

@misc{ultra_compact_twin_stars,
  title = {Ultra-compact twin stars with hybrid equations of state from bosonic dark matter — data release},
  author = {Rather, Ishfaq Ahmad and Pitz, Sarah Louisa and Schaffner-Bielich, Jürgen},
  year = {2026},
  publisher = {GitHub},
  url = {https://github.com/ishfaq333/ultracompact-twin-stars-dm}
}
\appendix
\section{Results for $\boldsymbol{c^2_\text{QM}=0.7}$}
\label{app:sos07}

\begin{figure}
    \centering
    \includegraphics[width=\columnwidth]{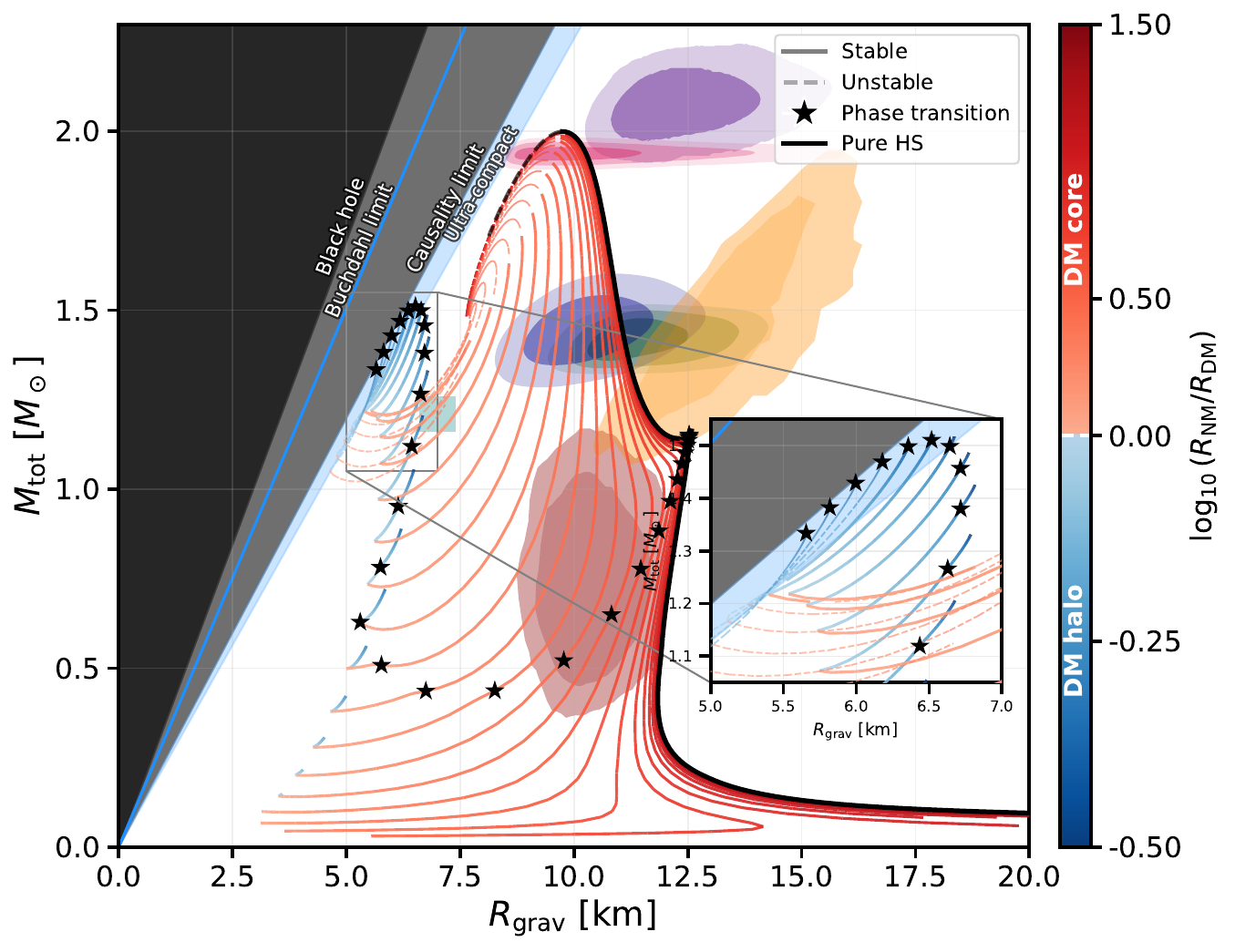}
    \captionsetup{justification=justified, singlelinecheck=false}
    \caption{Mass--gravitational radius sequences for $(P_t,\Delta\varepsilon)=(30,250)\,\text{MeV/fm}^3$, $c_\text{QM}^2=0.7$, $(m_b,n)=(300\,\text{MeV},40)$. Colour encodes $\log_{10}(R_{\rm NM}/R_{\rm DM})$ (red: DM core, blue: DM halo). The inset zooms into the UCO region.}
    \label{fig:MR_07}
\end{figure}

\begin{figure}
    \centering
    \includegraphics[width=\columnwidth]{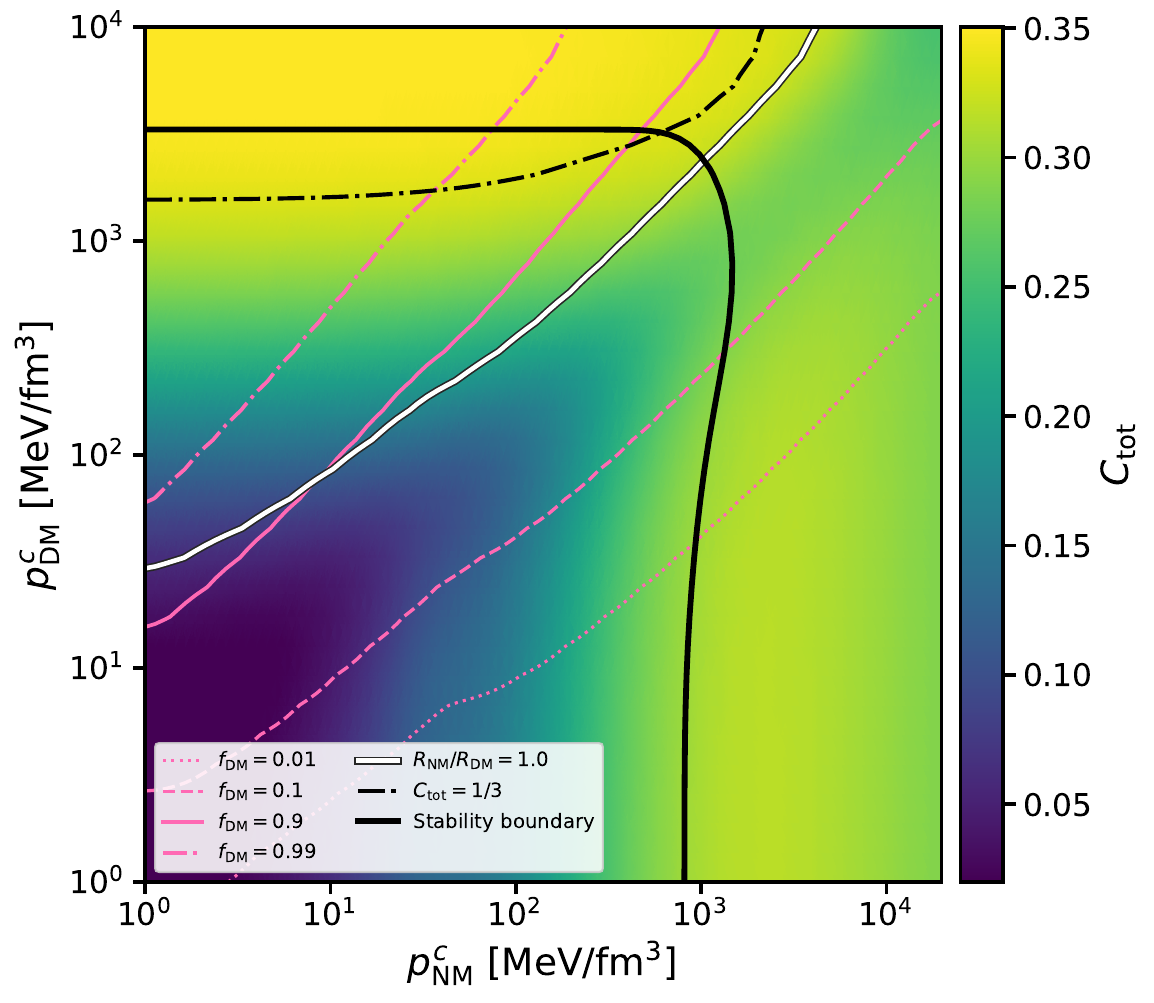}
    \captionsetup{justification=justified, singlelinecheck=false}
    \caption{Total compactness $C_{\rm tot}$ in the $(p^c_{\rm NM},\,p^c_{\rm DM})$ plane for $(P_t,\Delta\varepsilon)=(30,250)\,\text{MeV/fm}^3$, $c_\text{QM}^2=0.7$, $(m_b,n)=(300\,\text{MeV},40)$. Pink iso-lines show $R_{\rm NM}/R_{\rm DM}=0.5,\,1.0,\,2.0$. The $C=1/3$ boundary (dash-dot) and stability boundary (solid) are as in Fig.~\ref{fig:contour1}.}
    \label{fig:contour_07}
\end{figure}

\cref{fig:MR_07,fig:contour_07} show the MR sequences and compactness contour map for the representative case $(P_t,\Delta\varepsilon)=(30,250)\,\text{MeV/fm}^3$ with stiff DM $(m_b,n)=(300\,\text{MeV},40)$. Reducing $c^2_\text{QM}$ from 1.0 to 0.7 softens the quark-matter branch, with two immediate consequences. First, the maximum mass on the hybrid branch drops from $\sim2.3\,M_\odot$ to $\sim2.0\,M_\odot$, driven by the lower pressure support in the quark phase. Second, the twin-star radius gap $\Delta R_{\rm NM}$ narrows, as the quark-matter branch is pulled to smaller radii while the hadronic branch is unchanged. In the contour map (\cref{fig:contour_07}), the $C=1/3$ boundary shifts: the DM-halo UCO region at high $p^c_{\rm DM}$ persists and is essentially unchanged; DM self-gravity alone controls this population and is insensitive to $c^2_\text{QM}$, but the DM-core UCO island in the lower-right (already absent for the $c^2_\text{QM}=1$, $P_t=30$ MeV/fm$^3$ case) remains absent. This confirms that DM-core UCOs require both an early-onset phase transition and a sufficiently stiff quark EoS.

The morphological structure (core-halo transition, stability-boundary notch from the phase transition) is qualitatively identical to the $c_\text{QM}^2=1$ results. All key qualitative features discussed in the main text, DM-halo UCO branch, dark dopplegangers, stabilization of the phase transition, survive the reduction in $c_\text{QM}^2$. Quantitatively, the observable signatures are reduced in amplitude: dark doppleganger $\Lambda$ spreads are somewhat smaller, and the maximum redshift in the DM-halo UCO region is slightly lower. Lowering $c_\text{QM}^2$ further toward the conformal limit $1/3$ would progressively suppress the quark-matter branch: twin stars require $\Delta\varepsilon > \Delta\varepsilon_{\rm Seidov}(P_t)$, which at $c_\text{QM}^2\to1/3$ is increasingly hard to satisfy at low $P_t$, and the hybrid branch would eventually merge back with the hadronic sequence. The $c^2_\text{QM}=1$ and $c^2_\text{QM}=0.7$ results therefore bracket the phenomenologically relevant range.

\end{document}